\documentclass[a4paper,12pt]{article}
\usepackage[
bookmarks=true,bookmarksnumbered=true,
colorlinks = true,anchorcolor= blue, 
linkcolor = darkred, citecolor = darkgreen
]{hyperref}
\usepackage{amsfonts,amsthm,amsmath,amssymb,graphicx}
\usepackage{braket}
\usepackage{bm}
\usepackage{float}
\usepackage{mathrsfs}
\usepackage{booktabs}
\usepackage{makeidx}
\usepackage{color}
\usepackage{tikz}
\usepackage{here}
\numberwithin{equation}{section}
\usepackage{color}
    \definecolor{darkgreen}{rgb}{0,0.5,0}
     \definecolor{darkblue}{rgb}{0,0,0.6}
    \definecolor{purple}{rgb}{0.4,0.2,0.7}
    \definecolor{darkred}{rgb}{0.7,0,0}
    \definecolor{airforceblue}{rgb}{0.36, 0.54, 0.66}
    \definecolor{cyan}{rgb}{0.0, 1.0, 1.0}
  	\definecolor{cyan(process)}{rgb}{0.0, 0.72, 0.92}

\usepackage{tocloft}
\hypersetup{linktoc=all}

\makeatletter
\renewcommand{\cftsecpagefont}{\HyColor@HyperrefColor{darkred}\@linkcolor}
\renewcommand{\cftsubsecpagefont}{\HyColor@HyperrefColor{darkred}\@linkcolor}
\renewcommand{\cftsubsubsecpagefont}{\HyColor@HyperrefColor{darkred}\@linkcolor}
\makeatother

\usepackage{comment}

\newcommand{\be}{\begin{equation}}
\newcommand{\ee}{\end{equation}}
\newcommand{\ba}{\begin{eqnarray}}
\newcommand{\ea}{\end{eqnarray}}
\newcommand{\f}{\frac}
\newcommand{\s}{\sqrt}

\DeclareMathOperator{\Tr}{Tr}

\def\ba#1\ea{\begin{align}#1\end{align}}

 \def\vep{\varepsilon}

 \def\f {\frac}

\usepackage{geometry}
\begin{document}

\newcommand{\hiduke}[1]{\hspace{\fill}{\small [{#1}]}}
\newcommand{\aff}[1]{${}^{#1}$}
\renewcommand{\thefootnote}{\fnsymbol{footnote}}

\begin{titlepage}
\begin{flushright}
{\footnotesize MIT-CTP/5259}
\end{flushright}
\begin{center}
{\Large\bf
Four coupled SYK models and Nearly AdS$_2$ gravities: Phase Transitions in Traversable wormholes and in Braket wormholes
}\\
\bigskip\bigskip
\bigskip\bigskip
{\large Tokiro Numasawa\footnote{\tt numasawa@mit.edu}} \\ 
\bigskip\bigskip
{\small
\it Center for Theoretical Physics,
Massachusetts Institute of Technology, Cambridge, MA 02139, USA
}\\
\end{center}
\bigskip
\bigskip
\begin{abstract}
We study four coupled SYK models and nearly AdS$_2$ gravities.
In the SYK model side, we construct a model that couples two copies of two coupled SYK models.
In nearly AdS$_2$ gravity side, we entangle matter fields in two copies of traversable wormholes.
In both cases, the systems show first order phase transitions at zero temperature by changing couplings, which is understood as the exchange of traversable wormhole configurations.
In nearly AdS$_2$ gravity cases, by exchanging the role of space and time the wormholes are interpreted as braket wormholes.
In Lorentzian signature, these braket wormholes lead to two closed universes that are entangled with each other as well as matter fields in the flat space without dynamical gravity.
We study the effect of projection or entangling operation for matters on flat spaces and they cause phase transitions in braket wormholes, which leads to the pair annihilation of closed universes.
Using these braket wormholes, we discuss the way to embed states  in  2d holographic CFTs into Hilbert space  of many 2d free fields.

\end{abstract}

\bigskip\bigskip\bigskip

\end{titlepage}

\renewcommand{\thefootnote}{\arabic{footnote}}
\setcounter{footnote}{0}

{\hypersetup{colorlinks=true, linkcolor=black, filecolor = magenta, urlcolor=cyan}
\tableofcontents
}

\newpage

\section{Introduction }
The wormholes are interesting configurations of spacetime.
They are closely related to quantum entanglement \cite{Israel:1976ur,Maldacena:2001kr,Maldacena:2013xja,VanRaamsdonk:2010pw}.
In quantum mechanics, only having entangle states is not enough to send information.
Similarly, only having wormholes does not imply that we can send a message through a wormhole.
Rather, the average null energy condition (ANEC) \cite{Hartman:2016lgu,Faulkner:2016mzt} does not allow to make such a traversable wormhole.
One way to violate the ANEC is to use quantum effects like the Casimir energy.
Actually, some controllable examples of traversable wormholes are found \cite{Gao:2016bin,Maldacena:2017axo,Maldacena:2018gjk,Fu:2018oaq} recently.
For example,  we can make traversable wormholes by introducing the direct couplings between two asymptotic boundaries of the eternal black holes in AdS \cite{Gao:2016bin,Maldacena:2017axo} , which violates the ANEC.
Interestingly, we can also find traversable wormhole solutions in four dimensions \cite{Maldacena:2018gjk,Fu:2018oaq}.
In four dimensions, these direct coupling can arise from the local and causal dynamics.
The solutions have a nontrivial topology that can be detectable from the  observers who are in the asymptotical infinity.
These configurations are prohibited from the topological censorship \cite{Friedman:1993ty,Galloway:1999bp,Galloway:1999br} at classical level, but quantum effects enable us to construct such a configuration.

We can also realize a state that is similar to the eternal traversable wormholes \cite{Maldacena:2017axo,Maldacena:2018lmt} in the Sachdev-Ye-Kitaev model.
The Sachdev-Ye-Kitaev (SYK) model \cite{PhysRevLett.70.3339, KitaevTalk} is a strongly interacting quantum mechanical model but is still solvable in the large $N$ limit.
One of the remarkable properties of this model is that at low energies the theory is described by the conformal symmetry that is broken explicitly and spontaneously \cite{KitaevTalk,Maldacena:2016hyu}. 
The low energy effective action is known as the Schwartzian action \cite{Maldacena:2016hyu, Maldacena:2016upp}.
This action also appears as a low energy description of near extremal black holes \cite{Maldacena:2016upp,Kitaev:2018wpr,Yang:2018gdb}.
The theory which describes the low energy description of near extremal black holes is known as the nearly AdS$_2$ gravity \cite{Maldacena:2016upp}.
Especially, eternal traversable wormholes in nearly AdS$_2$ can be constructed based on this Schwartzian action \cite{Maldacena:2018lmt}.
The two cite SYK model with a sort of double trace deformation and its ground state is also analyzed exactly in the same manner with the eternal traversable wormholes in nearly AdS$_2$ \cite{Maldacena:2018lmt}, which motivates us to call the ground state "SYK traversable wormhole" \cite{Maldacena:2019ufo,Nosaka:2020nuk}. 

Euclidean wormholes, which are other kinds of wormholes in Euclidean signature that connect more than two asymptotic boundaries, also play important roles.
Recently Euclidean wormholes in the calculation of entanglement entropy, which are known as  replica wormholes \cite{Penington:2019kki,Almheiri:2019qdq}, play an important role to reproduce the Page curve \cite{Page:1993df} from semiclassical gravity calculation \cite{Almheiri:2019psf,Penington:2019npb,Anegawa:2020ezn,Hashimoto:2020cas,Hartman:2020khs,Hartman:2020swn,Balasubramanian:2020xqf,Balasubramanian:2020xqf,Almheiri:2020cfm,Krishnan:2020fer,Krishnan:2020oun}.
Euclidean wormholes are sometimes confusing objects in the AdS/CFT correspondence \cite{Maldacena:1997re} because they give correlations between partition functions and the factorization is not manifest \cite{Maldacena:2004rf,ArkaniHamed:2007js}.
There is an old discussion on the connection between Euclidean wormholes and ensemble averages \cite{Coleman:1988cy,Giddings:1988cx} and recently it was found that 2d pure dilaton gravity is equivalent to quantum mechanics with random Hamiltonians \cite{Saad:2019lba}.
There are further recent discussions on the ensemble average and quantum gravity \cite{Engelhardt:2020qpv,Marolf:2020xie,Balasubramanian:2020jhl,Belin:2020hea,McNamara:2020uza,Stanford:2020wkf,Bousso:2020kmy,Peng:2020rno} and average of conformal field theory \cite{Maloney:2020nni,Afkhami-Jeddi:2020ezh,Cotler:2020ugk,Perez:2020klz}.

\begin{figure}[ht]
\begin{center}
\includegraphics[width=10cm]{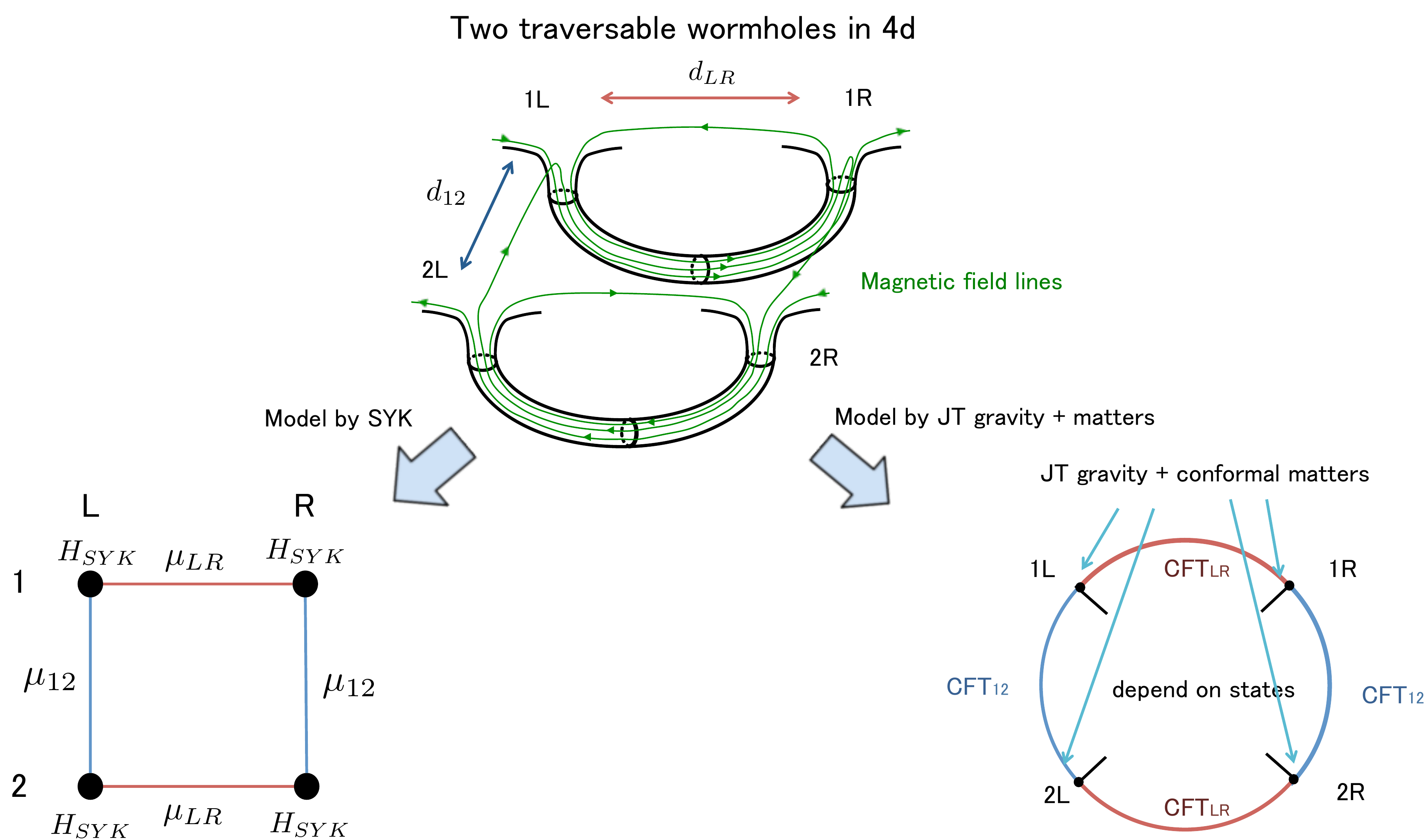}
\caption{The schematic form of two traversable wormholes in four dimensions.
Each traversable wormhole is the one in \cite{Maldacena:2018gjk,Maldacena:2020sxe}.
We model this 4d setup both in SYK models and Nearly AdS$_2$ gravities.
} 
\label{fig:TwoWormholesSetup}
\end{center}
\end{figure}

In this paper, we study the coupling of two traversable wormholes both in the SYK models and nearly AdS$_2$ gravities.
The motivation is to study what happens when we entangle matter fields in different spacetimes.
We model the situation of two traversable wormholes in four dimensions both in the SYK model and in Nearly AdS$_2$ gravity, see figure \ref{fig:TwoWormholesSetup}.
This is achieved by further introducing the double trace deformation between two traversable wormholes.
In nearly AdS$_2$ gravities, we consider the matter fields a part of which is living on a traversable wormhole and another part is living on a different traversable wormhole.   
In nearly AdS$_2$ gravity setup, by exchanging the role of space and time in this configuration, we can think of the configuration as two braket wormholes, which is introduced in \cite{Page:1986vw} and studied further in  \cite{Giddings:2020yes,Anous:2020lka,Chen:2020tes}. 
In Loretzian signature, two braket wormholes lead to two closed universes that are entangled with each other as well as other matters in flat spaces without dynamical gravity.
We vary the pattern of entanglement and study how spacetime changes.
As a bonus, the braket wormhole configuration gives a way to embed 2d holographic CFT state into the Hilbert space of many free CFTs.

Both in the SYK models and Nearly AdS$_2$ gravities, we found first order phase transitions at zero temperature when we change the couplings, which causes the change of entanglement.
These first order phase transitions are caused by the exchange of the dominant wormhole configurations.
They lead to a $\mathbb{Z}_2$ symmetry breaking by each wormhole configuration at a special point of coupling constants, which is characterized by an order parameter.

After exchanging the role of time and space, the above transitions are interpreted as the transition in braket wormholes.
By entangling operations or partial projections, braket wormholes can annihilate in the Euclidean regime before reaching $t = 0$ slice and disappear from the Lorentzian geometries.
This disappearance of braket wormholes plays an important role to keep the unitarity for the matter fields in the no gravity regions.

This paper is organized as follows.
In section \ref{sec:Review}, we review traversable wormholes in the SYK model and in nearly AdS$_2$ gravity.
In section \ref{sec:FourCoupledSYK}, we construct four coupled SYK models and derive the large $N$ saddle point equations.
Using these equations, we study the phase structure at zero temperature.
In section \ref{sec:MoreOnTraversableJT}, we study the property of traversable wormholes coupled to 2d CFT in nearly AdS$_2$ gravity.
This section includes the calculation of entanglement entropy using the island formula.
We also study traversable wormholes with partial couplings.
In section \ref{sec:FourCoupledJT}, we construct four coupled nearly AdS$_2$ gravities and derive solutions for their equation of motion.
We study the phase transitions by varying the boundary conditions outside the wormholes.
We also interpret traversable wormholes as braket wormholes by exchanging the role of Euclidean time and space.
The phase transitions in the context of braket wormholes are studied.
Section \ref{sec:DiscussionSummary} contains a brief summary and discussion of our results and we also discuss possible future directions.
In Appendix \ref{sec:2dCFTformula}, we collect some formulas in 2d CFT which is used in the main parts of our paper.


\section{Review of the SYK model and Nearly AdS$_2$ gravity \label{sec:Review}}

\subsection{SYK model and Nearly AdS$_2$ gravity}
\subsubsection{The SYK model}
Let us first consider $N$ Majorana fermions in $0+1$ dimensions that obey the anti-commutation relation $\{\psi^i, \psi^j\} = \delta_{ij}$.
The Hamiltonian of the SYK model \cite{KitaevTalk,Maldacena:2016hyu,Kitaev:2017awl} is  
\be
H_{SYK} = i^{\f{q}{2}} \sum_{i_1 < \cdots <i_q } J_{i_1\cdots i_q} \psi^{i_1}\cdots \psi^{i_q}, \label{eq:SYKHamiltonian}
\ee
with mean $\braket{J_{i_1\cdots i_q}} = 0$ and variance $\braket{J_{i_1\cdots i_q}^2} = \f{J^2}{N^{q-1}}(q-1)!  = \f{1}{q} \f{\mathcal{J}^2 (q-1)!}{(2N)^{q-1}}$.

In the large $N$ limit, we get the Schwinger-Dyson equation
\be
\partial_{\tau}G(\tau, \tau') - \int d\tau '' \Sigma(\tau,\tau'') G(\tau'',\tau' ) = \delta(\tau - \tau'),\qquad \Sigma(\tau,\tau') = \f{J^2}{q}(G(\tau,\tau'))^{q-1}, \label{eq:SDeqSYK}
\ee
for the Euclidean correlator 
\be
G(\tau,\tau') = \f{1}{N}\sum_{i=1}^N \braket{T_{\tau} (\psi^i(\tau)\psi^i(\tau'))} = \f{1}{N}\sum_{i=1}^N \big( \braket{ \psi^i(\tau)\psi^i(\tau')}\theta(\tau-\tau') -  \braket{ \psi_i(\tau')\psi_i(\tau)}\theta(-\tau+\tau')\big).
\ee
This Schwinger-Dyson equation is obtained as the equation of motion of the large $N$ effective action:
\ba
Z &= \int \mathcal{D}G\mathcal{D}\Sigma   e^{-NS_{eff}(G,\Sigma)} \notag \\
 &=\int  \mathcal{D}G\mathcal{D} \Sigma \exp \Bigg[N \Big\{  \log \text{Pf} (\partial_{\tau} - \Sigma) - \f{1}{2} \int  d \tau d \tau' \Big[ \Sigma(\tau,\tau')G(\tau,\tau') - \f{J^2}{q} G(\tau,\tau')^{q} \Big] \Big\}\Bigg].
\ea
In the long time limit $1 \ll \mathcal{J}(\tau -\tau')  \ll N $, we can ignore the derivative term  $\partial_{\tau} G(\tau,\tau')$  in (\ref{eq:SDeqSYK})
and can have an analytical solution
\be
G(\tau,\tau') \approx G_c(\tau,\tau') \equiv \f{c_{\Delta}}{|\mathcal{J}(\tau-\tau')|^{2\Delta}} \text{sgn}(\tau-\tau'),  \qquad {c_{\Delta}} = \f{1}{2} \Bigg[ \Big(1 -2\Delta\Big) \f{\tan \pi \Delta}{\pi \Delta} \Bigg]^{\Delta}\label{eq:zeroTsolSYK}
\ee
which is scale invariant.
Actually in this limit the conformal transformation of the scale invariant solution (\ref{eq:zeroTsolSYK}) 
\be
G(\tau,\tau') = [f(\tau) f(\tau')] ^{\Delta}G(f(\tau),f(\tau')), \qquad \Sigma(\tau,\tau') = [f(\tau)  f(\tau')] ^{1- \Delta}\Sigma(f(\tau),f(\tau'))
\ee
is also a solution of the Schwinger-Dyson equation (\ref{eq:SDeqSYK}).
Therefore, the system have an emergent conformal symmetry.
This conformal symmetry is spontaneously broken by each solution spontaneously and also explicitly broken by the derivative term  $\partial_{\tau} G(\tau,\tau')$.
The effect of this symmetry breaking is summarized in the so called Schwartzian action \cite{Maldacena:2016hyu,Kitaev:2017awl} 
\be
S =  -\f{N \alpha_S}{\mathcal{J}}  \int d\tau \{ f(\tau), \tau \} ,\qquad \{ f(\tau) , \tau \} \equiv \f{f'''(\tau)}{f'(\tau)}  -\f{3}{2} \Big(\f{f''(\tau)}{f'(\tau)} \Big)^2. \label{eq:SchActionSYK}
\ee
The coefficient can be determined numerically. For example, $\alpha_S \approx 0.00709$  for $q=4$ \cite{Cotler:2016fpe}.
For large $q$ limit it scales as $\alpha_S \sim \f{1}{4q^2}$.

At low temperature, the energy and the entropy become
\ba
E &= E_0 + \f{c}{2}T^2 + \cdots, \notag \\
S &= S_0 + c T + \cdots. 
\ea
where $c$ is the specific heat. 
The specific heat is given by the coefficient of the Schwartzian action \eqref{eq:SchActionSYK} as $c = \f{2\pi^2\alpha_S N }{\mathcal{J}}$.
Therefore the Schwartzian action captures the corrections to the ground state energy and entropy.
The zero temperature entropy $S_0$ is given by 
\be
S_0/N = \log \text{Pf} (\Sigma (\tau)) = \f{1}{2} \log 2 -\pi \int_0 ^\Delta \Big( \f{1}{2} - x\Big) \tan \pi x \ dx.
\ee
On the other hand, the ground state energy is calculated numerically through 
\be
E/N = \f{1}{q} \partial _{\tau} G(\tau)\Big|_{\tau \to 0} = \f{\mathcal{J}^2}{2q^2} \int _0^{\beta} (2G(\tau))^q.
\ee
For $q = 4$ case, the ground state energy is calculated as $E_0 \approx -0.0574 \mathcal{J}$.

\subsubsection{Nearly AdS$_2$ gravity}
Here we briefly summarize the results in nearly AdS$_2$ gravity.
The action of nearly AdS$_2$ gravity \cite{Maldacena:2016upp, Almheiri:2014cka, Kitaev:2018wpr}, or Jackiw-Teitelboim (JT) gravity \cite{JACKIW1985343, Teitelboim:1983ux} is given by
\ba
S &= \f{\phi_0}{16\pi G_N} \int dx^2 \s{-g}R  + \f{\phi_0}{8\pi G_N} \int dx^2 \s{-h}K  \notag \\
 & \qquad + \f{1}{16 \pi G_N} \int dx^2 \s{-g}\phi (R+2)  + \f{1}{8 \pi G_N} \int dt \s{-h}\phi (K+1)    + S_{mat}[\chi,g]. \label{eq:JTaction}
\ea
The first term is the two dimensional Einstein Hilbert action, which is topological.
It does not affect the equation of motion of the system but is important to take into account the extremal entropy.
We consider the AdS$_2$ with cut off at finite distance and impose the boundary condition at the boundary
\be
ds^2 |_{bdy} = -\f{du^2}{\epsilon^2}, \qquad \phi|_{bdy} = \phi_b = \f{\bar{\phi}_r}{\epsilon}, \label{eq:bdycondNAdS2}
\ee
and finally we take $\epsilon \to 0$ limit.
The equation of motion of this system is derived in a straighten forward way by taking the variation with respect to the metric $g_{\mu\nu}$ and the dilaton $\phi$.
The result is 
\ba
 \f{\delta S}{\delta \phi}  = 0 & \qquad \to \qquad  R+2 = 0, \notag \\
 \f{\delta S}{\delta g^{\mu\nu}}= 0 &\qquad \to \qquad  \nabla_{\mu}\nabla_{\nu}\phi - \nabla^2 \phi + g_{\mu\nu} \phi = \f{1}{8\pi G_N} \braket{T_{\mu\nu}^{mat}}. \label{eq:JTeomcov}
\ea
where $\braket{T_{\mu\nu}^{mat}}$ is the matter stress energy tensor expectation value.
The first equation simply set the metric to be that of AdS$_2$.
Here we set the matter stress tensor $\braket{T_{\mu\nu}^{mat}}$ to be zero.
Finally the system reduces to the motion of the boundary $t_P(u)$. Here $u$ is the boundary time (\ref{eq:bdycondNAdS2}) and $t_P$ is the time in Poincare coordinate in AdS$_2$.
The action for this function $t_P(u)$ is again by the Schawrtzian action \cite{Maldacena:2016upp,Kitaev:2018wpr}
\be
S = -\f{\bar{\phi}_r}{8 \pi G_N} \int du \{t_P(u), u \}.
\ee
Taking a conformal gauge $ds^2 = - e^{2\omega(x^+,x^-)}dx^+ dx^-$, the equation of motion \eqref{eq:JTeomcov} becomes 
\ba
2 \partial_{x^+} \partial_{x^-} \phi + \phi e^{2\omega} &= 16 \pi G_N \braket{T_{x^+ x^-}^{\text{mat}}}, \notag \\
-e^{2\omega} \partial _{x^+} (e^{-2\omega} \partial_{x^+} \phi) &= 8 \pi G_N \braket{T_{x^+ x^+}^{\text{mat}}} \notag \\
-e^{2\omega} \partial _{x^-} (e^{-2\omega} \partial_{x^-} \phi) &= 8 \pi G_N  \braket{T_{x^- x^-}^{\text{mat}}} \label{eq:dialtonEOM1}
\ea 
Let us briefly discuss the higher dimensional setup that are described by Nearly AdS$_2$ gravity.
Nearly AdS$_2$ gravity describes the low-energy dynamics of near extremal charged black holes \cite{Maldacena:2016upp,Maldacena:2018gjk, Maldacena:2020skw}.
The mass and the near extremal entropy are 
\ba
M &= \f{r_e}{G_N} + \f{2\pi^2 r_e^3}{G_N} T^2 + \cdots  \notag \\
S &=\f{\pi r_e^2}{G_N} + \f{4\pi^2 r_e^3}{G_N} T+ \cdots.
\ea
where $r_e$ is the radius of the $S^2$ in the near horizon geometry AdS$_2 \times S^2$ and $T$ is the Hawking temperature.   
The parameters in JT gravity are given by \cite{Almheiri:2019yqk} 
\be
\phi_0 = 4\pi r_e^3, \qquad \bar{\phi}_r = 8 \pi r_e^3. \label{eq:4dMagExBH}
\ee
If these extremal black holes are  magnetically charged, $r_e$ is given by 
\be
r_e^2 = \f{\pi q^2 G_N}{g^2}, 
\ee 
where $g$ is the $U(1)$ gauge coupling and $q$ is the magnetic charge of black holes.
In the presence of the magnetic fields, we have $q$ fermion zero modes on $S^2$  from a single massless $4d$ Dirac fermion \cite{ Ambjorn:1992ca}, which leads to the $q$ 2d Dirac fermions \cite{Maldacena:2018gjk,Maldacena:2020skw}.
 In other words, we have the free fermion CFT with central charge $c = q$.
These large number of two dimensional CFTs enhance the quantum effect from the matter fields.

\subsection{Traversable wormholes in SYK and in Nearly AdS$_2$}

\subsubsection{Two coupled SYK model}
Here we briefly describe the two coupled  SYK model \cite{Maldacena:2018lmt}, which ground state shares many properties with traversable wormholes in Nearly AdS$_2$.
First we prepare two SYK models which is decoupled. We denote the left fermions (right fermions) by $\psi_L^i$ ($\psi_R^i$).
The random couplings (\ref{eq:SYKHamiltonian}) of left and right SYK ($J_{j_1 \cdots j_q}^L$ and $J_{j_1 \cdots j_q}^R$)
are the same up to the sign : $J_{j_1 \cdots j_q}^L  = (-1)^{\f{q}{2}} J_{j_1 \cdots j_q}^R$.
The Hamiltonian of the two coupled SYK model (Maldacena-Qi model) is 
\be
H = H_{SYK}^L + H_{SYK}^R + H_{int}, \qquad H_{int} = i \mu \sum_{i=1}^N \psi_L^i \psi_R^i. \label{eq:MQHamiltonian}
\ee
In $\mu \ll \mathcal{J}$ limit, we can still use the low energy, Schwartzian action and the effective action for the coupled system is 
\be
S = N \int du \Bigg\{ - \f{\alpha_S}{\mathcal{J}} \Big( \Big\{ \tan \f{t_l(u)}{2}, u \Big\} + \Big\{ \tan \f{t_l(u)}{2}, u \Big\}\Big)  + \mu \f{c_{\Delta}}{(2\mathcal{J})^{2\Delta}} \Bigg[\f{t'_l(u)t'_r(u)}{\cos^2 \f{t_l(u)- t_r(u)}{2}} \Bigg]^{\Delta}\Bigg\}. \label{eq:TwoSYKEffAct}
\ee

Another comment is that we can consider  deformations like $S_{int}$ 
$\sim$ 
$i^p gN^{1-p}$ $ \sum(\psi_L^{i_1} \cdots \psi_L^{i_p})$ $(\psi_R^{i_1} \cdots \psi_R^{i_p})$. 
This only change the dimension $\Delta \to p \Delta$. 
For even $p$, we can keep the time reversal symmetry that exists for even $q/2$.
We can also realize a marginal deformation for $p = q$ case, which is a similar situation to JT gravity with 2d conformal matters.

\subsubsection{Traversable wormholes in Nearly AdS$_2$:  double trace deformations}
We introduce the double trace deformation \cite{Gao:2016bin,Maldacena:2017axo} that directly couples the two side of AdS$_2$:
\be
S_{int} = g\sum_{i = 1}^N\int  du O_L^i(u)O_R^i(u). \label{eq:DoubleTrace1}
\ee
where $O^i$ are a set of $N$ operators with dimension $\Delta$.
In gravity language, we impose the boundary conditions for dual fields $\phi(t,z)$ such that the two sides are directly coupled.
We consider the theory with $N$ matter fields with $N$ that is comparable with $\f{\phi_r}{G_N}$ so that we can balance matter quantum effects and the classical JT gravity.
Such a situation arise from the 4d magnetically charged near extremal black holes \cite{Maldacena:2020skw}.

For small $g$ , we can approximate the effect of the double trace deformation (\ref{eq:DoubleTrace1}) as 
\be
\braket{e^{ig \sum_i \int du O_L^i(u)O_R^i(u)}} \sim e^{ig \sum_i \int du \braket{O_L^i(u)O_R^i(u)}}.
\ee
This amounts to the resummation of a bulk ladder type diagram, which dominates in the large $N$, small $g$ limit with $Ng$ kept fixed.
Then, we couple this to the gravity mode, which is achieved by performing a reparametrization of the left and right times.
These reparametrizations are expressed as the map from the proper time $u$ to the global time coordinates $t_l(u), t_r(u)$ at the two boundaries of the AdS$_2$.
In this way, we finally obtain the effective action 
\be
S = N \int du \Bigg\{ - \phi_r \Big( \Big\{ \tan \f{t_l(u)}{2}, u \Big\} + \Big\{ \tan \f{t_l(u)}{2}, u \Big\}\Big)  + \f{g N}{2^{2\Delta}} \Bigg[\f{t'_l(u)t'_r(u)}{\cos^2 \f{t_l(u)- t_r(u)}{2}} \Bigg]^{\Delta}\Bigg\}.
\ee
Therefore, we obtain the same effective action with that in the two coupled SYK model (\ref{eq:TwoSYKEffAct}).
Here we put $8 \pi G_N = 1$ for simplicity.
 
\subsubsection{Low energy analysis}
Starting form both of the SYK model and the nearly AdS$_2$ gravity, we obtain the coupled Schwartzian action
\be
S = N \int d\tilde{u} \Bigg\{ -  \Big( \Big\{ \tan \f{t_l(\tilde{u})}{2}, \tilde{u} \Big\} + \Big\{ \tan \f{t_l(\tilde{u})}{2}, \tilde{u} \Big\}\Big)  + \eta \Bigg[\f{t'_l(\tilde{u})t'_r(\tilde{u})}{\cos^2 \f{t_l(\tilde{u})- t_r(\tilde{u})}{2}} \Bigg]^{\Delta}\Bigg\}, \label{eq:rescaledcoupled}
\ee 
where the relation between the parameters in the previous actions is 
\be
\tilde{u} = \f{\mathcal{J}}{\alpha_S} u = \f{N}{\phi_r}  u, \qquad \eta \equiv \f{\mu \alpha_S}{\mathcal{J}} \f{c_{\Delta}}{(2\alpha_S)^{2\Delta}} = \f{g}{2^{2\Delta}} \Big(\f{N}{\phi_r} \Big)^{2\Delta-1}.
\ee 
This action should be supplemented by $SL(2,\mathbb{R})$ constraints and the total $SL(2,\mathbb{R})$ charges vanish \cite{Maldacena:2016upp,Harlow:2018tqv}.
This system (\ref{eq:rescaledcoupled}) has a classical, static solution of the form $t_l (\tilde{u}) = t_r(\tilde{u}) = t' u$ with 
\be
(t') ^{2(1-\Delta)} = \eta \Delta , \qquad  \Big(\f{1}{\mathcal{J}} \f{dt}{du}  \Big) ^{2(1-\Delta)} = \f{\mu \Delta}{2 \mathcal{J}\alpha_S} \f{2c _{\Delta}}{2^{2\Delta}}. \label{eq:WHlengthLowEnergy}
\ee
The parameter $t'$ is important because  the time to traverse the interior, or the gap of the system, is determined by this $t'$.
The SYK correlation function in  the low energy limit, or the boundary propagator of the nearly AdS$_2$, is 
\be
\braket{\mathcal{O}(t_l)\mathcal{O}(t_r)} = \Bigg[\f{1}{\cos \f{t_l - t_r}{2}}\Bigg]^{2\Delta}  \rightarrow \braket{\mathcal{O}(\tilde{u}_1)\mathcal{O}(\tilde{u}_2)} = \Bigg[\f{1}{\cos \f{t'(\tilde{u}_1 - \tilde{u}_2) }{2}}\Bigg]^{2\Delta} .
\ee
In the Euclidean signature $u = - i \tilde{\tau}$,   the correlation function in the conformal limit becomes 
\be
\braket{\mathcal{O}(\tilde{\tau}_1)\mathcal{O}(\tilde{\tau}_2)} = \Bigg[\f{1}{\cos \f{t'(\tilde{\tau}_1 - \tilde{\tau}_2) }{2}},\Bigg]^{2\Delta}
\ee 
which decays exponentially for large relative Euclidean time $\tilde{\tau}_1 -\tilde{\tau}_2 \gg 1$. 
From this we can read off the energy gap of the system. This becomes 
\be
E_{\text{gap}, \tilde{u}} = t' \Delta, \qquad  E_{\text{gap}, u} = \f{\mathcal{J}}{\alpha_S} t' \Delta = \f{N}{\phi_r} t' \Delta. \label{eq:MQEgapConformal}
\ee
$E_{gap,\tilde{u}}$ is the energy gap with respect to the rescaled time $\tilde{u}$ whereas $E_{gap,u}$ is that for the physical time $u$.
In the  $q=4$ model, the energy gap scales as $E_{gap} \sim \mu^ \f{2}{3}$.

\subsubsection{Two couple SYK beyond small coupling }
Here we summarize further on the two coupled SYK model.
First, when $q=4k \ (k \in \mathbb{N})$ the Hamiltonian (\ref{eq:MQHamiltonian}) has a $\mathbb{Z}_4$ symmetry \footnote{Even when $q=4k +2 \ (k \in \mathbb{N})$, the Hamiltonian (\ref{eq:MQHamiltonian}) is invariant under the combination of $\mathbb{Z}_2$ with the generator $w : \psi_L \to \psi_R, \ \psi_R \to \psi_L$ and the time reversal $\mathcal{T}$ that satisfies $\mathcal{T}\psi_A^i \mathcal{T}^{-1} = \psi_A^i$ for $A =L,R$ i.e. $(w \mathcal{T})H(w \mathcal{T})^{-1} = H$.
Using this symmetry we can do basically the same arguments with $q =4k$ cases.
For simplicity, we focus on the $q = 4k$ cases with the $\mathbb{Z}_4$ symmetry.}  which is given by 
\be
\psi_L \to \psi_R, \qquad \psi_R \to -\psi_L. \label{eq:Z4symMQ}
\ee
The mass term is directly related to the thermofield double state.
Actually, the SYK thermo field double state is defined as 
\be
\ket{TFD(\beta)} = Z_{\beta}^{-\f{1}{2}} e^{-\f{\beta}{2}(H_L+H_R)}\ket{I},
\ee
where $\ket{I}$ is the infinite temperature thermo field double state 
\be
(\psi_L^j + i \psi_R^j) \ket{I} = 0, \qquad \text{for} \ \ j = 1 ,\cdots ,N. \label{eq:MaxEntanglementSYK} 
\ee
This means that the $\ket{I}$ is characterized as the state that is annihilated by the annihilation operator  $f_j = \psi_L^j + i \psi_R^j$.
Note that the $\mathbb{Z}_4$ symmetry (\ref{eq:Z4symMQ}) acts on these annihilation operators as $f_j \to -i f_j$ and the thermofield double is invariant under the $\mathbb{Z}_4$ symmetry.
By multiplying both sides by $\psi_L^j$, we obtain $S^j\ket{I} = \ket{I}$ for $S^j = -2i \psi_L^j\psi_R^j$. 
$S^j$ is the spin operator with eigenvalues $\pm 1$ since $(S^j)^2 = 1$.
The mass term is actually a sum of these spin operators.
Therefore, the infinite temperature thermo field double state is characterized as the ground state of the mass term Hamiltonian, in which  all the spins are up.
In other words, in the $\mu \to \infty$ limit the ground state of the two coupled model (\ref{eq:MQHamiltonian}) is perfectly agrees with the infinite temperature TFD state.
On the other hand, in the $\mu \to 0$ limit the ground state is the two copies of the SYK ground states.
This is also the TFD state at zero temperature. 
This suggest that even for generic $\mu $ the ground state $\ket{G(\mu)}$ of (\ref{eq:MQHamiltonian}) is close to the thermafield double state.
Actually we can directly study the maximal overlaps between $\ket{G(\mu)}$ and $\ket{TFD(\beta)}$ when we vary the inverse temperature $\beta$  at finite $N$ \cite{Garcia-Garcia:2019poj,Alet:2020ehp}, and they are very close to $1$.
This means that the variational approximation by the $TFD$ state works well \footnote{Similar variational approximations by "SYK boundary states" are also applied to the ground state in a mass deformed SYK model \cite{Kourkoulou:2017zaj,Nosaka:2019tcx,Numasawa:2019gnl}. }.

It will be useful to study how the state deviate from the maximally entangled state \eqref{eq:MaxEntanglementSYK}.
Since the maximally entangled state is annihilated by the fermion annihilation operator $f_i$, the expectation value of the occupation number operator 
\be
\f{\braket{\psi|f_i^{\dagger}f_i|\psi}}{\braket{\psi|\psi}} = \braket{|\psi_L+i\psi_R|^2}.
\ee
will characterize how the state is close to the maximally entangled state $\ket{I}$.
This is related to the spin operator expectation value 
\be
\braket{|\psi_L+i\psi_R|^2} = 1 - \braket{f_i ^{\dagger}f_i} = 1 - \braket{S_i}.
\ee
When the condensation $\braket{S_i}$ is big, the state becomes close to the maximally entangled state.

\begin{figure}[ht]
\begin{center}
\includegraphics[width=7cm]{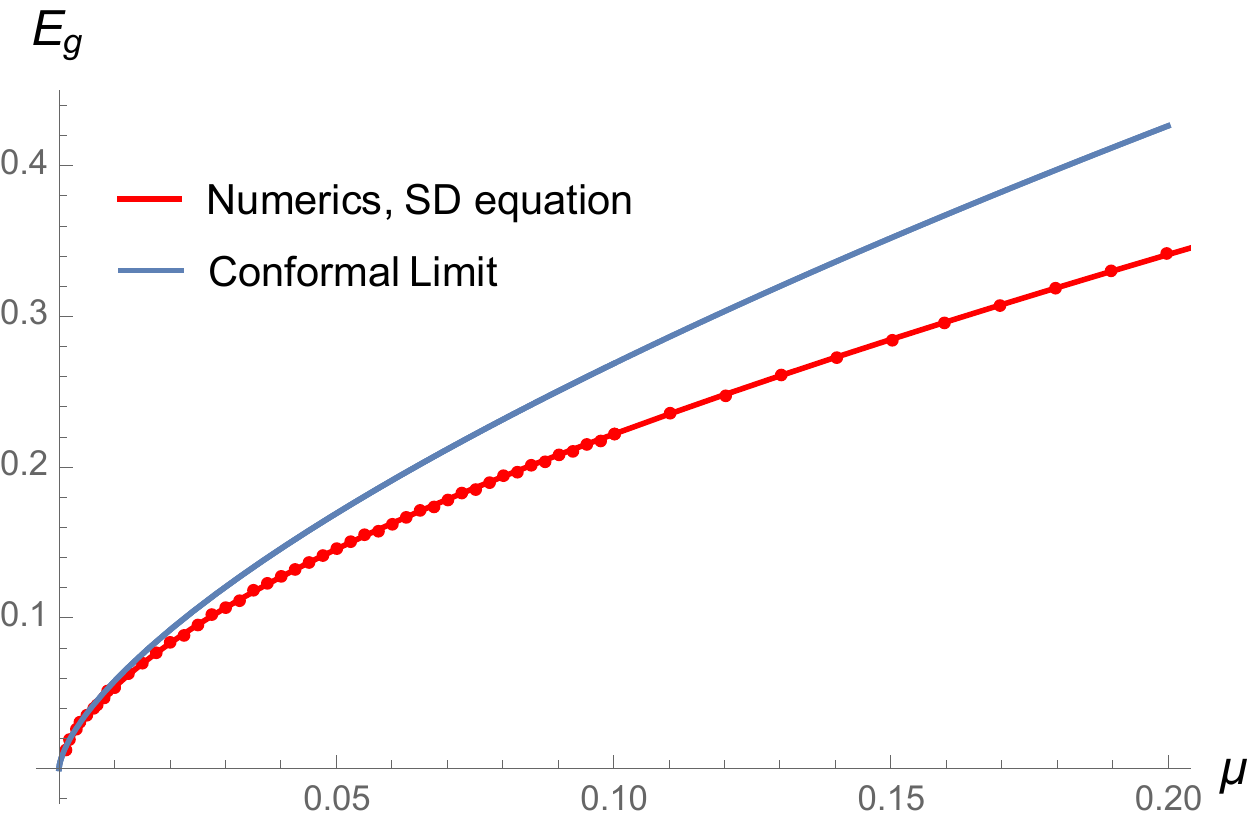}
\includegraphics[width=7cm]{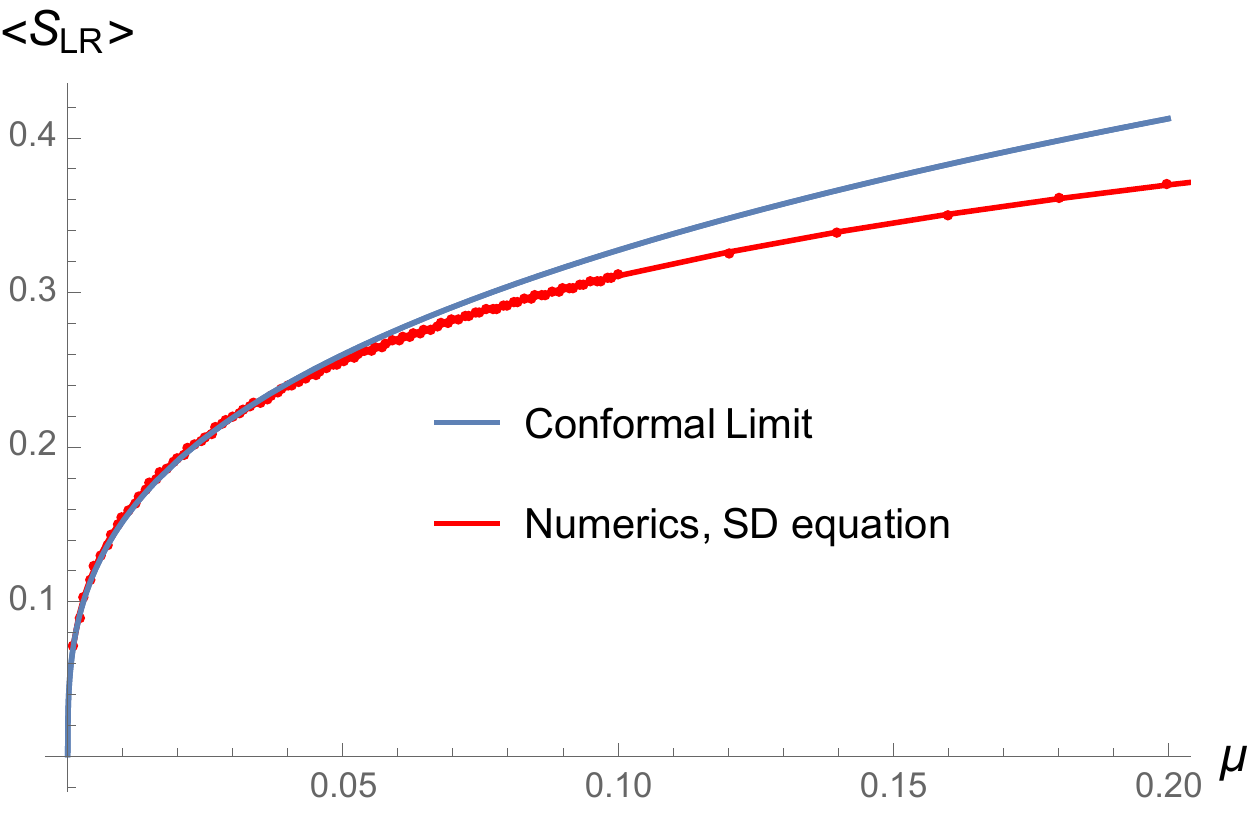}
\caption{We plot the $E_g$ and the spin operator expectation value $\f{1}{2}\braket{S_{LR}} = |\braket{\psi_L\psi_R}|$ for several $\mu$ and compare with the conformal limit results (\ref{eq:MQEgapConformal}).
Here we take the temperature to be $T=0.001$, which is very low and essentially the system at zero temperature.
The conformal limit is a good approximation for small $\mu$.
The spin operator, which is given by $\f{1}{2}\braket{S_{LR}} = -i\braket{\psi_L(0)\psi_R(0)} =-iG_{LR}(0)$, behaves as $-iG_{LR}(0) \approx c_{\Delta}(\f{t'}{2})^{2\Delta}$ in the conformal limit.
\label{fig:MQgroundState1}
} 
\end{center}
\end{figure}

In this model, we can actually write down the large $N$ effective action in $G, \Sigma$ variables.
We introduce the correlators $G_{ab} (\tau_1,\tau_2) = \braket{\psi_a(\tau_1) \psi_b(\tau_2)}$ for $a,b = L,R$ that means that the fermion belongs to the left SYK cluster or the right cluster. Similarly, we introduce the self energy variable $\Sigma_{ab}(\tau_1,\tau_2)$.
Then, the effective action becomes \cite{Maldacena:2018lmt} 
\ba
-S_E/N &= \log \text{Pf} (\partial_{\tau} \delta_{ab} - \Sigma_{ab}) - \f{1}{2} \int d \tau_1 d\tau_2 \sum_{a,b} \Bigg[ \Sigma_{ab}(\tau_1,\tau_2) G_{ab}(\tau_1,\tau_2) - s_{ab} \f{\mathcal{J}^2}{2 q^2} [2 G_{ab}(\tau_1,\tau_2)]^q \Bigg] \notag \\
& \ + i\f{\mu}{2} \int d\tau [-G_{LR}(\tau_1,\tau_2) +G_{RL}(\tau_1,\tau_2) ] .
\ea
Then, the Schwinger Dyson equation becomes 
\ba
&\partial_{\tau_1} G_{LL}(\tau_1,\tau_2) - \int d\tau_3 \Sigma_{LL}(\tau_1,\tau_3) G_{LL}(\tau_1,\tau_2) -  \int d\tau_3 \Sigma_{LR}(\tau_1,\tau_3) G_{RL}(\tau_1,\tau_2)  = \delta(\tau_1-\tau_2), \notag \\
&\partial_{\tau_1} G_{LR}(\tau_1,\tau_2) - \int d\tau_3 \Sigma_{LL}(\tau_1,\tau_3) G_{LR}(\tau_1,\tau_2) -  \int d\tau_3 \Sigma_{LR}(\tau_1,\tau_3) G_{RR}(\tau_1,\tau_2)  = 0.  \notag \\
&\Sigma_{LL}(\tau_1,\tau_2) = \f{\mathcal{J}^2}{q} [2G_{LL}(\tau_1,\tau_2)]^{q-1}, \qquad \Sigma_{LR}(\tau_1,\tau_2) = (-1)^{\f{q}{2}}\f{\mathcal{J}^2}{q} [2G_{LR}(\tau_1,\tau_2)]^{q-1} - i\mu \delta(\tau_1-\tau_2) .\label{eq:SDMaldacenaQi1}
\ea
This equation allows us to study the two coupled model beyond the low energy effective action (\ref{eq:rescaledcoupled}) and is also useful to check that the Schwartzian analysis gives the correct answer.
The comparison of the numerics from the SD equation (\ref{eq:SDMaldacenaQi1}) and the results from the conformal limit (\ref{eq:MQEgapConformal}) are shown in figure \ref{fig:MQgroundState1}.
The low energy limit (\ref{eq:rescaledcoupled}) shows a good agreement with the exact numerical results from the SD equation (\ref{eq:SDMaldacenaQi1}) for small $\mu$.

\subsubsection{Traversable wormholes in JT gravity with conformal matters}
Here we discuss more on the traversable wormholes in nearly AdS$_2$ gravity.
We can consider the traversable wormholes in JT gravity, especially the bulk fields are 2d CFT \cite{Maldacena:2018lmt,Chen:2019qqe} and gives a direct interaction on them at two asymptotic boundaries.
In this case, we can explicitly construct the dilaton  profile.
Imagine that we have $N$ massless Majorana fermions $\psi(t,\sigma) = (\psi_+(t,\sigma), \psi_-(t,\sigma))^T$ with the twisted boundary condition 
\ba
&\psi_{+}(t,0) = \cos \pi \epsilon\ \psi_-(t, 0)- \sin \pi \epsilon\ \psi_-(t,\pi) \notag \\
&\psi_{-}(t,\pi) = -\cos \pi \epsilon\ \psi_+(t, \pi) - \sin \pi \epsilon \ \psi_-(t, 0).
\ea
$\epsilon = 0$ corresponds to a usual Majorana fermion on a strip of width $\pi$ with boundary conditions.
Non zero $\epsilon$ introduce a non zero transparency and gives a direct coupling between two asymptotic boundaries.
$\epsilon = \f{1}{2}$ corresponds to the anti periodic boundary conditions for a Majorana fermion i.e. NS sector.
On the other hand, $\epsilon = -\f{1}{2}$ corresponds to periodic boundary conditions i.e. R sector.
The expectation value of the energy momentum tensor on AdS$_2$ becomes 
\be
\braket{T_{tt}^{\text{mat}}(t,\sigma)} = \braket{T_{\sigma \sigma}^{\text{mat}}(t,\sigma)} = -\f{N}{4} \epsilon (1- \epsilon) \label{eq:EMfermion1}
\ee
The energy per fermion on cylinder is
\be
E (\epsilon) = -\f{1}{48} [1+12\epsilon(1-\epsilon)].
\ee 
By solving (\ref{eq:dialtonEOM1}), with the energy momentum tensor (\ref{eq:EMfermion1}) gives the dilaton profile under the assumption of $\phi = \phi(\sigma)$
\be
\phi = N \f{\epsilon (1- \epsilon)}{4\pi} \Big[\f{\f{\pi}{2}-\sigma}{\tan \sigma} + 1 \Big] + \f{N}{48 \pi} =  \f{c}{2\pi}\epsilon (1- \epsilon) \Big[\f{\f{\pi}{2}-\sigma}{\tan \sigma}+ 1 \Big] + \f{c}{24 \pi} .
\ee
To satisfy the boundary condition $ds^2 = \f{du^2 }{\vep^2}, \phi_b = \f{\bar{\phi}_r}{\vep}$, we consider the rescaling 
\be
ds^2 = \f{t'^2 (-dt^2 + d\sigma^2)}{\sin^2( t' \sigma)}, \qquad \phi(\sigma) = \f{N}{4\pi} \epsilon(1-\epsilon)\Big[\f{\f{\pi}{2}-(t'\sigma)}{\tan( t'\sigma)}+ 1 \Big] + \f{N}{48 \pi},
\ee
where $0 < \sigma < \f{\pi}{t'}$. 
Then, near the boundary $\sigma = \vep$, we obtain 
\be
\phi_b \approx \f{c}{2\pi} \epsilon(1-\epsilon) \f{\f{\pi}{2}}{t' \vep} = \f{c}{4t'}  \epsilon(1-\epsilon) \f{1}{\vep} .
\ee
Matching with the boundary condition $\phi_b = \f{\bar{\phi}_r}{\vep}$, we obtain
\be
t'= \f{c}{4} \f{\epsilon(1-\epsilon)}{\bar{\phi}_r}.
\ee
Using $\eta = 4\epsilon \ll 1$ and $c = \f{N}{2}$, we obtain 
\be
t' = \f{N \eta}{2 \bar{\phi}_r},
\ee
which reproduce the results in the low energy limit (\ref{eq:WHlengthLowEnergy}) for  $\eta \ll 1$ and $\Delta = \f{1}{2}$.

The ADM energy for one side is calculated as\footnote{We absorb the constant shift $\f{c}{24\pi}$ in the dilaton field $\phi$ to the constant $\phi_0$ which accounts the extremal entropy.}
\ba
M &= \f{1}{8\pi G_N}\s{h} [\phi_b - \partial_n\phi] \notag \\
&=\lim_ {\epsilon \to 0}  \f{1}{8\pi G_N}\f{t' }{\sin t' \sigma}\Bigg[ \Big( \f{2\bar{\phi}_r t'}{\pi} \Big[\f{\f{\pi}{2} - t'\sigma}{\tan (t' \sigma)}+1 \Big] \Big)+ \f{\sin (t' \sigma)}{t'} \partial_{\sigma} \Big( \f{2\bar{\phi}_r t'}{\pi}\Big[ \f{\f{\pi}{2} - t'\sigma}{\tan (t' \sigma)} + 1 \Big]\Big) \Bigg] \Bigg|_{\sigma = \epsilon} \notag \\
&= - \f{\bar{\phi}_r  t'^2}{16\pi G_N} .
\ea
The total ADM energy is the twice of this, one from the left (we denote it by $M_L$) and the one from the right boundary $(M_R)$.
This becomes 
\be
M= M_{L} + M_R = -\f{\bar{\phi}_r  t'^2}{8\pi G_N}.
\ee
To compare with Maldacena-Qi notation, we set $8\pi G_N = 1$ and use $t' = \f{N \eta}{2 \bar{\phi}_r}$.
Then, this becomes 
\be
\f{M}{N}  = -\f{\bar{\phi}_r}{N} (t')^2 = -\f{\bar{\phi}_r}{N} \f{N^2 \eta^2}{4 \bar{\phi}_r^2} = -\f{N}{\bar{\phi}_r} \f{\eta^2}{4}.
\ee
which reproduce the Maldacena-Qi results of the ground state energy shift for $\Delta = \f{1}{2}$.

Beyond small $\epsilon$ limit, we can also consider finite $\epsilon$ in this model.
$\epsilon = \f{1}{2}$ case is specially interesting because this gives a perfectly transparent boundary condition between two side.
When fermion boundary conditions are anti periodic, then this becomes
\be
\phi(\sigma) =  \f{c}{8\pi} \bigg[\f{\f{\pi}{2}-\sigma}{\tan \sigma}+ 1 \bigg] + \f{c}{24 \pi}, 
\ee
with $c = N/2$ for fermions. 
This results for perfectly transparent boundary conditions are written only using the central charge, which is a universal quantity in CFT.
Actually we can impose this type of boundary condition in any CFT, even in holographic 2d CFT.
Another interesting property is that we can also insert the no gravitating outside region between two boundaries and can study physics on this non gravitating region.
This situation especially comes from traversable wormholes in 4 dimensions \cite{Maldacena:2018gjk} where the the wormhole throat region is described by the JT gravity with CFT described  below (\ref{eq:4dMagExBH}).
In the four coupled model, we use this setup as well as we analyze some properties in two coupled model through non gravitating region.

\section{ 4 coupled SYK models \label{sec:FourCoupledSYK}}
\subsection{The Hamiltonian}
Here we consider the model which couples  four cites SYK models.
First, we prepare four decoupled SYK models. We label these fermions by $\psi^j_{\alpha A}$ with $A=L,R$ , $\alpha = 1, 2$ and $ j = 1 ,\cdots, N$. 
The Hamiltonian of the $4$ coupled SYK model is 
\be
H_{4} = H_{1L} +  (-1)^{\f{q}{2}}H_{1R}+  (-1)^{\f{q}{2}} H_{2L} +  H_{2R} + H_{int}{}^{11}_{LR} + H_{int}{}^{22}_{LR} + H_{int}{}^{12}_{LL} +  H_{int}{}^{12}_{RR}. \label{eq:4coupledHamiltonian}
\ee
Here $H_{\alpha A}$ is 
\be
H_{\alpha A} =  i^{\f{q}{2}}\sum_{j_1<  \cdots <j_q} J_{j_1\cdots j_q} \psi^{j_1}_{ \alpha A}\cdots \psi^{j_q}_{\alpha A}.
\ee
and $H_{int}{}^{\alpha \beta}_{AB}$ is 
\be
 H_{int}{}^{\alpha \beta}_{AB} = i \Big[ \delta^{\alpha \beta}\mu_{AB} \sum_{k=1}^{N}\psi_{ \alpha A}^k\psi_{\beta B}^k +  \mu^{\alpha \beta} \sigma^z _{AB}\sum_{k=1}^{N}\psi_{\alpha A}^k\psi_{\beta B}^k\Big],
\ee
where $\sigma^z =\begin{pmatrix} 1 & 0 \\ 0 & -1 \end{pmatrix}$ is the Pauli matrix.
Or, more explicitly we can write 
\be
H_{int}{}^{11}_{LR} = i \mu_{LR}\sum_{k=1}^{N} \psi_{1L}^k\psi_{1R}^k, \qquad H_{int}{}^{22}_{LR} = i \mu_{LR}\sum_{k=1}^{N} \psi_{2L}^k\psi_{2R}^k,
\ee
and \footnote{Here we include the minus sign in $H_{int}{}^{12}_{RR}$ so that $H_{int}{}^{12}_{RR}$ becomes the complex conjugate of $H_{int}{}^{12}_{LL}$. }
\be
H_{int}{}^{12}_{LL} = i \mu_{12} \sum_{k=1}^{N}\psi_{1L}^k\psi_{2L}^k, \qquad H_{int}{}^{12}_{RR} = -i \mu_{12}\sum_{k=1}^{N} \psi_{1R}^k\psi_{2R}^k.
\ee
Another description of this model is the two coupled Maldacena-Qi Hamiltonian, which means the coupling of two coupled SYK models;
\ba
H_{4} &= (H_{1L} +  (-1)^{\f{q}{2}} H_{1R} + H_{int}{}^{11}_{LR}) + (  (-1)^{\f{q}{2}}H_{2L} +  H_{2R} +H_{int}{}^{22}_{LR}) +  H_{int}{}^{12}_{LL} +  H_{int}{}^{12}_{RR} \notag \\
&= H_{MQ_1} + H_{MQ_2}+  H_{int}{}^{12}_{LL} +  H_{int}{}^{12}_{RR} .
\ea
Here we make a pair between $L$ and $R$ and defined $H_{MQ_1} = H_{1L} +  (-1)^{\f{q}{2}} H_{1R} + H_{int}{}^{11}_{LR}$, $H_{MQ_2} =   (-1)^{\f{q}{2}}H_{2L} +  H_{2R} +H_{int}{}^{22}_{LR}$ .
We can also make a pair in $1$ and $2$ direction, and then we can describe as 
\ba
H_{4} &= (H_{1L} +  (-1)^{\f{q}{2}} H_{2L} + H_{int}{}^{12}_{LL}) + ( (-1)^{\f{q}{2}}H_{1R} +  H_{2R} +H_{int}{}^{12}_{RR}) + H_{int}{}^{11}_{LR} + H_{int}{}^{22}_{LR}  \notag \\
&= H_{MQ_L} + H_{MQ_R}  + H_{int}{}^{11}_{LR} + H_{int}{}^{22}_{LR} .
\ea

The ground state of the mass term Hamiltonian $H_{int}{}^{11}_{LR} +  H_{int}{}^{22}_{LR}$ is given by
\be
(\psi_{1L}^k + i\psi_{1R}^k) \ket{I_{LR}} = 0 ,\qquad (\psi_{2L}^k + i\psi_{2R}^k) \ket{I_{LR}} = 0.
\ee
The state satisfies 
\be
H_{1L} \ket{I_{LR}} = (-1)^{\f{q}{2}}H_{1R}, \qquad H_{2L} \ket{I_{LR}} = (-1)^{\f{q}{2}}H_{2R}
\ee
and 
\be
H_{int}{}^{11}_{LR} \ket{I_{LR}} = -H_{int}{}^{22}_{LR} \ket{I_{LR}}
\ee
Therefore, the state satisfies 
\be
H_{MQ_L} \ket{I_{LR}} = H_{MQ_R} \ket{I_{LR}},
\ee
and $\ket{I_{LR}}$ is interpreted as the infinite temperature thermofield  double state of the two coupled SYK model.

Similarly, the ground state of the mass term Hamiltonian $H_{int}{}^{12}_{LL} +  H_{int}{}^{12}_{RR}$ is given by
\be
(\psi_{1L}^k + i\psi_{2L}^k) \ket{I_{12}} = 0 ,\qquad (\psi_{1R}^k - i\psi_{2R}^k) \ket{I_{12}} = 0.
\ee
By changing the sign $\psi_{2R}^k = -\tilde{\psi}_{2R}^k$, we obtain the same relation with the $LR$ direction:
\be
(\psi_{1L}^k + i\psi_{2L}^k) \ket{I_{12}} = 0 ,\qquad (\psi_{1R}^k + i\tilde{\psi}_{2R}^k) \ket{I_{12}} = 0.
\ee
with the opposite sign in front of the mass term 
\be
H_{int}{}^{11}_{LR} = i \mu_{LR}\sum_{k=1}^{N} \psi_{1L}^k\psi_{1R}^k, \qquad H_{int}{}^{22}_{LR} = -i \mu_{LR}\sum_{k=1}^{N} \psi_{2L}^k\tilde{\psi}_{2R}^k,
\ee
Therefore, by the same calculation with the $LR$ infinite temperature thermofield double case, we obtain 
\be
H_{MQ_1}\ket{I_{12}}  = H_{MQ_2}\ket{I_{12}}
\ee
Again, $\ket{I_{12}}$ is the infinite temperature thermofield double state of the two coupled SYK model but in $12$ direction.

\begin{figure}[ht]
\begin{center}
\includegraphics[width=12cm]{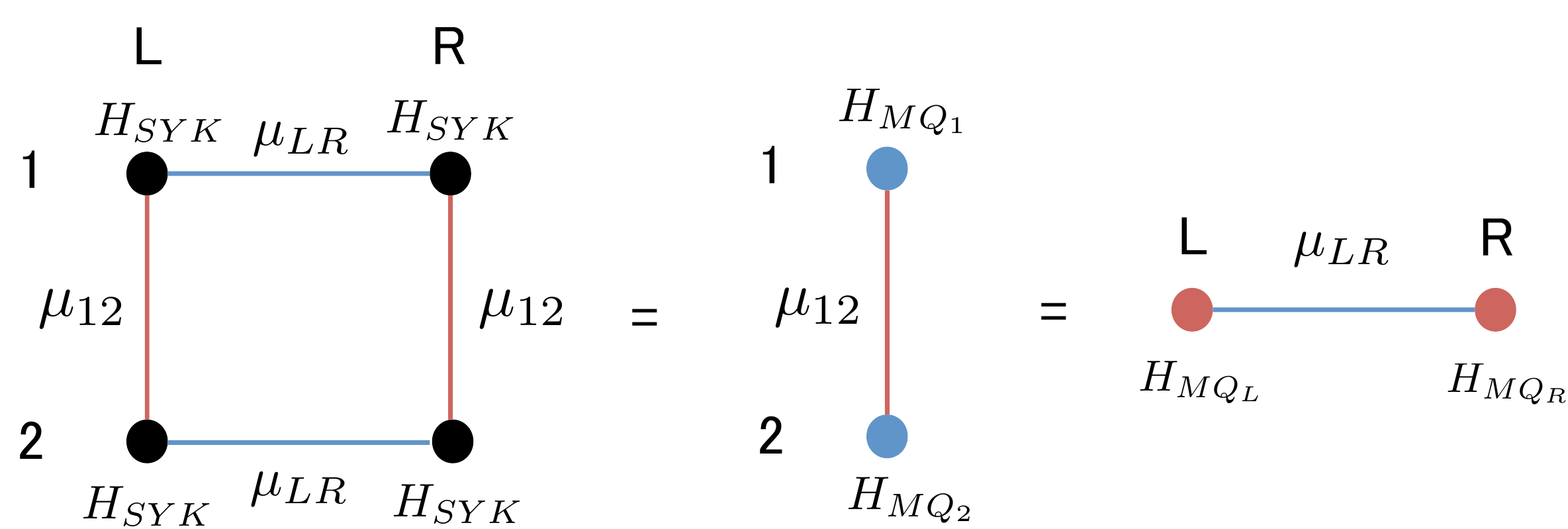}
\caption{The schematic form of the interaction.
{\bf Left:} On each dot we have a copy of the SYK model. The blue line is the interaction between $L$ and $R$.
The red line is the interaction between $1$ and $2$.
{\bf Middle:} We can think of the 4 coupled model as the two coupled Maldacena-Qi model.
{\bf Right:} Other way to think of the 4 coupled model as the two coupled Maldacena-Qi model.
} 
\label{fig:4cinteraction}
\end{center}
\end{figure}

There is a duality transformation, which plays an important role in our analysis.
The transformation is given by
\be
\psi_{1R}^k \to \psi_{2L}^k  , \qquad \psi_{2L}^k \to \psi_{1R}^k, \qquad  \psi_{2R}^k \to - \psi_{2R}^k. \label{eq:z2symmetrySYK}
\ee
that exchange the couplings $(\mu_{LR},\mu_{12}) \to ( \mu_{12},\mu_{LR})$.
Therefore this duality transformation exchanges the role of L-R direction and 1-2 direction.
We can think of strong-weak duality that exchange the strongly coupled a two coupled MQ model to a weakly coupled MQ model.
Moreover, this becomes a $\mathbb{Z}_2$ symmetry in the four coupled system at the "self dual" point $\mu_{LR} = \mu_{12}$.
At the symmetric point, an order parameter is 
\be
S_{dif} = \f{1}{2} (S_{LR}^{11} + S_{LR}^{22} ) - \f{1}{2} (S_{LL}^{12} - S_{RR}^{12} )
\ee
where $S_{AB}^{\alpha \beta} = -2i \psi_{\alpha A}\psi_{\beta B}$ are spin operators constructed from fermions.
This operator transforms $S_{dif} \to -S_{dif}$ and plays an role of order parameters.
If the operator $S_{dif}$ have an expectation value, the $\mathbb{Z}_2$ symmetry at $\mu_{LR} = \mu_{12}$ point is broken.

\subsubsection{The analysis of the mass term Hamiltonian for $N=1$ case \label{sec:MassTermH}}
We analyze the mass term Hamiltonian for $N=1$ i.e. one fermion per each site.
In this case, using the Jordan-Wigner transformation we can explicitly realize the fermions as 
\ba
\psi_{1L} = \f{1}{\s{2}} \sigma_x\otimes \mathbb{I} , \qquad \psi_{1R} = \f{1}{\s{2}} \sigma_y \otimes \mathbb{I} ,\qquad \psi_{2L} = -\f{1}{\s{2}} \sigma_z\otimes \sigma_y , \qquad \psi_{2R} = \f{1}{\s{2}}\sigma_z\otimes \sigma_x.
\ea
Then, the spin operators $S_{AB}^{\alpha \beta} = -2i \psi_{\alpha A}\psi_{\beta B}$ become
\be
S_{LR}^{11} = \sigma_z\otimes \mathbb{I}, \qquad S_{LR}^{22} = \mathbb{I}\otimes \sigma_z, \qquad S_{LL}^{12} = \sigma_y\otimes \sigma_y, \qquad S_{RR}^{12} = \sigma_x\otimes \sigma_x.
\ee
The mass term Hamiltonian, which is a $4 \times 4$ matrix, is \footnote{We can think of this Hamiltonian as the two cite quantum XY model $H_{XY} = J_x \sigma_1^x\sigma_2^x+J_y \sigma_1^y\sigma_2^y - h(\sigma_1^z + \sigma_2^z)$ with $J_x = \f{\mu_{12}}{2}, J_y = -\f{\mu_{12}}{2}$ and $h = \f{\mu_{LR}}{2}$. }
\be
H_{int}^{LR} = - \f{1}{2}\mu_{LR} (\sigma_z\otimes \mathbb{I}+ \mathbb{I}\otimes \sigma_z), \qquad  H_{int}^{12} = -\f{1}{2}\mu_{12}(\sigma_y\otimes \sigma_y - \sigma_x\otimes \sigma_x).
\ee
We denote the eigenstate of $S_{LR}^{ii}$ ($i = 1,2$) by $\ket{\uparrow}$.
Then, the ground state of $H_M = \mu_{LR} H_{int}^{LR} +\mu_{12} H_{int}^{12}$ is 
\be
\ket{G(\mu_1,\mu_2)} = \cos \f{\theta}{2} \ket{\uparrow \uparrow} - \sin \f{\theta}{2} \ket{\downarrow \downarrow},  \qquad  \tan \theta = \f{\mu_{12}}{\mu_{LR}}.
\ee
The full spectrum are given by $-\s{\mu_{LR}^2 + \mu_{12} ^2}, 0 , 0 , \s{\mu_{LR}^2 + \mu_{12} ^2}$.
There is always a gap between ground state and the first excited state that is given 
\be
E_g = \s{\mu_{LR}^2 + \mu_{12} ^2}.
\ee
When we fix $\mu_{LR}$ and increase $\mu_{12}$ from $0$ to $\infty$, this ground state continuously interpolate $\ket{I_{LR}}$ and $\ket{I_{12}}$ without closing a gap.
Energy gap monotonically increases as we increase $\mu_{12}$.

If we do not have SYK term in (\ref{eq:4coupledHamiltonian}), the system is just the $N$ copies of the two spin system described above.
In this way, the mass term Hamiltonian is completely diagonalized analytically. 
The energy gap is given by $\s{\mu_{LR}^2 + \mu_{12} ^2}$ again.

\subsection{The large $N$ effective action}
As we did in the two coupled SYK model, we can write down the action for ($G ,\Sigma$) variables.
The effective action in Euclidean signature is 
\ba
-S_E/N  &= \log \text{Pf}(\partial_{\tau} \delta_{AB}\delta^{\alpha\beta} -  \Sigma_{AB}^{\alpha\beta}) \notag \\
& - \f{1}{2}\sum_{A,B=L,R}\sum_{\alpha,\beta = 1,2} \int d\tau_1 d\tau_2\Bigg [ G_{AB}^{\alpha \beta}(\tau_1,\tau_2) \Sigma_{AB}^{\alpha \beta}(\tau_1,\tau_2)- s_{AB}^{\alpha \beta} \f{\mathcal{J}^2}{2q^2}[2G_{AB}^{\alpha \beta}(\tau_1,\tau_2)]^q \Bigg]  \notag \\
&+ i \mu_{LR} \int d\tau_1[ G_{LR}^{11}(\tau_1,\tau_1) + G_{LR}^{22}(\tau_1,\tau_1)] + i \mu_{12} \int d\tau_1 [G_{LL}^{12}(\tau_1,\tau_1) - G_{RR}^{12}(\tau_1,\tau_1)]. \label{eq:4coupledEAlargeN}
\ea
Here $G_{AB}^{\alpha\beta}(\tau_1,\tau_2)$ stands for the fermion correlation function
\be
G_{AB}^{\alpha\beta}(\tau_1,\tau_2) = \f{1}{N} \sum_{i=1}^N\braket{\psi_{A\alpha}^i(\tau_1)\psi_{B\beta}^i(\tau_2)},
\ee
and $\Sigma_{AB}^{\alpha\beta}(\tau_1,\tau_2)$ is the self energy.
$s_{AB}^{\alpha\beta}$ is given by 
\ba
&s_{AB}^{\alpha\beta} = \hat{s}_{AB}\tilde{s}^{\alpha\beta}, \notag \\
&\hat{s}_{LL} = \hat{s}_{RR} = 1, \qquad  \hat{s}_{LR} = \hat{s}_{RL} = (-1)^{\f{q}{2}}, \notag \\
&\tilde{s}^{11} =\tilde{s}^{22} = 1, \qquad \tilde{s}^{12} =\tilde{s}^{21} = (-1)^{\f{q}{2}}.
\ea
which comes from the factor $(-1)^{\f{q}{2}}$ in \eqref{eq:4coupledHamiltonian}.

By taking the saddle point of \eqref{eq:4coupledEAlargeN}, we obtain the Schwinger-Dyson equation.
The Schwinger Dyson equation is 
\be
G^{\alpha\beta}_{AB}(\omega) = -\big[(i\omega +  \Sigma(\omega))^{-1}\big]^{\alpha\beta}_{AB}. \label{eq:4coupledSDmomentum}
\ee
and 
\ba
\Sigma_{AB}^{\alpha\beta}(\tau_1,\tau_2) = s_{AB}^{\alpha\beta}J^2 \Big(G_{AB}^{\alpha\beta}(\tau_1,\tau_2) \Big)^{q-1} - i  \delta^{\alpha \beta} \mu_{AB}\delta(\tau_1-\tau_2) - i  \mu^{\alpha \beta} \sigma^z _{AB}\delta(\tau_1-\tau_2). \label{eq:4coupledSDtime}
\ea
The first equation \eqref{eq:4coupledSDmomentum} is more explicitly written as 
\ba
&\begin{pmatrix}
G^{11}_{LL}(\omega_n) & G^{11}_{LR} (\omega_n) & G^{12}_{LR}(\omega_n) & G^{12}_{LL} (\omega_n) \\
G^{11}_{RL}(\omega_n) & G^{11}_{RR}(\omega_n) & G^{12}_{RR}(\omega_n) & G^{12}_{RL} (\omega_n) \\
G^{21}_{RL}(\omega_n) & G^{21}_{RR}(\omega_n) & G^{22}_{RR}(\omega_n) & G^{22}_{RL} (\omega_n) \\
G^{21}_{LL}(\omega_n) & G^{21}_{LR}(\omega_n) & G^{22}_{LR}(\omega_n) & G^{22}_{LL} (\omega_n)
\end{pmatrix} \notag \\
&= -
\begin{pmatrix}
i\omega_n +\Sigma^{11}_{LL}(\omega_n) & \Sigma^{11}_{LR} (\omega_n)& \Sigma^{12}_{LR}(\omega_n) & \Sigma^{12}_{LL} (\omega_n) \\
\Sigma^{11}_{RL}(\omega_n) & i\omega_n +\Sigma^{11}_{RR}(\omega_n)& \Sigma^{12}_{RR}(\omega_n) & \Sigma^{12}_{RL} (\omega_n) \\
\Sigma^{21}_{RL}(\omega_n) &\Sigma^{21}_{RR}(\omega_n)&  i\omega_n + \Sigma^{22}_{RR}(\omega_n) & \Sigma^{22}_{RL} (\omega_n) \\
\Sigma^{21}_{LL}(\omega_n) &\Sigma^{21}_{LR}(\omega_n)&  \Sigma^{22}_{LR}(\omega_n) & i\omega_n +  \Sigma^{22}_{LL} (\omega_n)
\end{pmatrix}^{-1}
\ea
We have the $\mathbb{Z}_4$ symmetry \eqref{eq:Z4symMQ} in the two coupled SYK.
Since the construction of the 4 coupled SYK model is based on the two coupled SYK, the 4 coupled Hamiltonian \eqref{eq:4coupledHamiltonian} also have symmetries that are inherited from the two coupled SYK.
There is a $\mathbb{Z}_4^{\text{L-R}}$ symmetry 
\be
\psi_{1L} \to -\psi_{1R} ,\qquad \psi_ {1R} \to \psi_{1L}, \qquad \psi_{2L} \to \psi_{2R} ,\qquad \psi_ {2R} \to -\psi_{2L}.
\ee
which is the $\mathbb{Z}_4$ symmetry that exchanges the left SYK models and right SYK models.
There is another $\mathbb{Z}_4$ symmetry 
\be
\psi_{1L} \to -\psi_{2L} ,\qquad \psi_ {2L} \to \psi_{1L}, \qquad  \psi_{1R} \to  -\psi_{2R} ,\qquad \psi_ {2R} \to \psi_{1R}.
\ee
which exchanges the system $1$ and the system $2$. 
We call this $\mathbb{Z}_4$ symmetry $\mathbb{Z}_4^{\text{1-2}}$.
For symmetric configurations, the equation of motion simplifies.
The Schwinger Dyson equation reduces to
\ba
&\begin{pmatrix}
G^{11}_{LL}(\omega_n) & G^{11}_{LR} (\omega_n) & G^{12}_{LR}(\omega_n) & G^{12}_{LL} (\omega_n) \\
-G^{11}_{LR}(\omega_n) & G^{11}_{LL}(\omega_n) & G^{12}_{LL} (\omega_n) & -G^{12}_{LR}(\omega_n) \\
-G^{12}_{LR}(\omega_n)  & -G^{12}_{LL} (\omega_n) & G^{11}_{LL}(\omega_n) & G^{11}_{LR} (\omega_n) \\
-G^{12}_{LL} (\omega_n) &  G^{12}_{LR}(\omega_n) & -G^{11}_{LR} (\omega_n) & G^{11}_{LL}(\omega_n)
\end{pmatrix} \notag \\
&= -
\begin{pmatrix}
i\omega_n +\Sigma^{11}_{LL}(\omega_n) & \Sigma^{11}_{LR} (\omega_n)& \Sigma^{12}_{LR}(\omega_n) & \Sigma^{12}_{LL} (\omega_n) \\
-\Sigma^{11}_{LR} (\omega_n) & i\omega_n +\Sigma^{11}_{LL}(\omega_n)& \Sigma^{12}_{LL} (\omega_n) & -\Sigma^{12}_{LR}(\omega_n) \\
-\Sigma^{12}_{LR}(\omega_n) &-\Sigma^{12}_{LL} (\omega_n)&  i\omega_n + \Sigma^{11}_{LL}(\omega_n)& \Sigma^{11}_{LR} (\omega_n) \\
-\Sigma^{12}_{LL} (\omega_n) &\Sigma^{12}_{LR}(\omega_n)& -\Sigma^{11}_{LR} (\omega_n) & i\omega_n +  \Sigma^{11}_{LL}(\omega_n)
\end{pmatrix}^{-1}
\ea
If we write down the independent equations explicitly, they become
\ba
G^{11}_{LL}(\omega_n) &= -\f{i\omega_n +  \Sigma^{11}_{LL}(\omega_n)}{(i\omega_n +  \Sigma^{11}_{LL}(\omega_n))^2 +( \Sigma^{11}_{LR} (\omega_n))^2 + (\Sigma^{12}_{LL} (\omega_n) )^2 + (\Sigma^{12}_{LR}(\omega_n))^2}, \notag \\
G^{11}_{LR}(\omega_n) &= \f{\Sigma^{11}_{LR}(\omega_n)}{(i\omega_n +  \Sigma^{11}_{LL}(\omega_n))^2 +( \Sigma^{11}_{LR} (\omega_n))^2 + (\Sigma^{12}_{LL} (\omega_n) )^2 + (\Sigma^{12}_{LR}(\omega_n))^2}, \notag \\
G^{12}_{LL}(\omega_n) &= \f{\Sigma^{12}_{LL}(\omega_n)}{(i\omega_n +  \Sigma^{11}_{LL}(\omega_n))^2 +( \Sigma^{11}_{LR} (\omega_n))^2 + (\Sigma^{12}_{LL} (\omega_n) )^2 + (\Sigma^{12}_{LR}(\omega_n))^2}, \notag \\
G^{12}_{LR}(\omega_n) &= \f{\Sigma^{12}_{LR}(\omega_n)}{(i\omega_n +  \Sigma^{11}_{LL}(\omega_n))^2 +( \Sigma^{11}_{LR} (\omega_n))^2 + (\Sigma^{12}_{LL} (\omega_n) )^2 + (\Sigma^{12}_{LR}(\omega_n))^2}.
\ea
and
\ba
\Sigma_{LL}^{11}(\tau_1,\tau_2) &= \f{\mathcal{J}^2}{q} (2G_{LL}^{11}(\tau_1,\tau_2))^{q-1} \notag \\
\Sigma_{LR}^{11}(\tau_1,\tau_2) &= (-1)^{\f{q}{2}}\f{\mathcal{J}^2}{q} (2G_{LR}^{11}(\tau_1,\tau_2))^{q-1}  - i \mu_{LR} \delta(\tau) \notag \\
\Sigma_{LL}^{12}(\tau_1,\tau_2) &=  (-1)^{\f{q}{2}}\f{\mathcal{J}^2}{q} (2G_{LL}^{12}(\tau_1,\tau_2))^{q-1}  - i \mu_{12} \delta(\tau) \notag \\
\Sigma_{LR}^{12}(\tau_1,\tau_2) &= \f{\mathcal{J}^2}{q} (2G_{LR}^{12}(\tau_1,\tau_2))^{q-1}.
\ea
As far as we observed, the solutions that we obtained are symmetric under $\mathbb{Z}_4^{\text{L-R}}$ and $\mathbb{Z}_4^{\text{1-2}}$.
The equation for symmetric configurations are simpler and they are convenient to reduce the numerical costs.

\subsection{Phases of the 4 coupled SYK models at zero temperature \label{sec:4SYKPhaseDiagram}}
\begin{figure}
\begin{center}
\includegraphics[width=7cm]{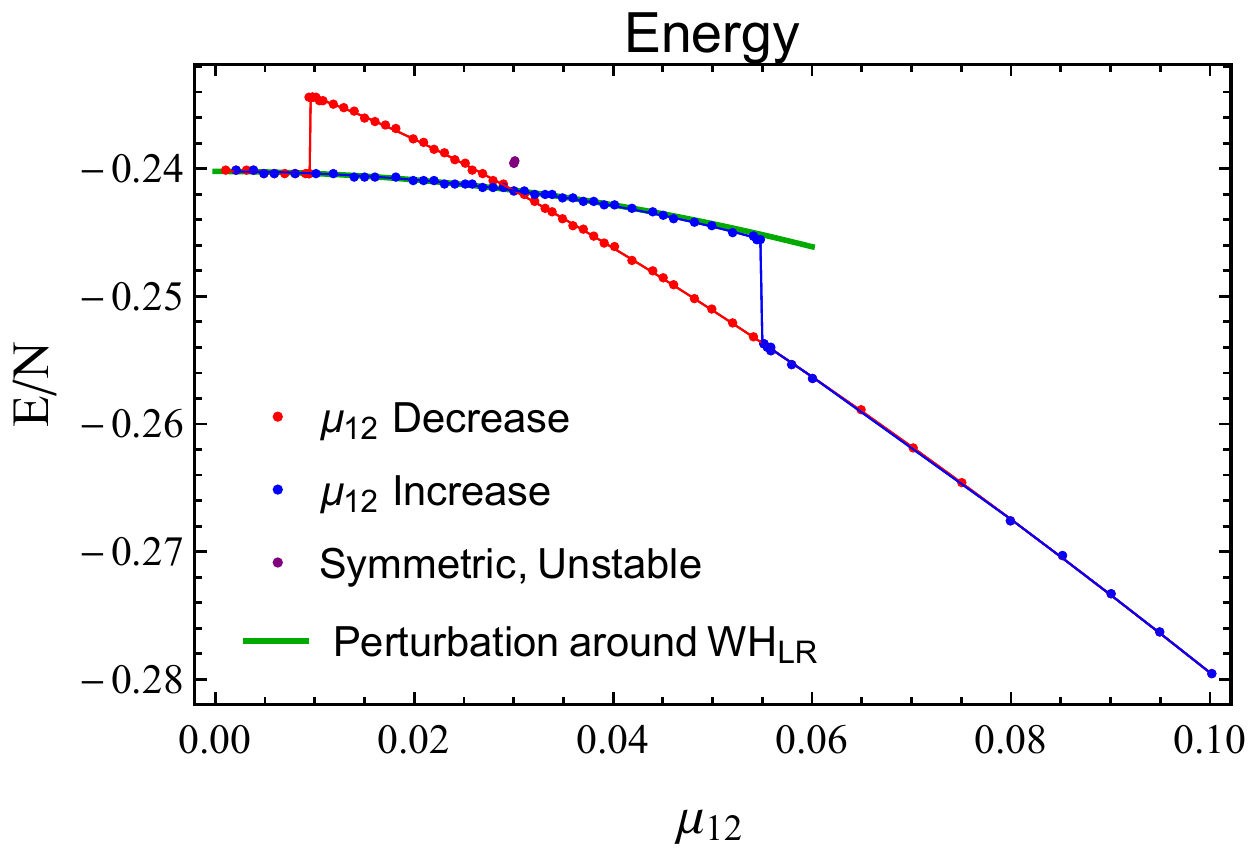}
\includegraphics[width=7cm]{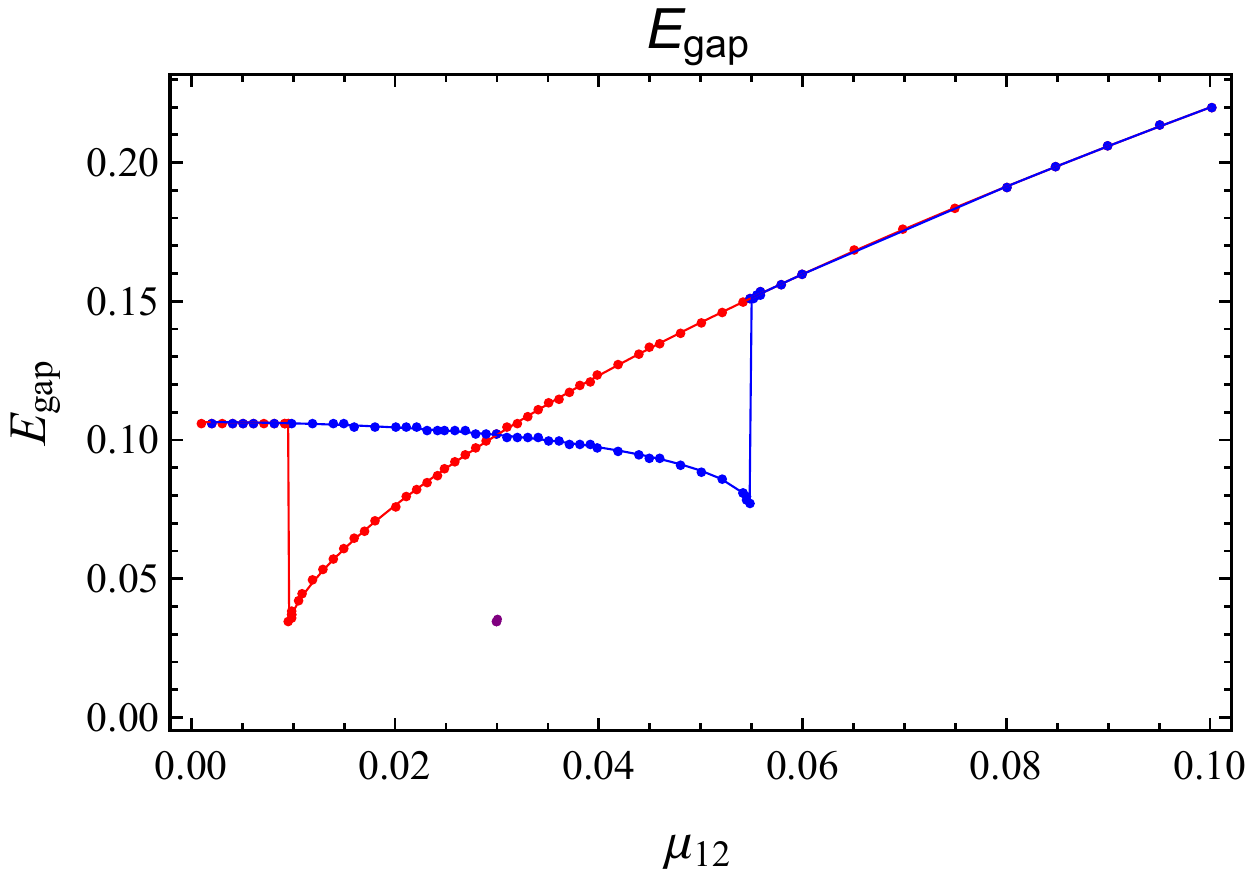}
\includegraphics[width=7cm]{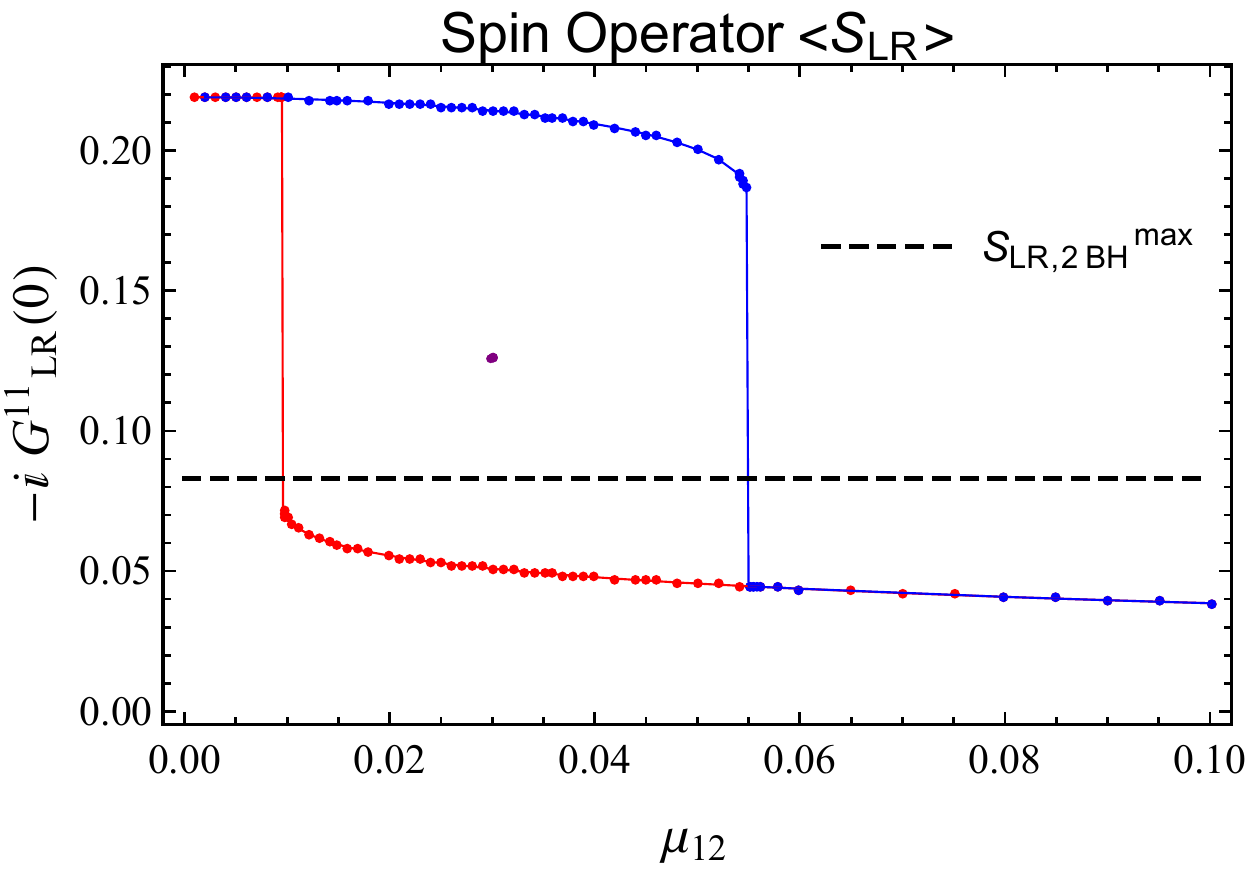}
\includegraphics[width=7cm]{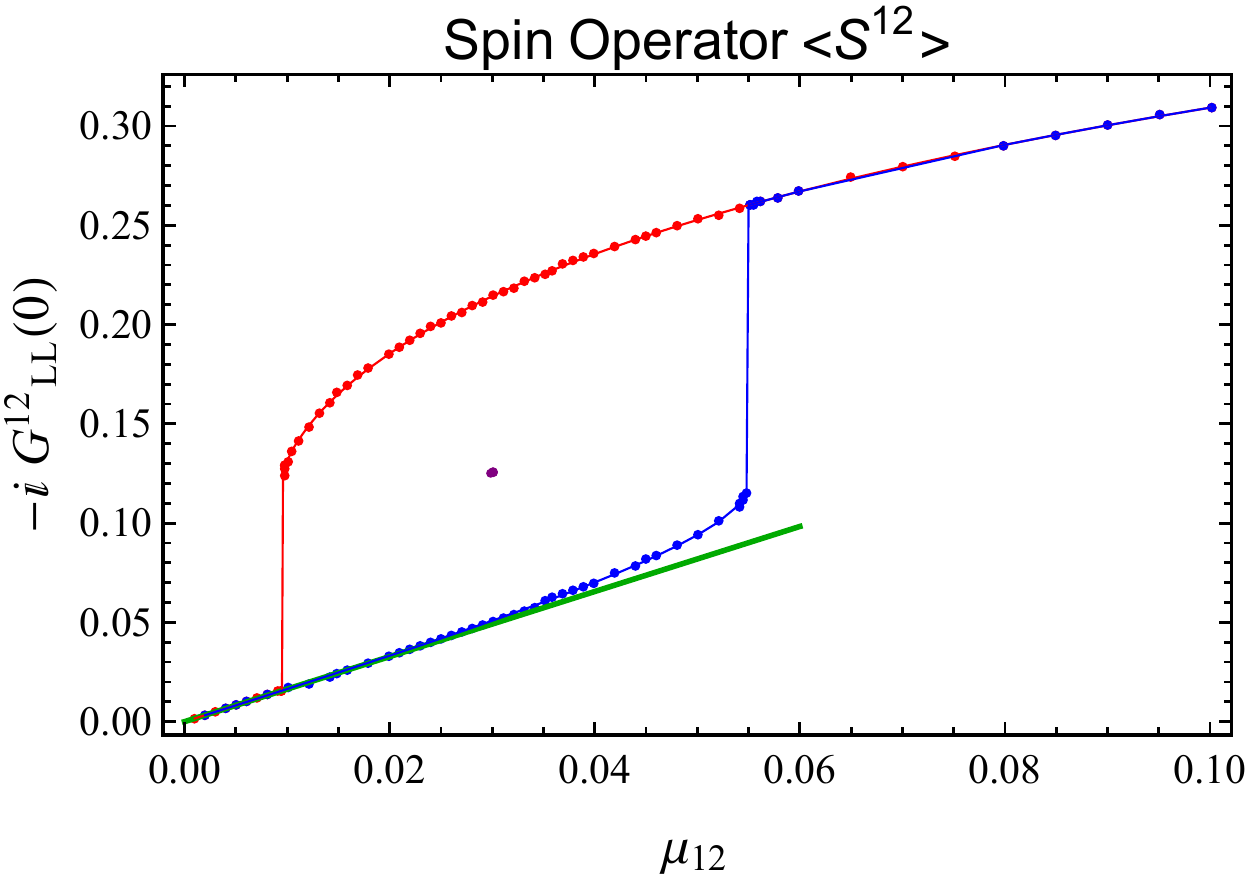}
\includegraphics[width=7cm]{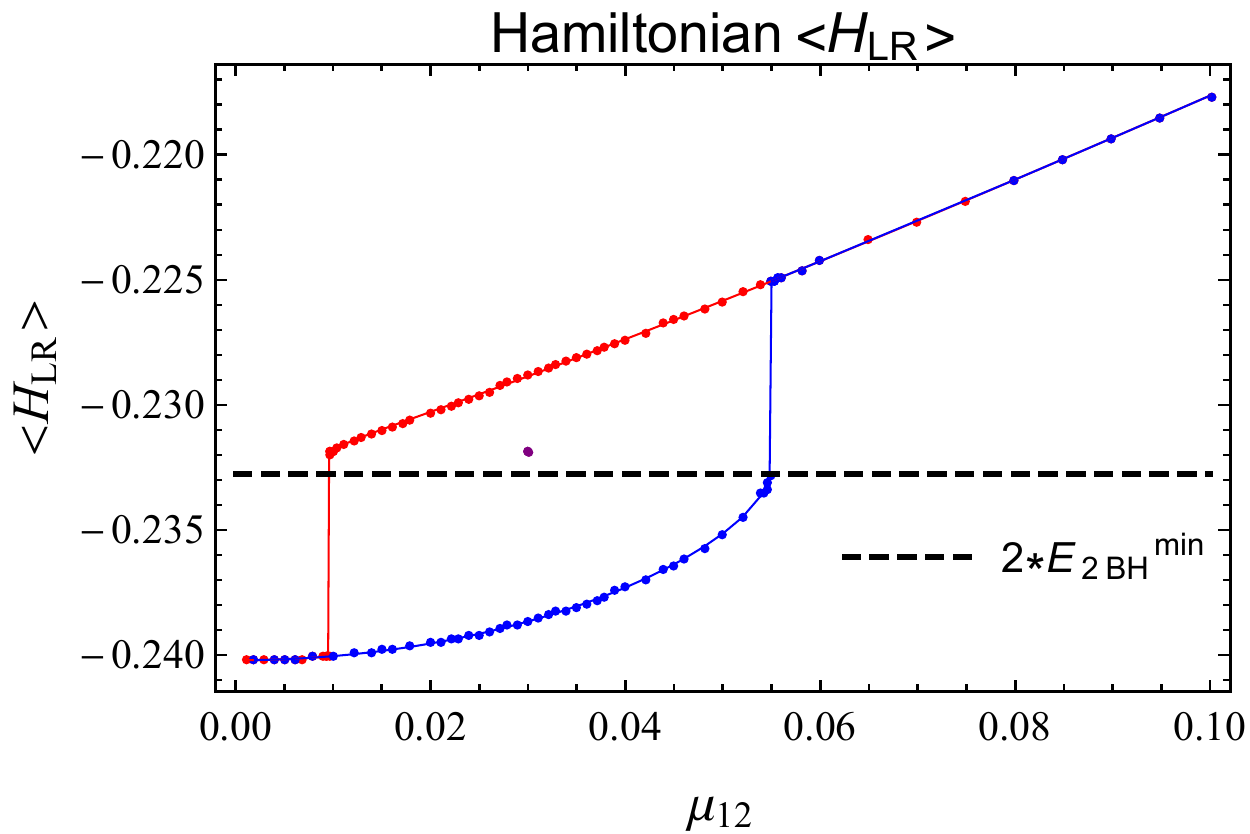}
\includegraphics[width=7cm]{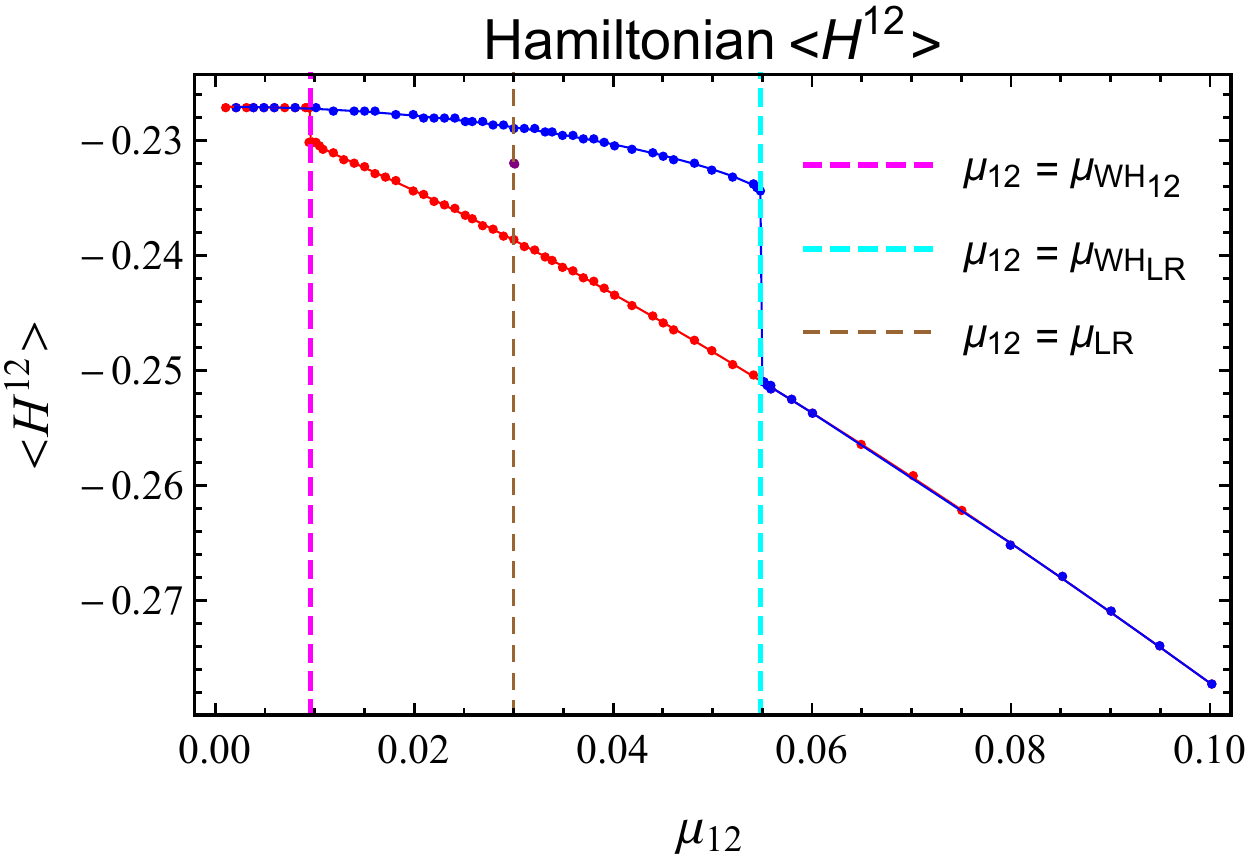}
\includegraphics[width=7cm]{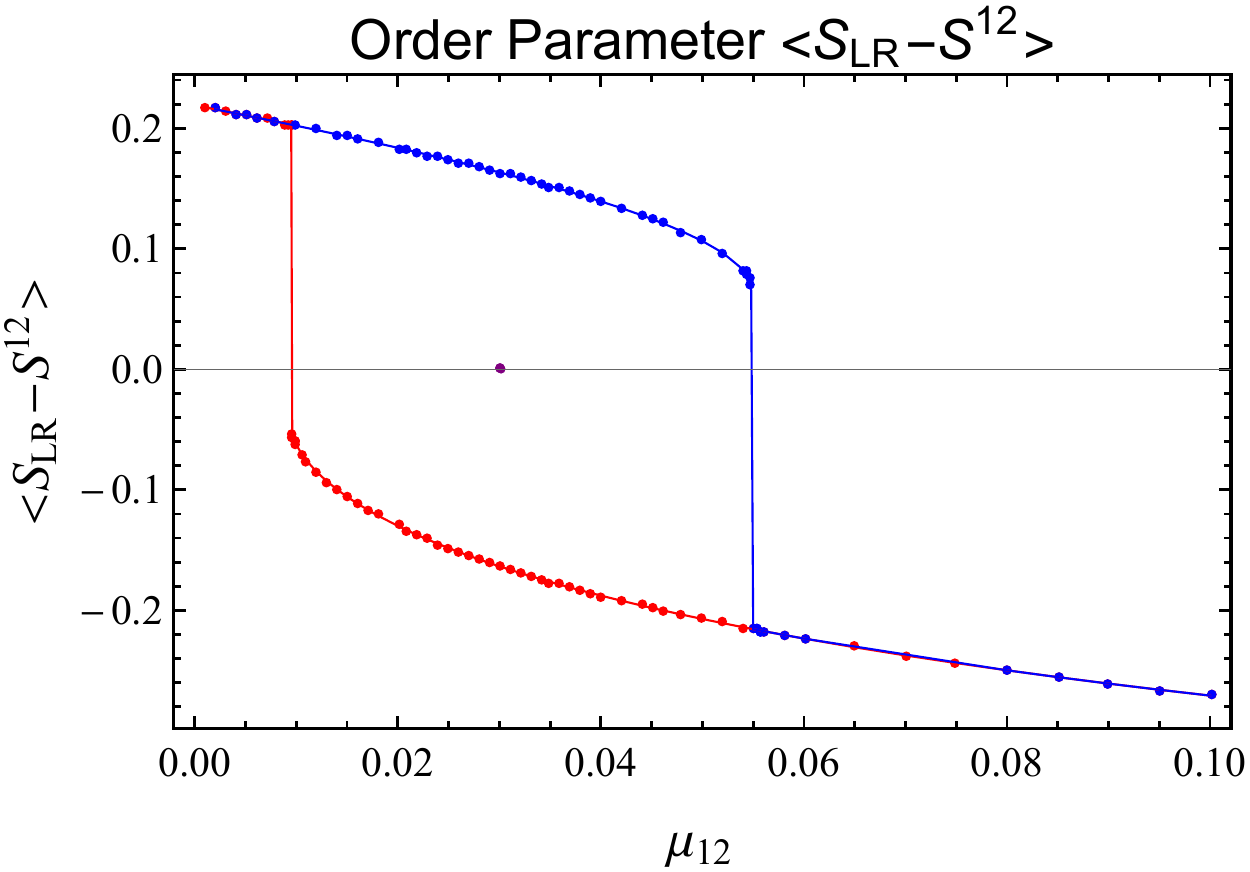}
\caption{The phase diagram of the four coupled model for $\mu_{LR}= 0.03$, $T=0.001$, and varying $\mu_{12}$.
} 
\label{fig:Mu300B1000}
\end{center}
\end{figure}

The Schwinger-Dyson equation \eqref{eq:4coupledSDmomentum}, \eqref{eq:4coupledSDtime} can be solved numerically.
We focus on the phase structure at zero temperature.
Using the correlation functions that are numerically found, we can evaluate several observables. 
We show the results in figure \ref{fig:Mu300B1000}, and we explain the results below.

\begin{figure}
\begin{center}
\includegraphics[width=6.5cm]{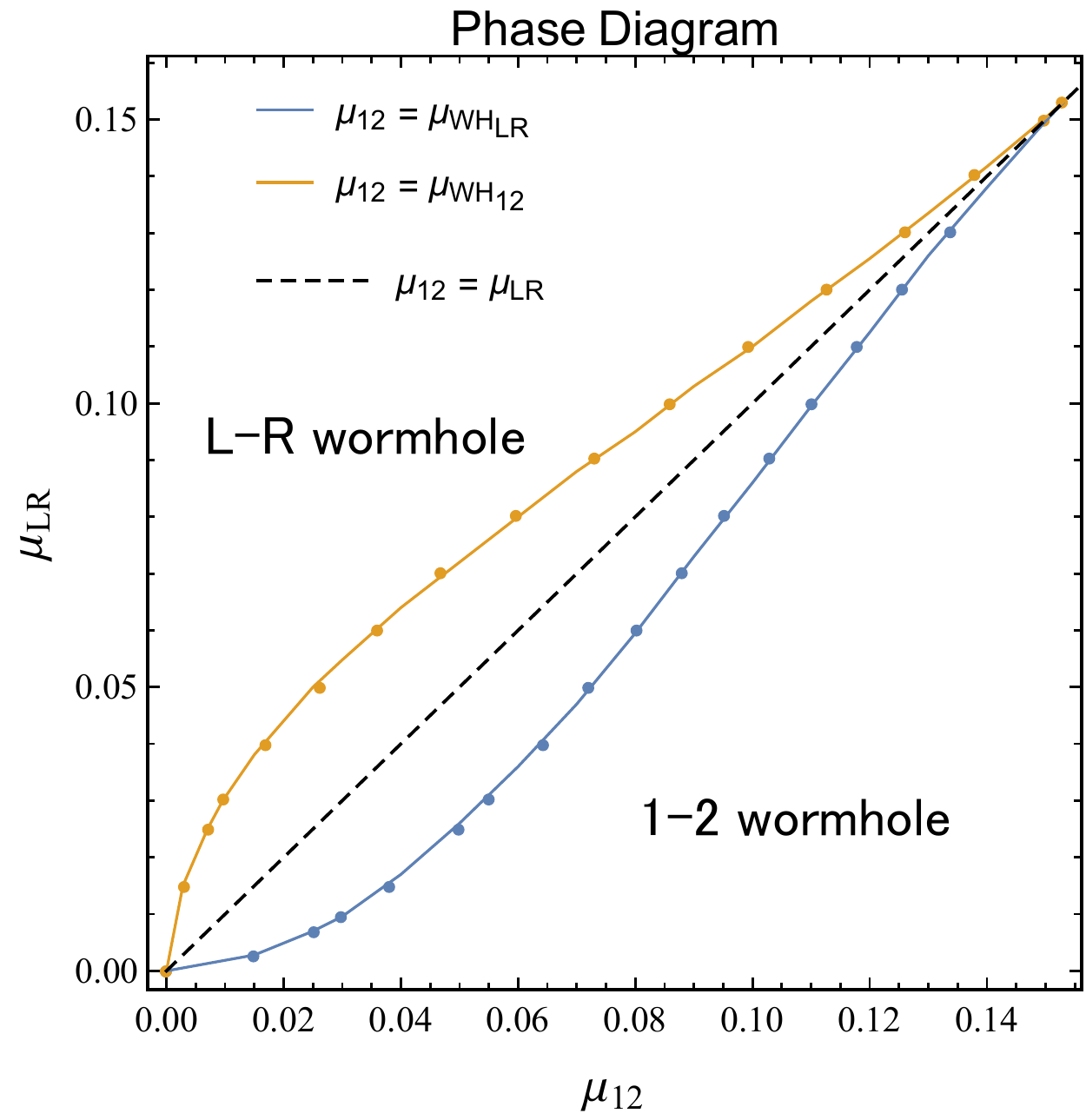}
\caption{The phase diagram at zero temperature in  the $\mu_{LR} - \mu_{12}$ plane.
The diagram is symmetric under the reflection along $\mu_{12} = \mu_{LR}$ line.
The blue line and the orange line meet at about $\mu_{12} = \mu_{LR} \approx 0.154$.
Beyond this critical point, the different wormhole phases are continuously connected.
} 
\label{fig:4coupledPhaseD}
\end{center}
\end{figure}

Here we explain the numerical results.
First we fix $\mu_{LR}$.
Within the range of the interaction $\mu_{12} \in [\mu_{WH_{12}} , \mu_{WH_{LR}}]$, there are two local minima.
One solution is understood as the wormhole in L-R direction, and the other is understood as the wormhole in 1-2 direction.
The 1-2 wormhole solution disappears at $\mu_{12} = \mu_{WH_{12}}$ whereas the L-R wormhole solution disappears at $\mu_{12} =\mu_{WH_{LR}} $.
$\mu_{WH_{12}}$ and $\mu_{WH_{LR}}$ are functions of $\mu_{LR}$.
The phase diagram is shown in figure \ref{fig:4coupledPhaseD}.
The dominance of the saddle is exchanged at $\mu_{LR} = \mu_{12}$ as we can see from the behavior of the system energy.
At $\mu_{LR} = \mu_{12}$ the expectation value of the spin operators $\braket{S_{LR}}$ and $\braket{S^{12}}$ jump \footnote{ Here $S_{LR}$ means $S_{LR}^{11}$ or $S_{LR}^{22}$, which will take the same value in our solutions. Similarly, by  Here $S^{12}$ we means either of $S_{LL}^{12}$ or  $-S_{RR}^{12}$. }.
At this point the Hamiltonian is  $\mathbb{Z}_2$ symmetric and the order parameter $S_{LR}- S^{12}$ has an expectation value, which means the spontaneous symmetry breaking of the $\mathbb{Z}_2$ symmetry.

In the L-R wormhole phase, increasing $\mu_{12}$, which is the strength of the interaction $ H^{12}_{LL}+ H^{12}_{RR}$ in 1-2 direction, actually decreases the energy gap of the system rather than increasing. 
This is different from what happens the four coupled model without the SYK interaction terms as we saw in section \ref{sec:MassTermH}.
In that case the gap is given by $\s{\mu_{12}^2 + \mu_{LR}^2}$ and we observe that the energy gap increases when we increase $\mu_{12}$.
Therefore, this decreasing behavior is a strongly coupled phenomenon.
In 
real time,
 the inverse of the energy gap plays the role of the wormhole length.
Therefore, we can interpret that  entangling bulk fields in two different wormholes increases the wormhole length.
In the bulk language, we expect that entangling bulk fields in different wormholes leads to bulk excitation, which leads to increase of the wormhole length.

Since we observed $\mathbb{Z}_2$ symmetry breaking at $\mu_{12} = \mu_{LR}$, we expect that we also have an symmetric solution which is unstable.
We can actually find a symmetric solution numerically imposing the symmetry in \eqref{eq:z2symmetrySYK}.
This symmetric saddle is unstable, which we can confirm numerically \footnote{For example, we can perturb the solution by some $G_{AB}^{\alpha \beta} \to G_{AB}^{\alpha \beta} + \delta  G_{AB}^{\alpha \beta} $ with small perturbation $\delta  G_{AB}^{\alpha \beta}$.
Then, the solution falls to the LR wormhole or the 12 wormhole depending on the perturbation. }.
We expect that this unstable solution exists even for $\mu_{12} \neq \mu_{LR}$ thought we can not reach it numerically because of instability.

The first order phase transition only exists for small $\mu_{LR}$.
We checked this numerically and the phase transition disappears around $\mu_{LR} \sim 0.154$.
Beyond this critical value, the L-R womrhole, 1-2 wormhole and the expected unstable solution meet.
At the symmetric point $\mu_{LR} = \mu_{12}$ actually we found that the phase transition disappears and we only have symmetric solution.

\begin{figure}
\begin{center}
\includegraphics[width=6.5cm]{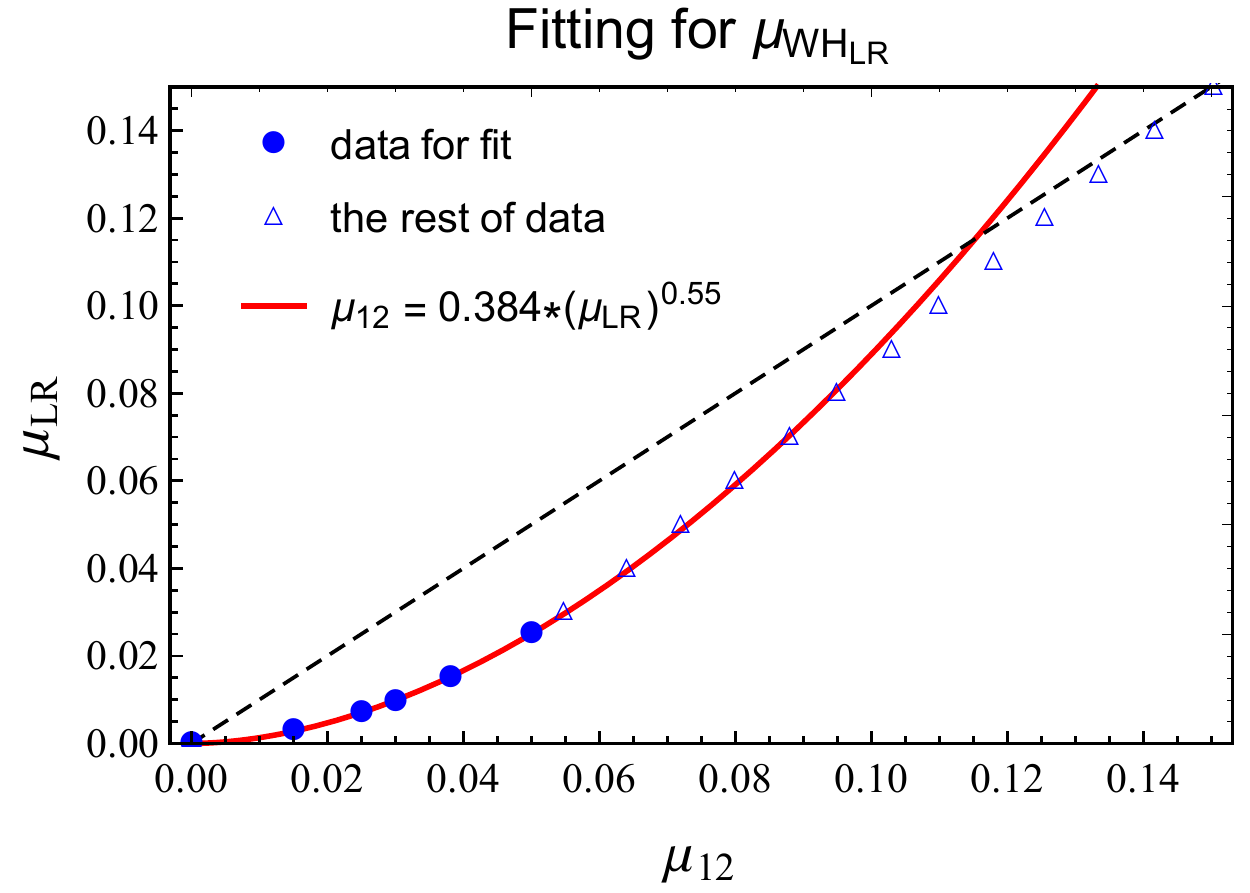}
\includegraphics[width=6.5cm]{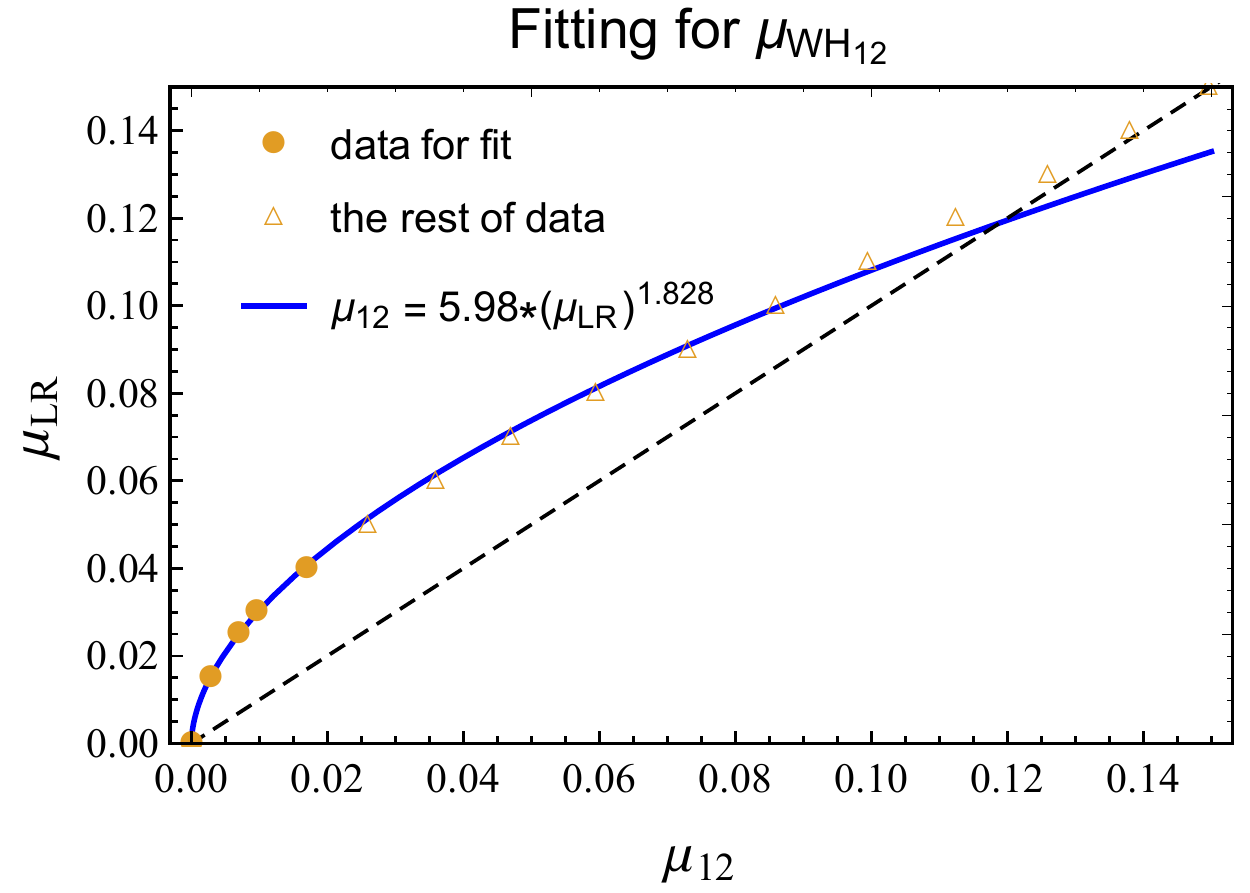}
\caption{The fitting of $\mu_{WH_{LR}}$ and $\mu_{WH_{12}}$. 
We use the data for small $\mu_{12}$ and fit the data assuming power low behavior.
The rest of date are well fitted when $\mu_{12}$ is sufficiently small.
{\bf Left:} The fitting of $\mu_{WH_{LR}}$.
{\bf Right:} The fitting of $\mu_{WH_{12}}$.
} 
\label{fig:MuBoundaryFit}
\end{center}
\end{figure}

We have also found a power low behavior for $\mu_{WH_{12}}$ and $\mu_{WH_{LR}}$ as functions of $\mu_{LR}$ at small $\mu_{LR}$, see Figure \ref{fig:MuBoundaryFit}.
$\mu_{WH_{LR}}$ behaves as $\mu_{WH_{LR}} \sim (\mu_{LR})^{0.55}$.
The power is close to $1/2$.
This power low that is smaller than $1$ means that the maximal value of $\mu_{12}$ with L-R wormhole saddle with fixed $\mu_{LR}$ is parametrically large compared to $\mu_{LR}$ when $\mu$'s are small.
This means that even for large $\mu_{12}$, the L-R wormhole saddle exists.
Roughly speaking, we can entangle bulk matter fields a lot in different wormholes through boundary interactions.

\subsection{small $\mu_{12}$ perturbation}
When $\mu_{12}$ is small, we can treat this term as an perturbation term and use (conformal) perturbation theory.
The $G_{LL}^{12}$ correlator becomes 
\ba
G_{LL}^{12}(\tau_1,\tau_2) &=  \braket{\psi_L^1 (\tau_1)\psi_L^2(\tau_2) ( i \mu_{12} \int_{-\infty}^{\infty} d\tau \psi_L^1(\tau)\psi_L^2(\tau) - i \mu_{12} \int_{-\infty}^{\infty} d\tau \psi_R^1(\tau)\psi_R^2(\tau) ) } \notag \\
&=   i \mu_{12}  \int _{-\infty}^{\infty}  d\tau G_{LL}^{11}( \tau - \tau_1)G_{LL}^{22}(\tau- \tau_2) + i \mu_{12} \int _{-\infty}^{\infty}  d\tau G_{LR}^{11}(\tau - \tau_1)G_{LR}^{22}(\tau - \tau_2) \notag \\
&=   i \mu_{12} \Bigg[ \int _{-\infty}^{\infty}  d\tau G_{LL}^{two}( \tau - \tau_1)G_{LL}^{two}(\tau- \tau_2) + \int _{-\infty}^{\infty}  d\tau G_{LR}^{two}(\tau - \tau_1)G_{LR}^{two}(\tau - \tau_2)\Bigg].  \label{eq:perturbation1}
\ea
We used the symmetry $G_{LL}^{11}(- \tau) = -G_{LL}^{11}(\tau) $ and  $G_{LR}^{11}(- \tau) = G_{LR}^{11}(\tau) $.
In the last line we use $G_{LL}^{two}$ and $G_{LR}^{two}$, which is the correlation function of the two coupled model \eqref{eq:SDMaldacenaQi1}, since we use perturbation theory around $\mu_{12} = 0$ and at this point we just have two decoupled Maldacena-Qi model.
This relation holds whether or not the conformal limit is applicable.
\begin{figure}[H]
\begin{center}
\includegraphics[width=10.0cm]{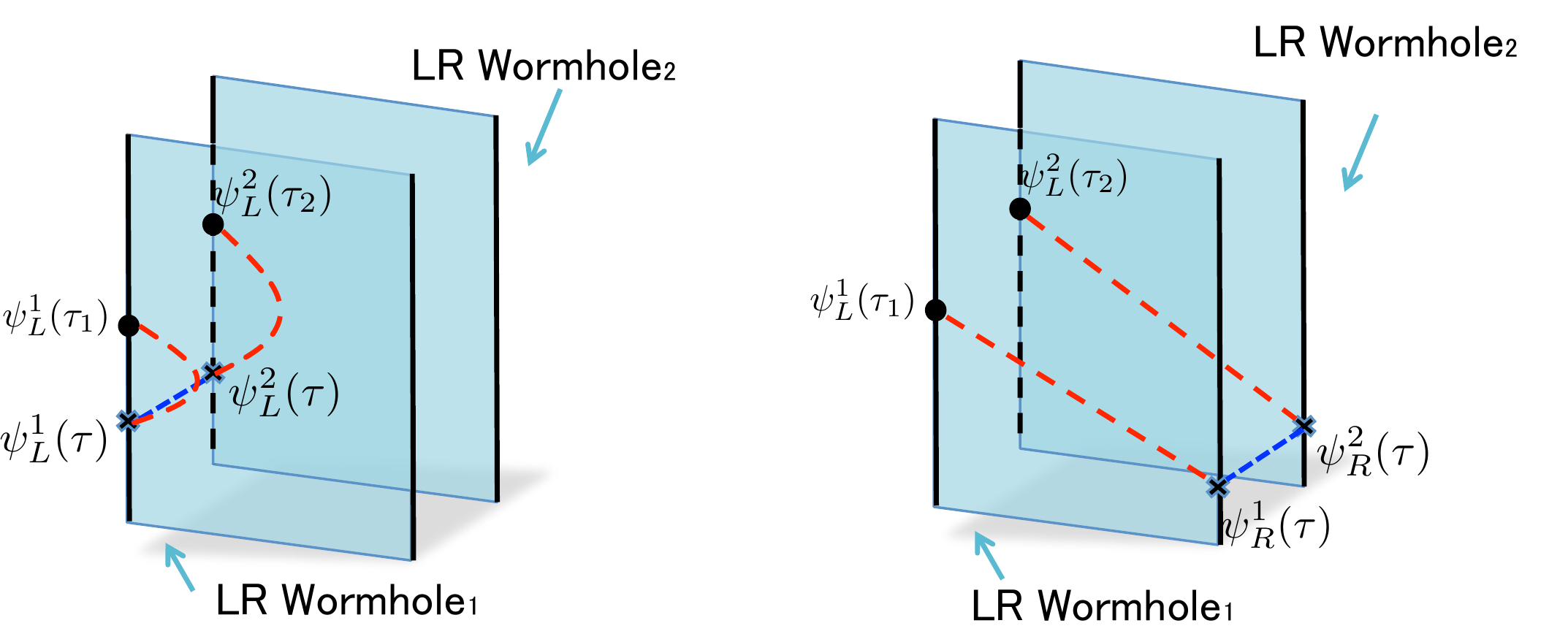}
\caption{A bulk picture of the perturbation in (\ref{eq:perturbation1}). 
} 
\label{fig:BulkPerturbation}
\end{center}
\end{figure}
Intuitively, these perturbation corresponds to taking into account the bulk diagrams in figure \ref{fig:BulkPerturbation}.
The first term in (\ref{eq:perturbation1}) comes from the direct interactions in the left boundaries and the second term comes from that in the right boundaries.
Once we obtain the correlation function $G_{LL}^{12}$ in this perturbation, we can calculate the physical quantities in perturbation in $\mu_{12}$.
For example, we can calculate the spin operator expectation value $\braket{S_{LL}^{12}} = -2iG_{LL}^{12}(0)$ perturbatively by   
\be
G_{LL}^{12}(0) =   i \mu_{12} \Bigg[ \int _{-\infty}^{\infty}  d\tau G_{LL}^{two}( \tau )^2 + \int _{-\infty}^{\infty}  d\tau G_{LR}^{two}(\tau )^2\Bigg]. 
\ee
Then we can also evaluate the change of the ground state energy of the Hamiltonian \eqref{eq:4coupledHamiltonian}.
We show this perturbation in figure \ref{fig:Mu300B1000} and for $\mu_{12}$ the results agree with the exact numerical results.

\subsection{Varying $\mu_{12}$ with $\mu_{LR} + \mu_{12}$ to be fixed }
\begin{figure}[H]
\begin{center}
\includegraphics[width=7cm]{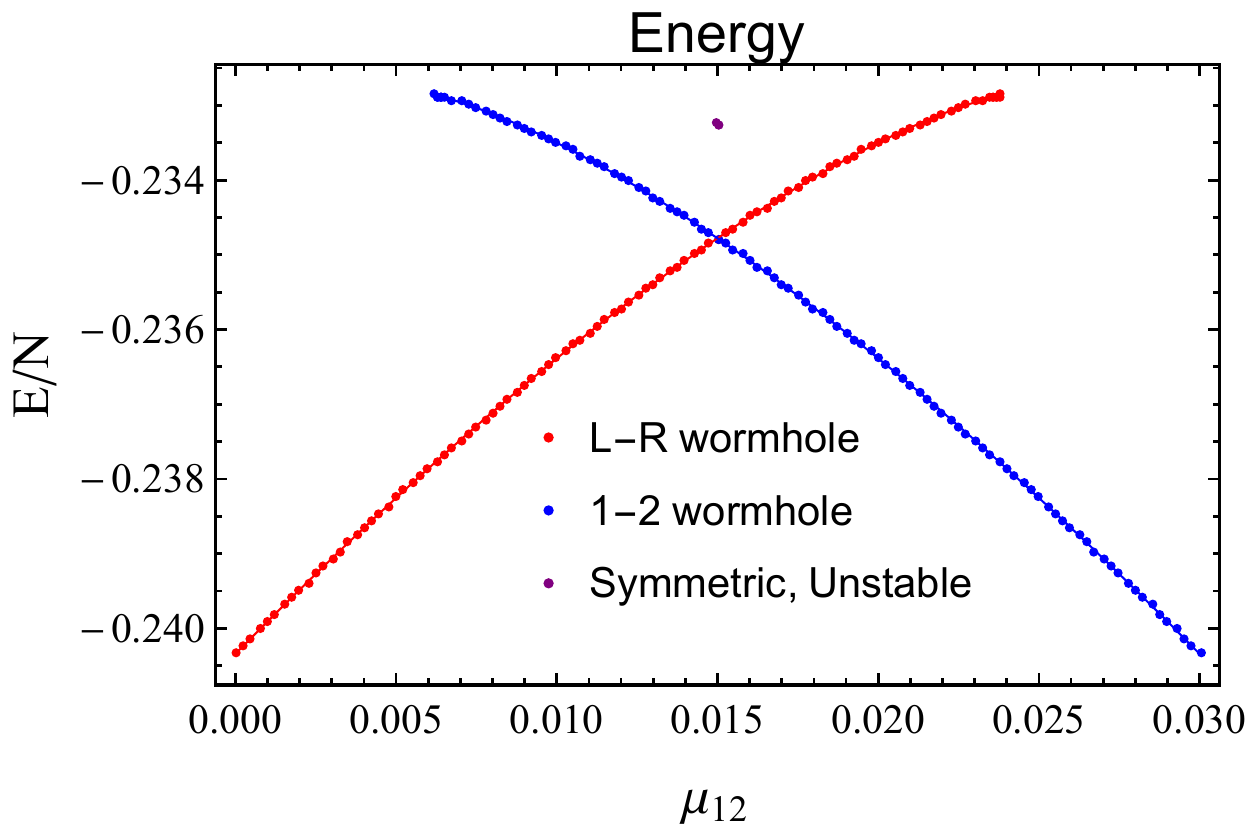}
\includegraphics[width=7cm]{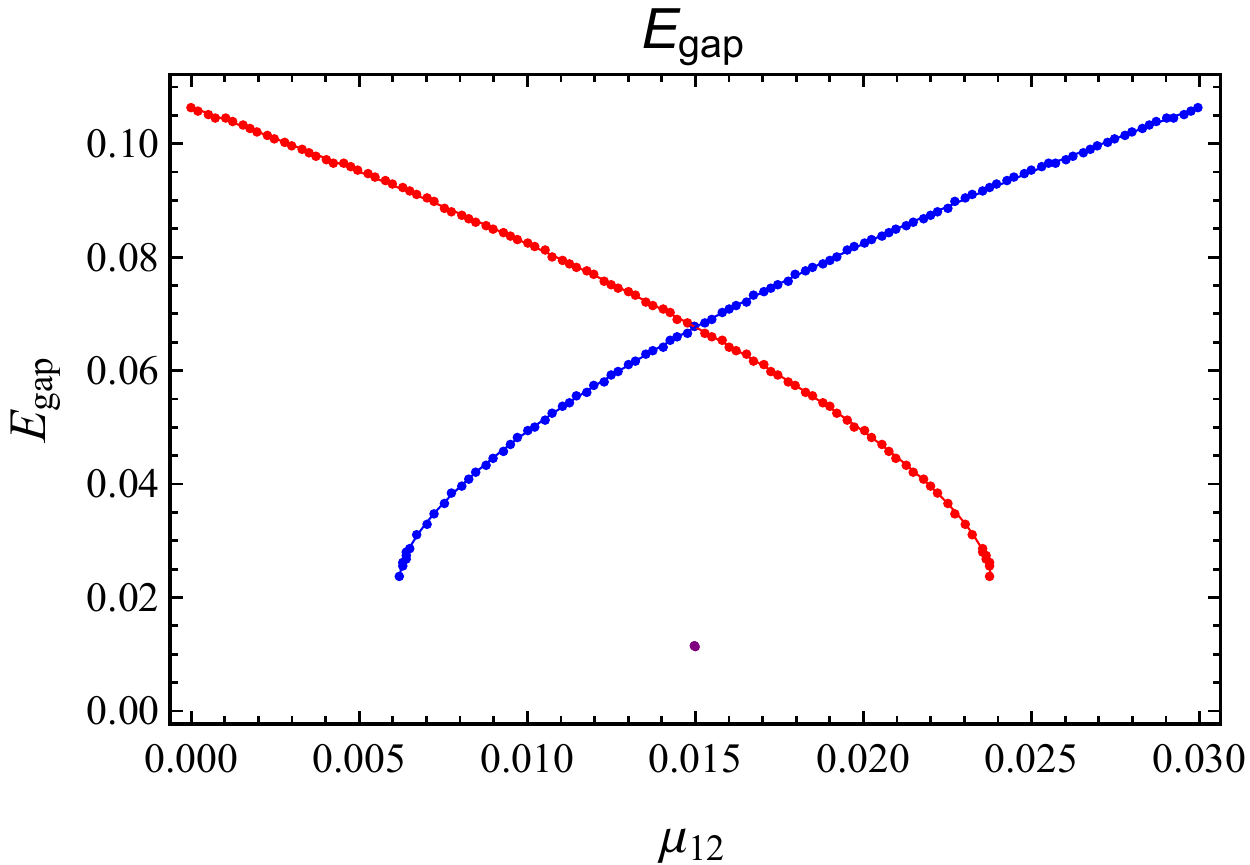}
\includegraphics[width=7cm]{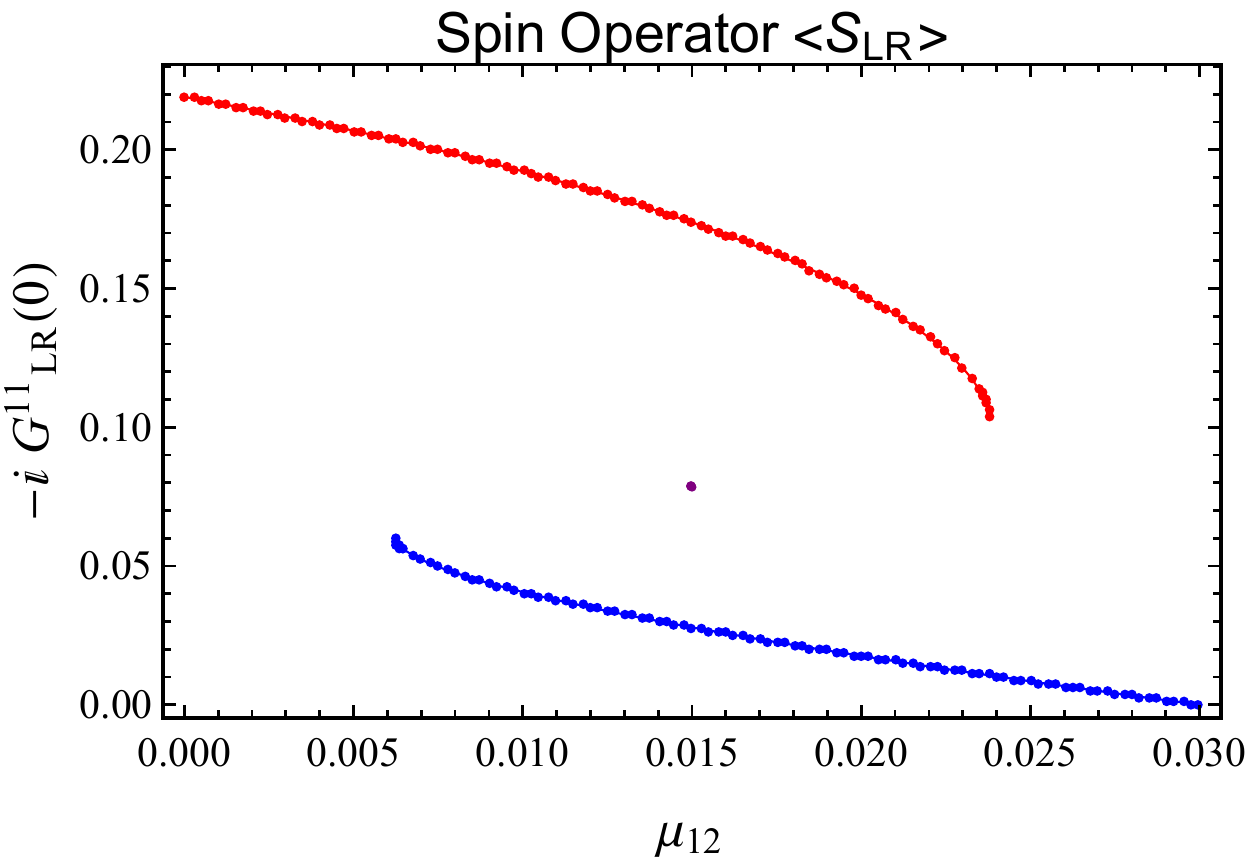}
\includegraphics[width=7cm]{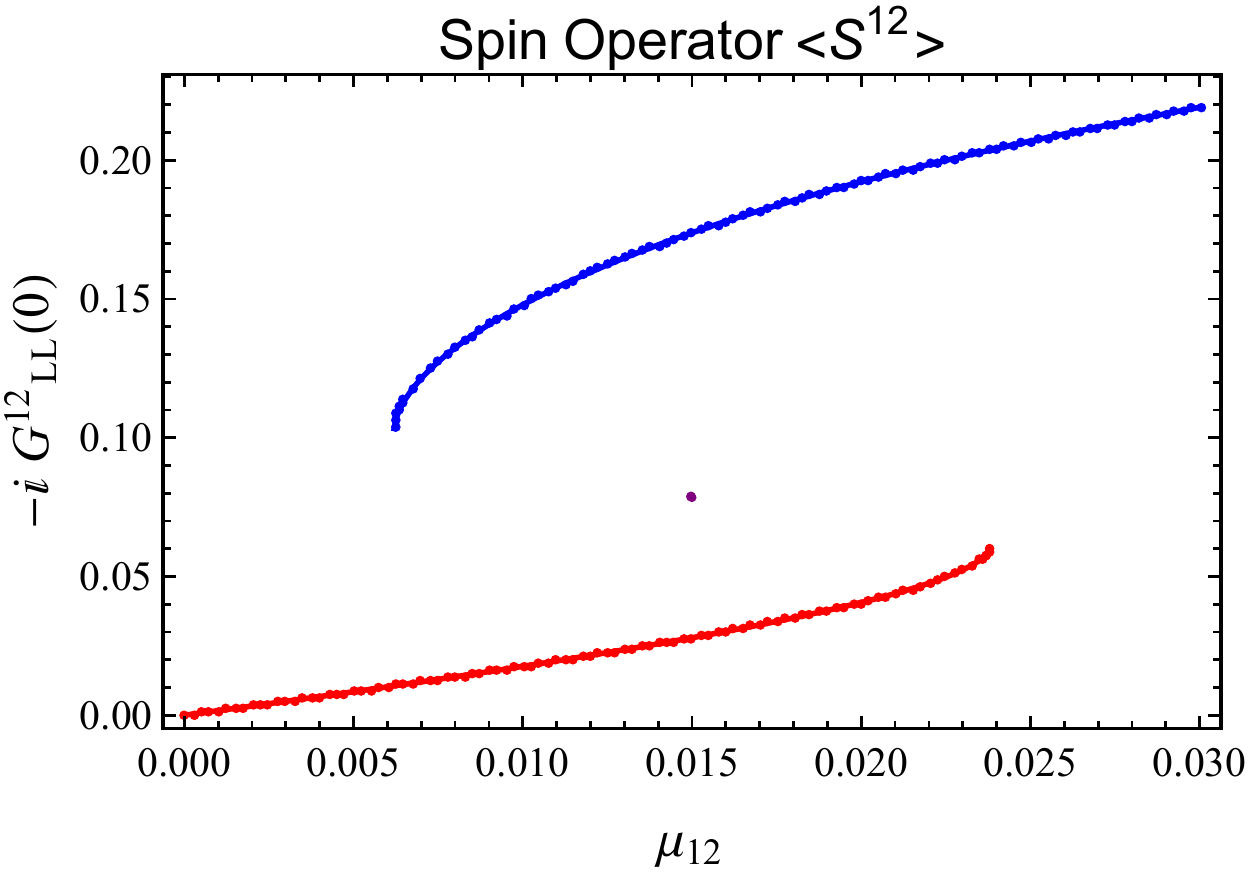}
\includegraphics[width=7cm]{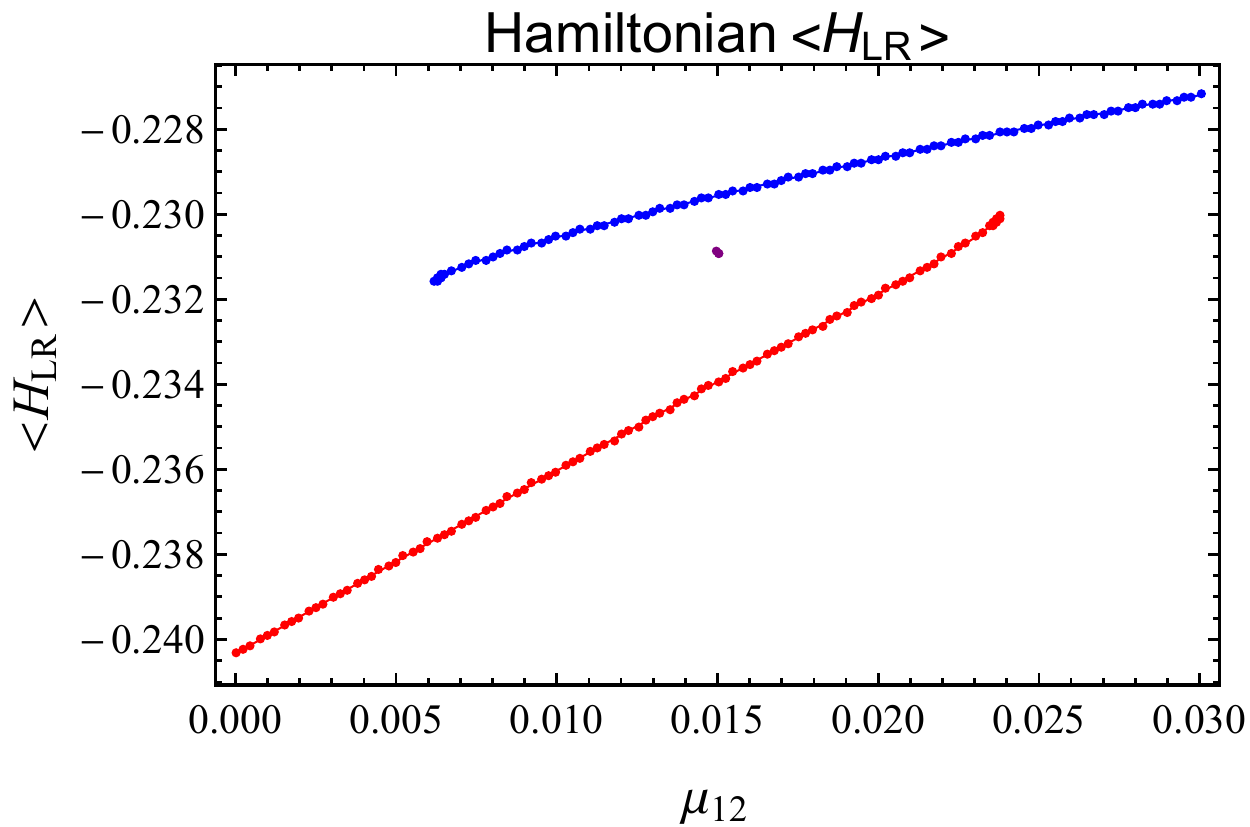}
\includegraphics[width=7cm]{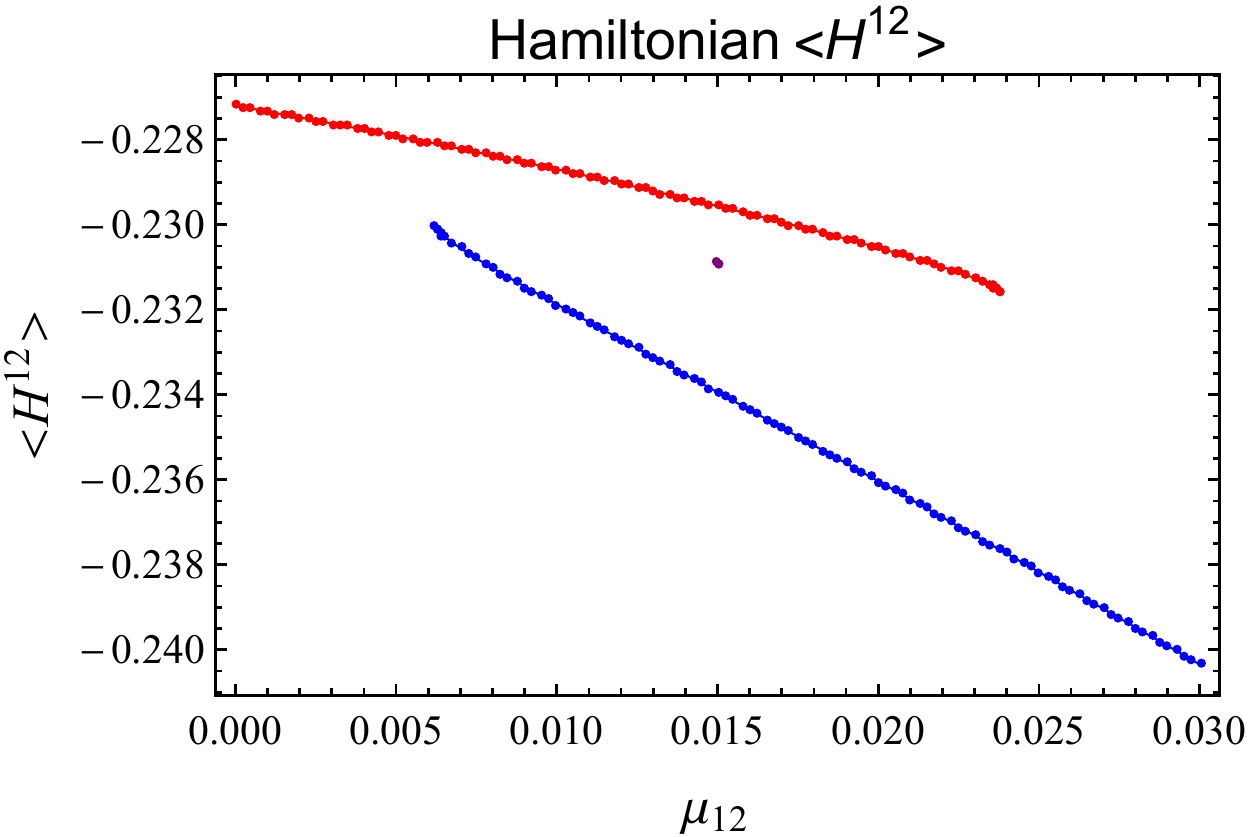}
\includegraphics[width=7cm]{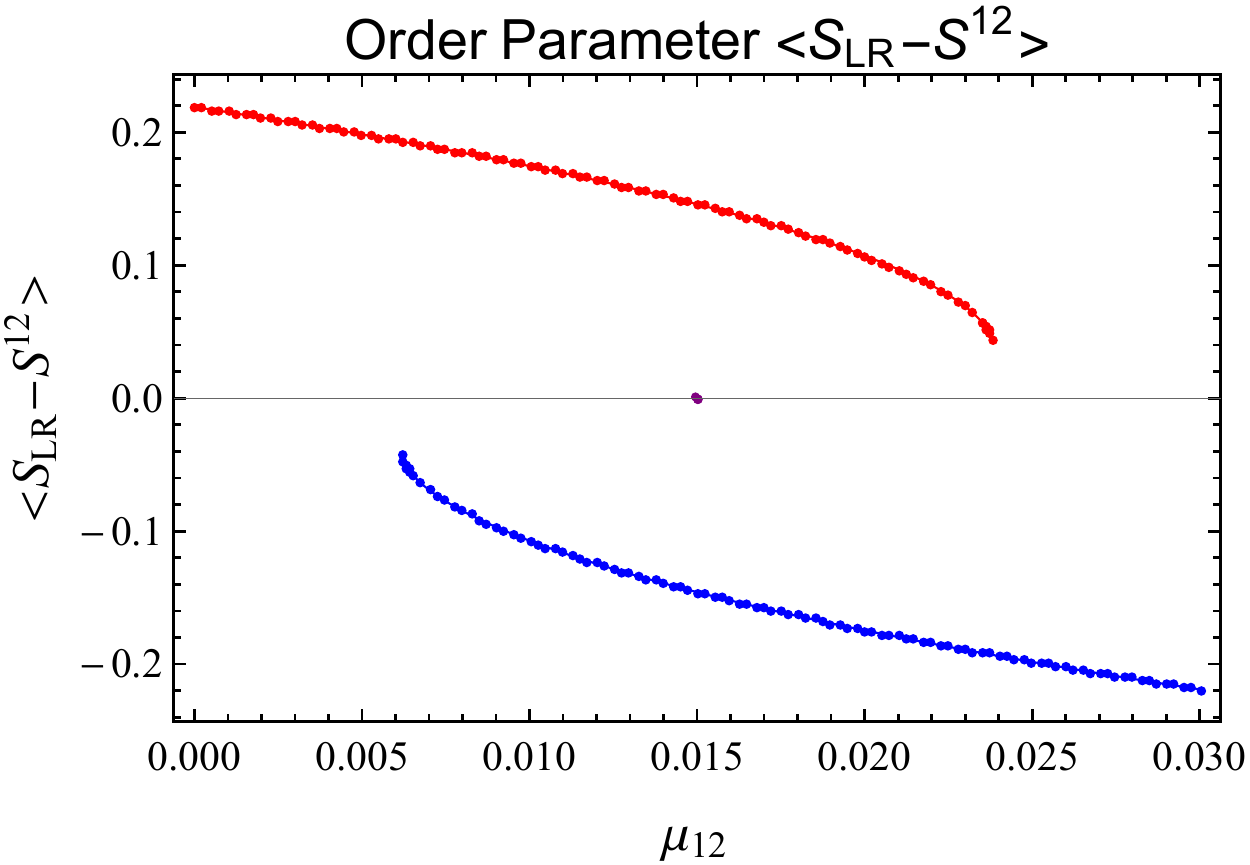}
\caption{he phase diagram of the four coupled model for $\mu_{LR} + \mu_{12} = 0.03$, $T=0.001$ with varying $\mu_{12}$
} 
\label{fig:MuTot300B1000}
\end{center}
\end{figure}

We can also study the phase diagram when we fix $\mu_{LR} + \mu_{12}$ to be a constant.
In other words, we change $\mu_{12}$ with fixing $\mu_{tot} \equiv \mu_{LR} +  \mu_{12}$.
The results are shown in figure \ref{fig:MuTot300B1000}.
Because we treat $\mu_{LR}$ and $\mu_{12}$ equally, the phase diagram becomes symmetric under the exchange of L-R direction and 1-2 direction.
If we compare with the classical Ising model, $\mu_{12} -\mu_{tot}/2$ is an analog of external magnetic field and the order parameter $\braket{S_{LR} - S^{12}}$  is an analog of magnetization.

In the analysis in the gravity side, we also study the same type of change of couplings and draw the similar behavior of the system energy.

\subsection{Effective potential}

\begin{figure}[ht]
\begin{center}
\includegraphics[width=8.5cm]{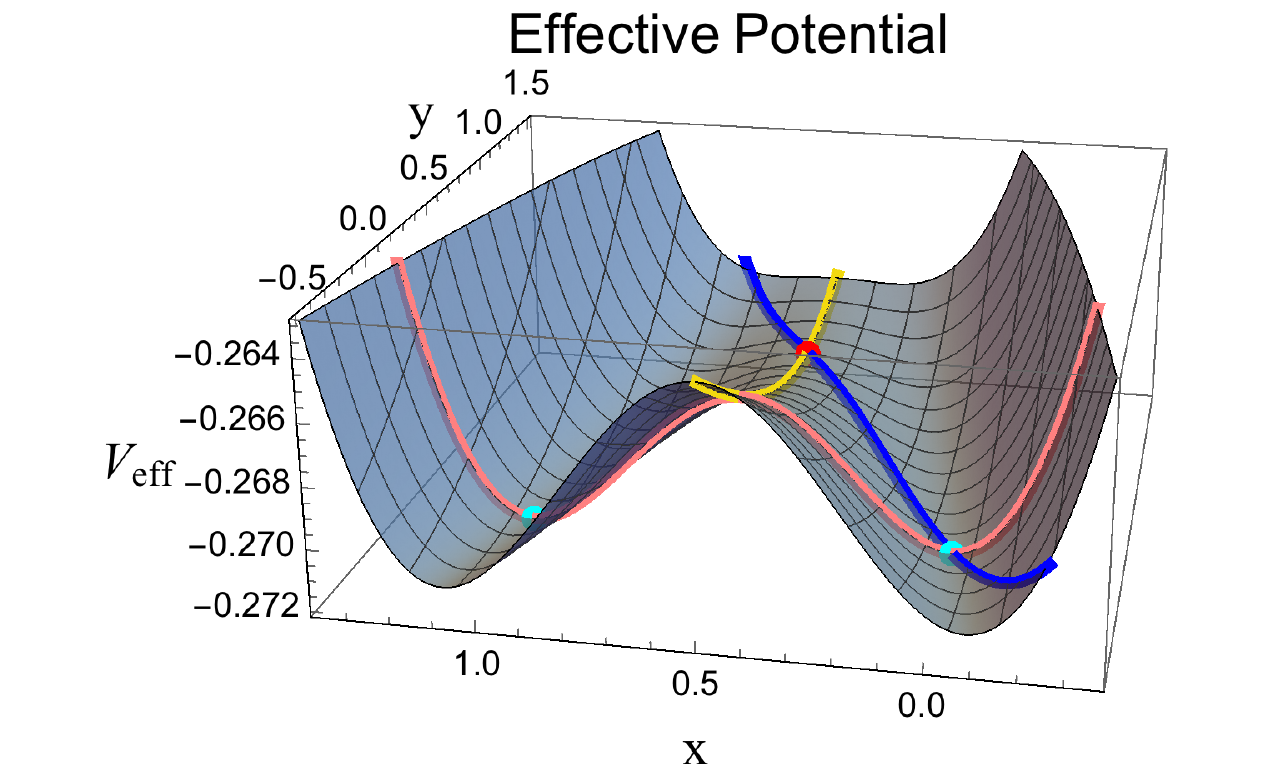} 
\includegraphics[width=6.6cm]{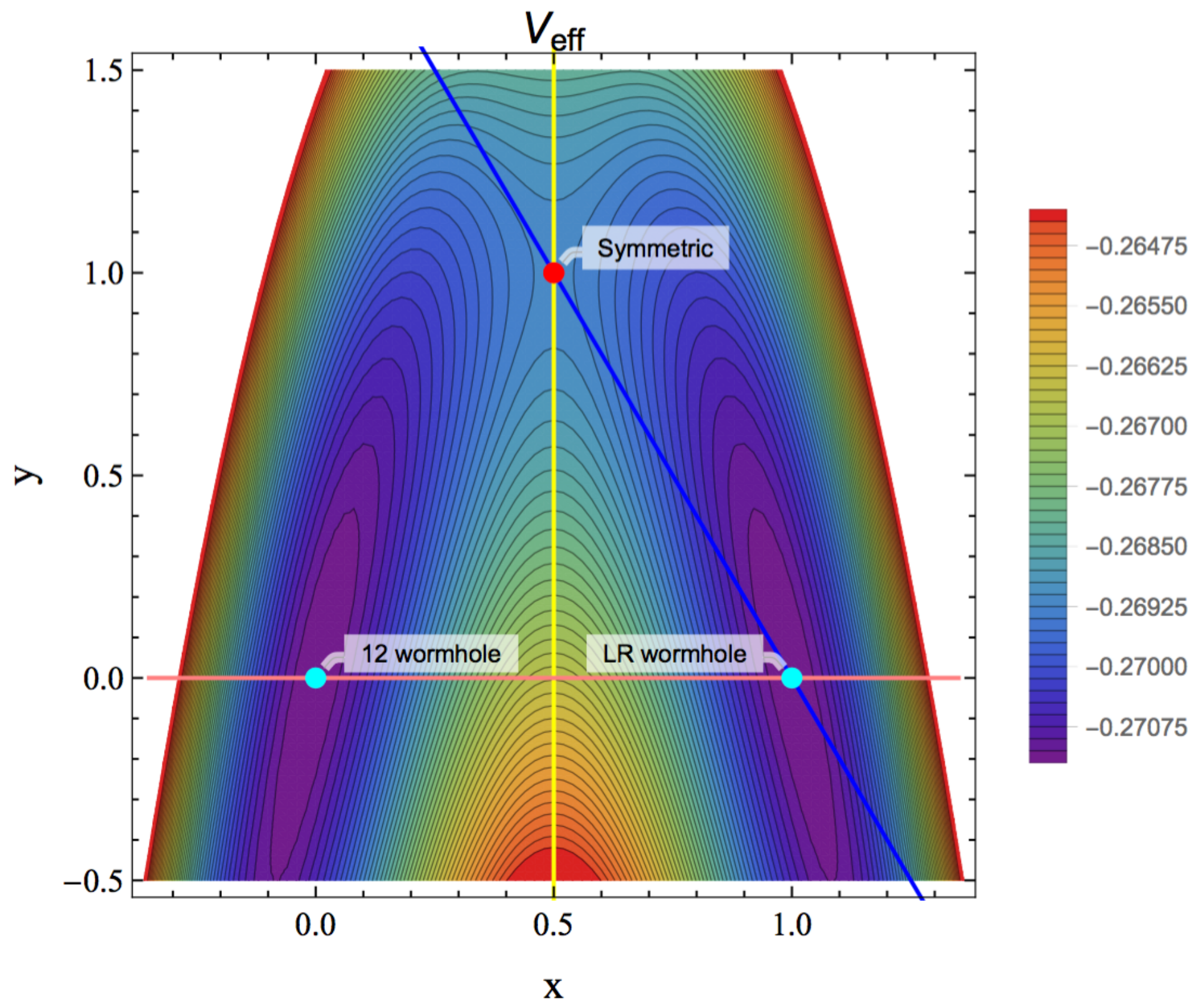} \notag \\
\includegraphics[width=5.2cm]{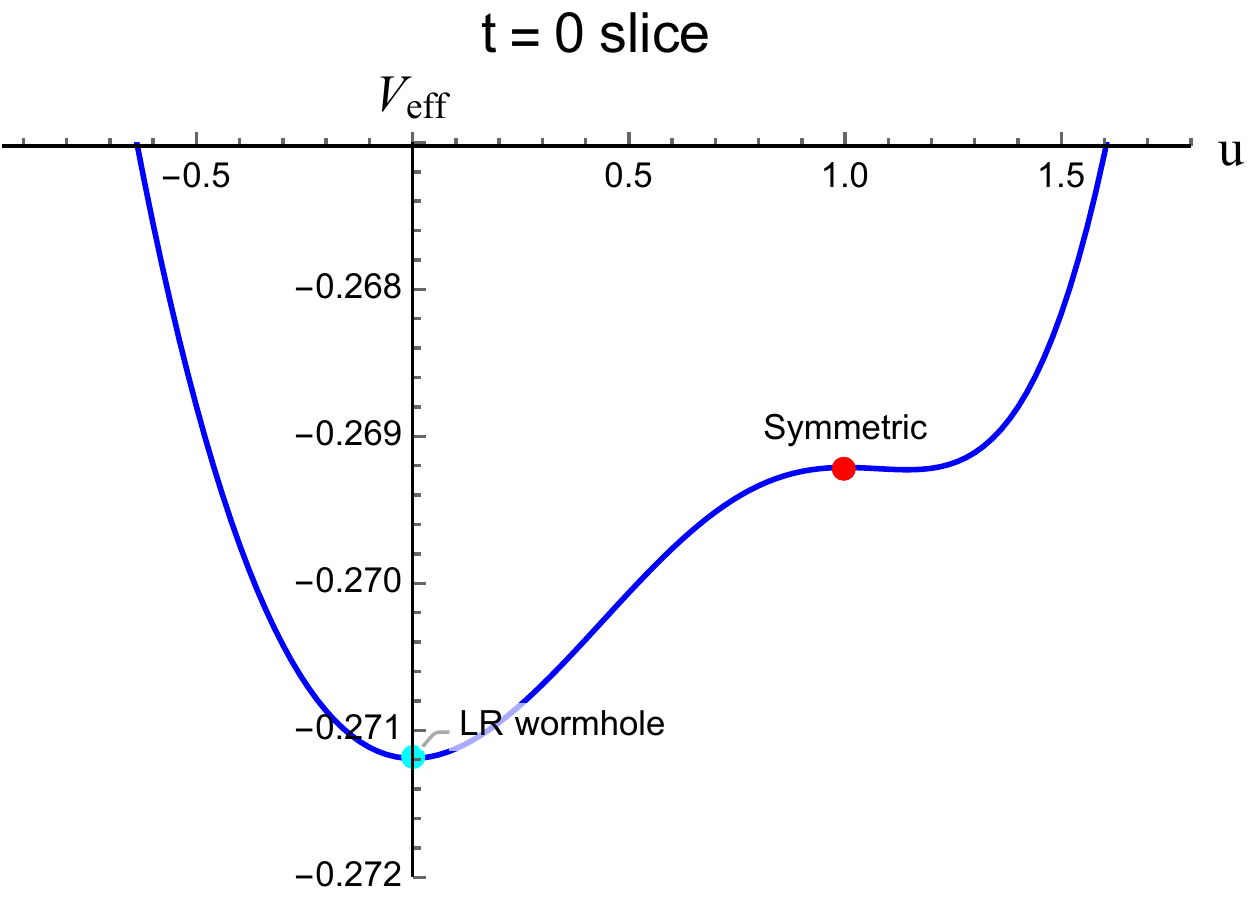}
\includegraphics[width=5.2cm]{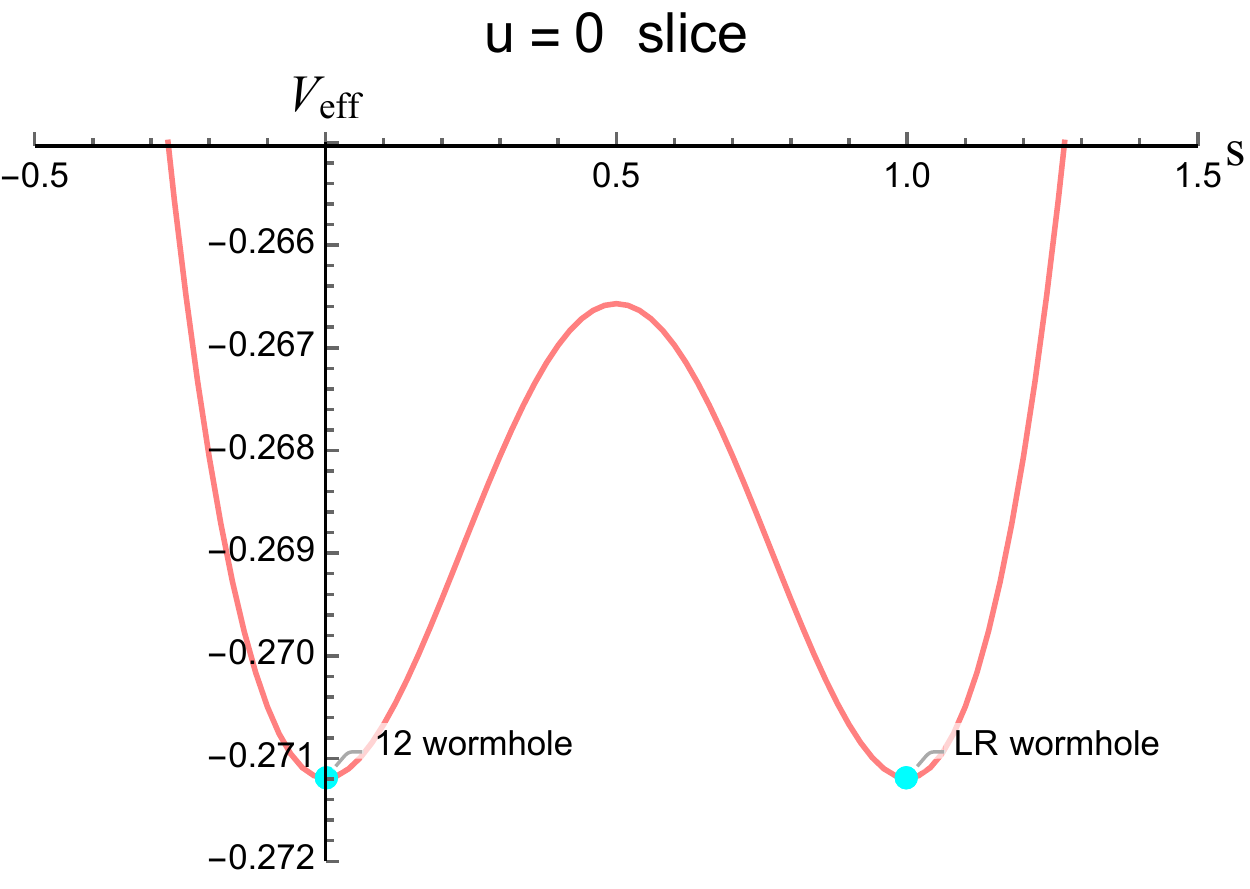}
\includegraphics[width=5.2cm]{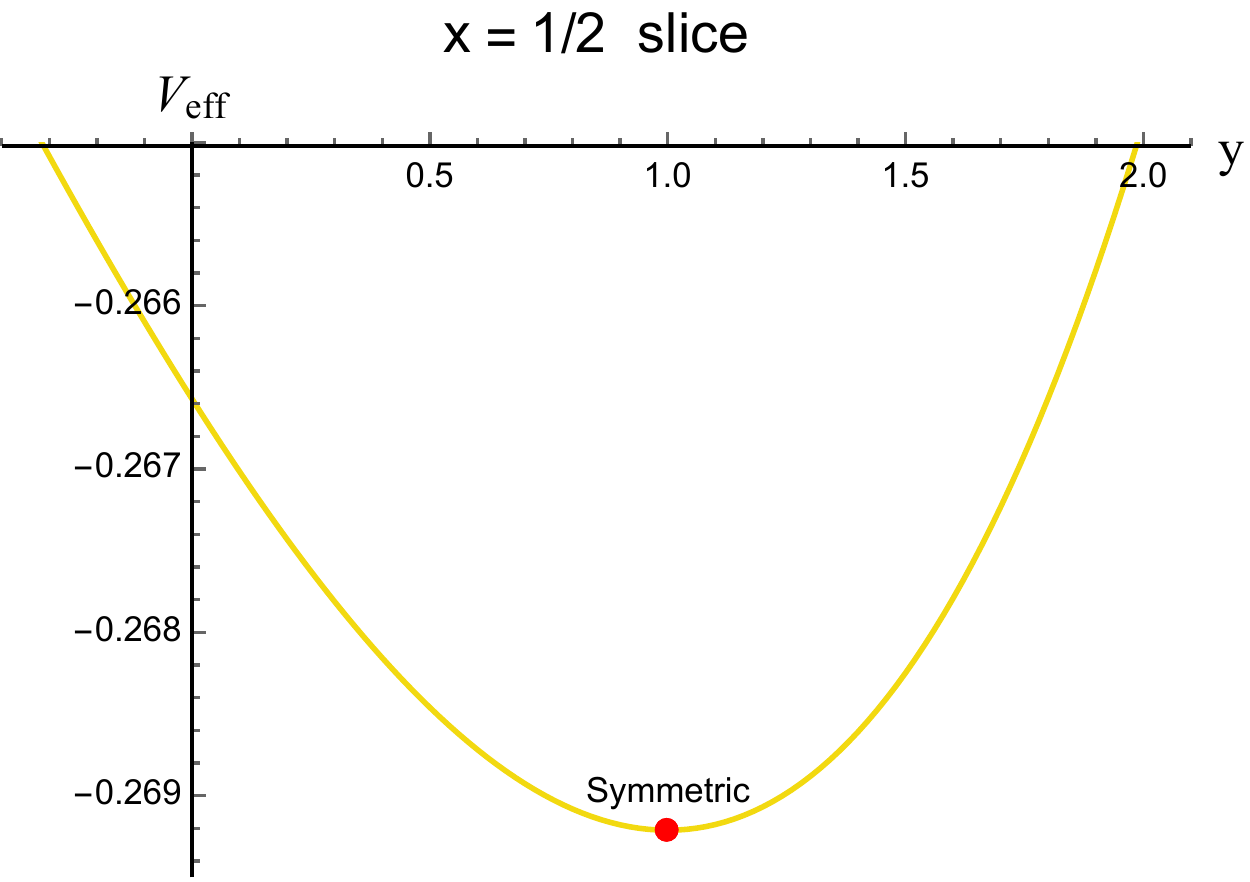}
\caption{The plots of the effective potential for $\mu_1 = \mu_2 = 0.075$, $T=0.001$ on the 2 dimensional slice (\ref{eq:2Dslice}).
{\bf Top Left:} The 3D plot of the effective potential as an function $V_{\text{eff}}(x,y)$.
{\bf Top Right:} The contour plot of the effective potential as an function $V_{\text{eff}}(x,y)$.
{\bf Bottom Left:} The $t=0$ slice of the $V_{\text{eff}}(x,y)$. In this slice, it is easy to see that the symmetric saddle point is an extremum rather than a minimum.
{\bf Bottom Middle:} The $u=0$ slice of the $V_{\text{eff}}(x,y)$. In this slice, the symmetry to exchange $WH_{LR}$ and $WH_{12}$ saddles.
{\bf Bottom Right:} The $x=1/2$ slice of the $V_{\text{eff}}(x,y)$. Along this direction, the symmetric saddle is a minimum.} 
\label{fig:Mu750B1000}
\end{center}
\end{figure}

Given the $G,\Sigma$ configuration, 
we can evaluate the effective action \eqref{eq:4coupledEAlargeN} even for off-shell configurations.
For the static configuration $G_{AB}^{\alpha\beta}(\tau_1,\tau_2) = G_{AB}^{\alpha\beta}(\tau_1-\tau_2)$, we can consider the effective potential 
\be
S_E/N \approx \beta V_{\text{eff}}(G(\cdot),\Sigma(\cdot)).
\ee
Here $\beta$ is the IR cutoff in the Euclidean time.
In the $4$ coupled SYK model, we have three saddle points: L-R wormholes $(G_{WH_{LR}}(\tau),\Sigma_{WH_{LR}}(\tau) )$, 1-2 wormholes $(G_{WH_{12}}(\tau),\Sigma_{WH_{12}}(\tau))$ and the symmetric solutions $(G_{sym}(\tau), \Sigma_{sym}(\tau) )$ \footnote{Here we suppress the index in $G_{AB}^{\alpha\beta}$ as $G$.}.
Then, using these saddles we can consider the slice of 
\ba
G(\tau) &= s G_{WH_{LR}}(\tau) + t G_{WH_{12}}(\tau)  + u G_{sym}(\tau), \qquad  s+t+u = 1, \notag \\
\Sigma(\tau) &= J^2 G(\tau)^{q-1}. \label{eq:2Dslice}
\ea
These define a 2 dimensional slice in $G, \Sigma$ configuration spaces.
We can parametrize $s,t,u$ by two parameters $x$ and $y$ as 
\be
s = x - \f{y}{2}, \qquad t = (1-x) - \f{y}{2} , \qquad u = y.
\ee

The effective potential on in this $(x,y)$ plane can be evaluated numerically and we show the plot of them in figure \ref{fig:Mu750B1000}.
In this $(x,y)$ plane, we have found that the L-R wormholes and 1-2 wormholes are local minima.
On the other hand, the symmetric saddle point is actually only an extremum and a maximum in one direction. 
Therefore, the plot shows that  the symmetric saddle point is unstable.
Another interesting thing is that in this slice the L-R wormhole and 1-2 wormhole are actually smoothly connected.
We expect that the effective potential are smoothly changed when we move from $\mu_{LR} = \mu_{12}$ point. 
Then we expect that these unstable saddle points are always exists when the system has both of L-R wormhole and 1-2 wormhole saddles.

\subsection{$\mu$ dependence of the symmetric solutions}

In this section we study the $\mu$ dependence of the  symmetric saddle points with comparison to wormhole solutions at the $\mu_{12} = \mu_{LR} \equiv \mu$ points.
Here we focus on the energy gap and the spin operator expectation values.

\begin{figure}[H]
\begin{center}
\includegraphics[width=7cm]{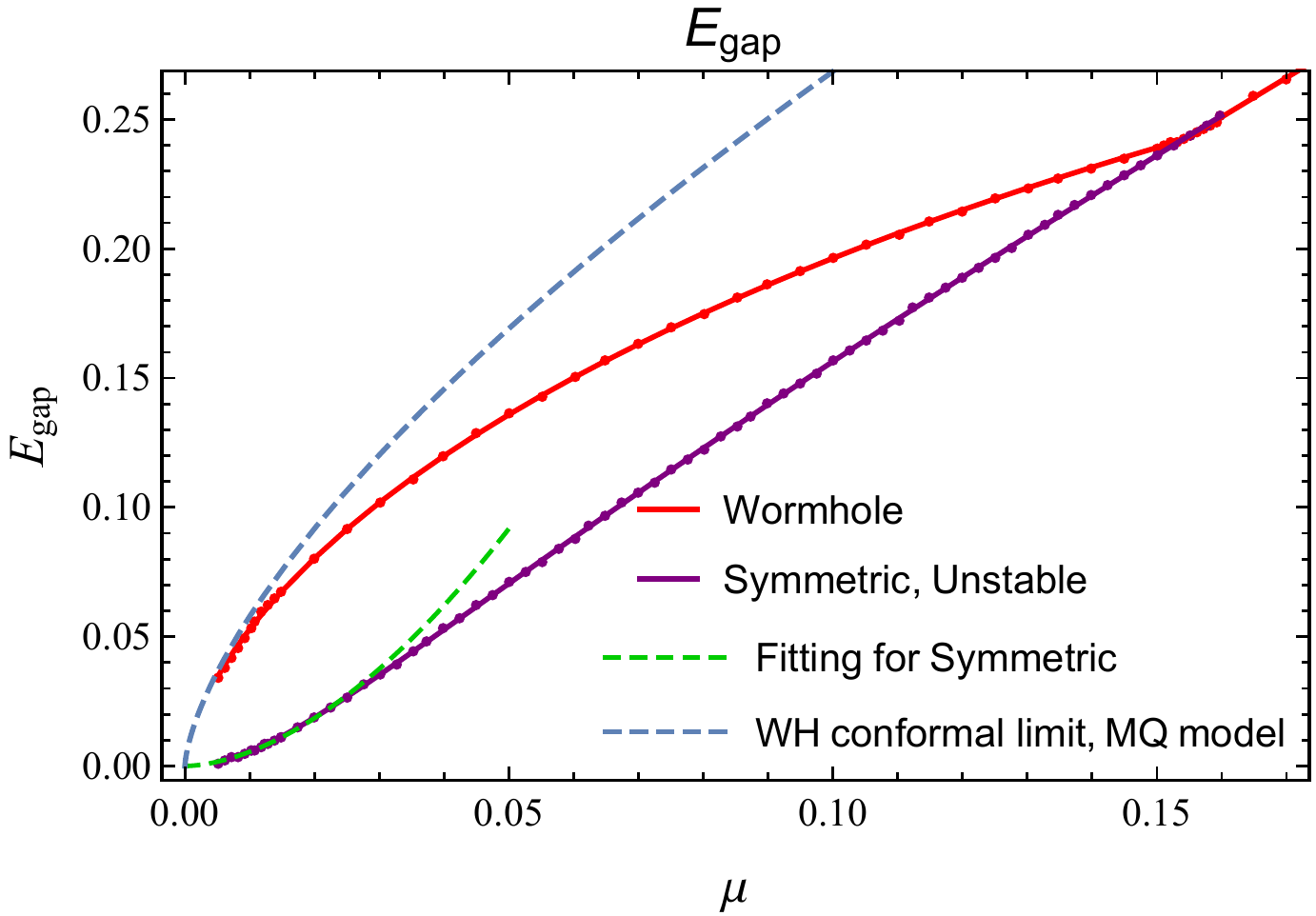}
\includegraphics[width=7cm]{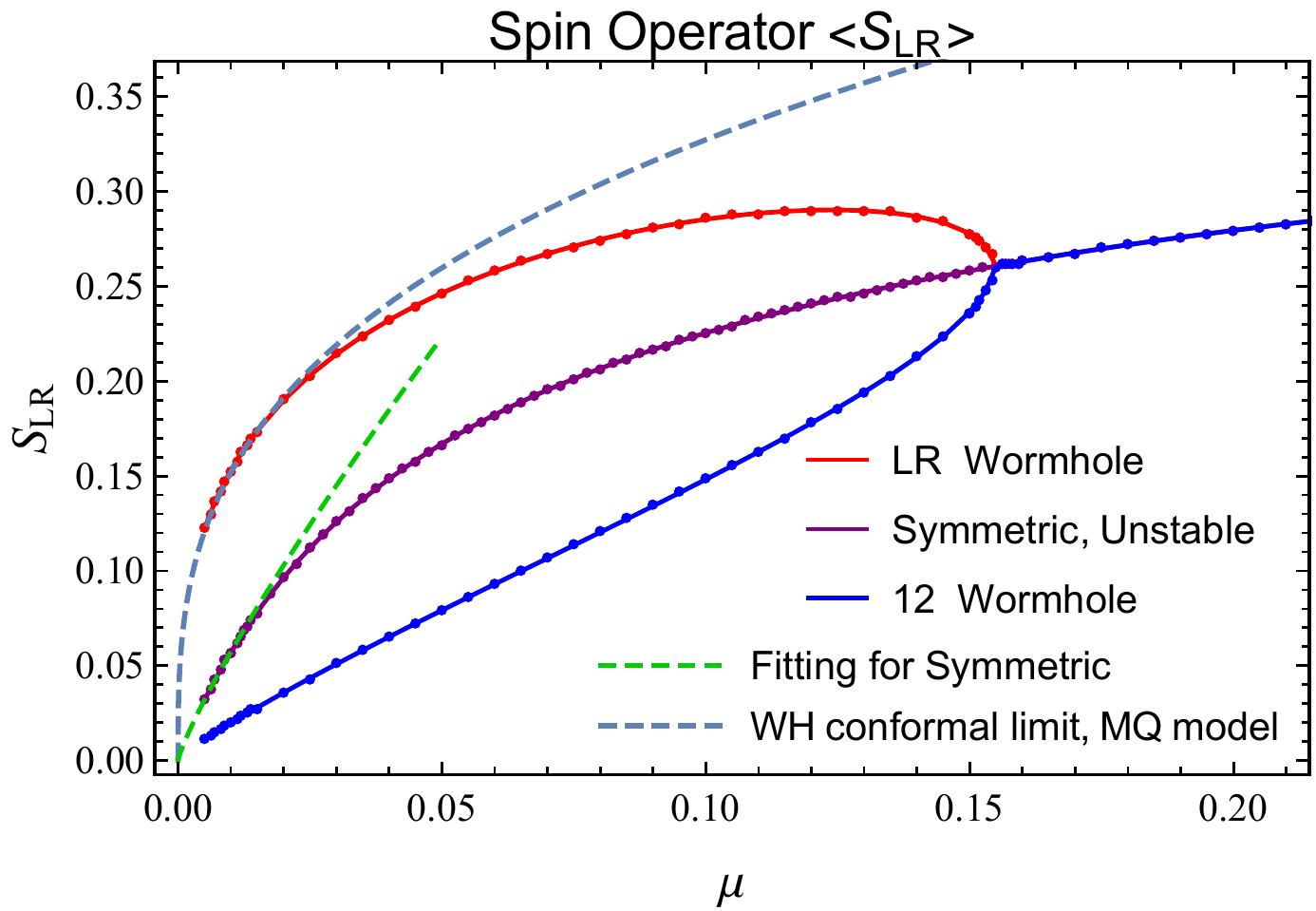}
\caption{The plots of $E_{gap}$ and $S_{LR}$ for $\mu_{LR} = \mu_{12} \equiv \mu$ cases i.e. the L-R coupling and the 1-2 coupling are the same.
We vary $\mu$ and plot $E_{gap}$ and $S_{LR}$ as a function of $\mu$.
{\bf Left:}  The plot of the $E_{gap}$. For fitting of the $E_{gap}$ in symmetric solutions, we use the data below $\mu < 0.01$.
{\bf Right:}  The plot of the spin operator $S_{LR}$.  For fitting of the $S_{LR}$ in symmetric solutions, we use the data below $\mu < 0.01$.
} 
\label{fig:mudepsymmetric}
\end{center}
\end{figure}

First we consider the gap.
A fitting for the numerical data (the green dashed line in the left of figure \ref{fig:mudepsymmetric}) gives 
\be
E_{gap} \approx 17.3054 \mu^{1.74896},
\ee
scaling of which is close to $E_{gap} \sim \mu ^{\f{7}{4}}$.
On the other hand, in both of  the L-R wormhole phases and the 1-2 wormhole phases, the energy gap $E_{gap}$ agrees with that in the two coupled model which is given by \eqref{eq:MQEgapConformal}.
This is because the ratio $\mu/E_{gap}^{MQ}$ is small, which controls the strength of the perturbation that couples two wormholes, where $\mu/E_{gap}^{MQ}$ is the gap in the two coupled model at the coupling $\mu$.
Comparing the $E_{gap} \sim \mu ^{1.74896}$ in the symmetric solution to the scaling of the $E_{gap} \sim \mu^{\f{2}{3}}$ in the wormhole phases, the power is greater than $1$ and this power low makes the energy gap smaller than the naive gap $\mu$. 
If we identify the energy gap with the wormhole length, this suggests that the wormhole length in the symmetric saddle is much larger than that in wormhole phases. 

The spin operator expectation value is plotted in figure \ref{fig:mudepsymmetric}.
The power law fitting  in the symmetric saddle (the green dashed line in the right of figure \ref{fig:mudepsymmetric}) is given by 
\be
S_{LR} \sim 2.80592 x^{0.84502},
\ee
which is close to $S_{LR} \sim \mu ^{\f{7}{8}}$.
This power of the scaling in symmetric solutions is much smaller than that in L-R wormhole solutions which behaves as $S_{LR} \sim \mu^{\f{1}{3}}$.
On the other hand, the scaling in the 1-2 wormhole phase is  $S_{LR} \sim \mu$ for generic $q$.
Therefore, the expectation value of $S_{LR}$ in the symmetric solution is much larger than that in the 1-2 solution.
As plot showed, the spin operator expectation value is between that of L-R wormhole and 1-2 wormhole.
The three solutions meet at $\mu \sim 0.154\mathcal{J}$ as we discussed.
Beyond that point, the solutions merges and we only have one solution.

\section{Further analysis on Traversable wormholes coupled with CFT$_2$ \label{sec:MoreOnTraversableJT}}
\begin{figure}[ht]
\begin{center}
\includegraphics[width=12cm]{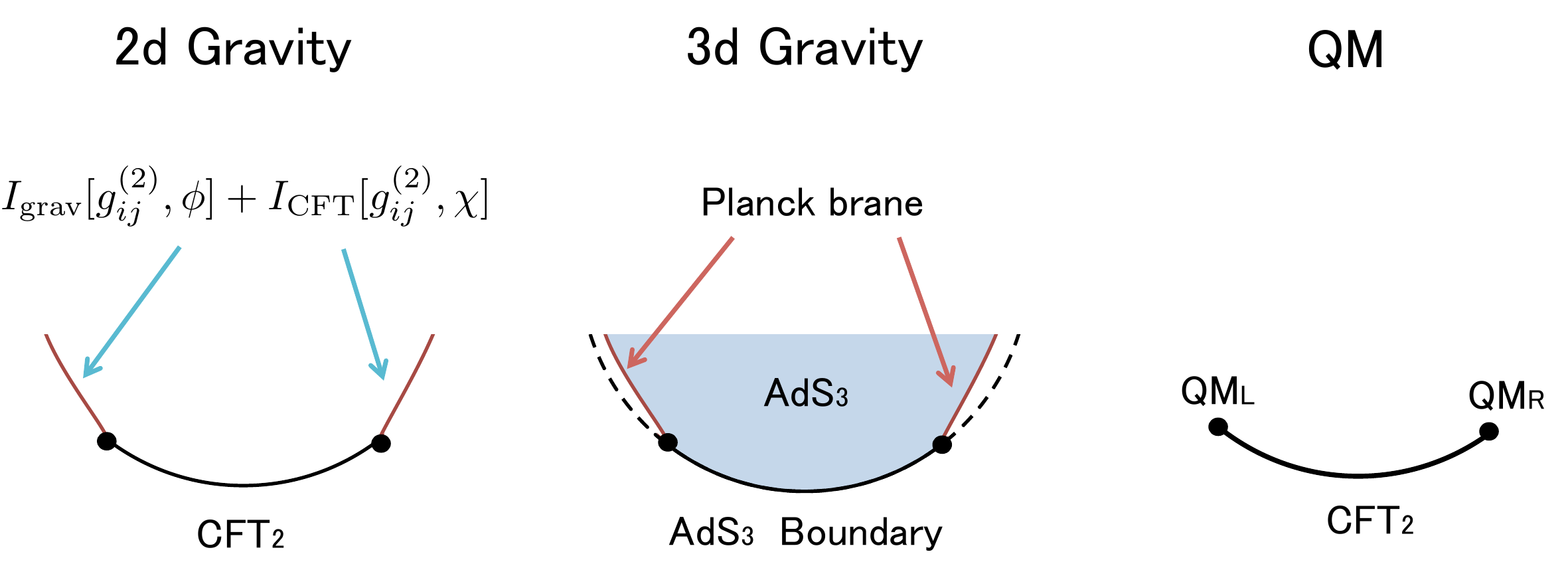}
\end{center}
\caption{3 descriptions of the same system. In gravity descriptions, the deep inside of the geometries are not depicted since they depend on states.
{\bf Left: }JT gravities coupled to 2d conformal matters. 
2 JT gravities are coupled to 2d conformal matters on a flat space (interval), which is the same with the bulk matter.
Bulk matters and CFT on the flat space are connected by the transparent boundary condition.
{\bf Middle:} 3d gravity description when the bulk matters have their holographic duals.
{\bf  Right:} Full quantum mechanical description. We take the holographic duals of JT gravities. 
} 
\label{fig:2coupledJT}
\end{figure}

Before going to the construction of four coupled JT gravities, we study the properties of two coupled JT gravities which we will use in the analysis of 4 coupled models, which include some new results on entanglement entropy and partially coupling in this setup. 
We consider the models of JT gravity with CFT$_2$ that are coupled to the same CFT$_2$ on with no dynamical gravity \cite{Almheiri:2019psf,Almheiri:2019hni}.
We imagine that JT gravity with CFT$_2$ is dual to a quantum mechanical system.
The situation we consider is that CFT$_2$ is coupled to two quantum mechanical systems with dual description by JT gravity + CFT$_2$ systems.
We basically use the saddle point approximation and the results only depend on the central charge of CFT$_2$.
Therefore, they can be holographic CFT with dual AdS$_3$ gravity description.
In figure \ref{fig:2coupledJT} we show the setup in this subsection with three descriptions \cite{Almheiri:2019hni,Almheiri:2019yqk}: 2d gravity description, 3d gravity description and the full quantum mechanical description. 
We can take the 3d gravity interpretation only when the bulk matter fields are holographic 2d CFTs.
In figure \ref{fig:2coupledJT}, we only describe the theory that we consider and the IR region is not described since they depend on states.

\subsection{The traversable wormhole solutions}

\begin{figure}[ht]
\begin{center}
\includegraphics[width=12cm]{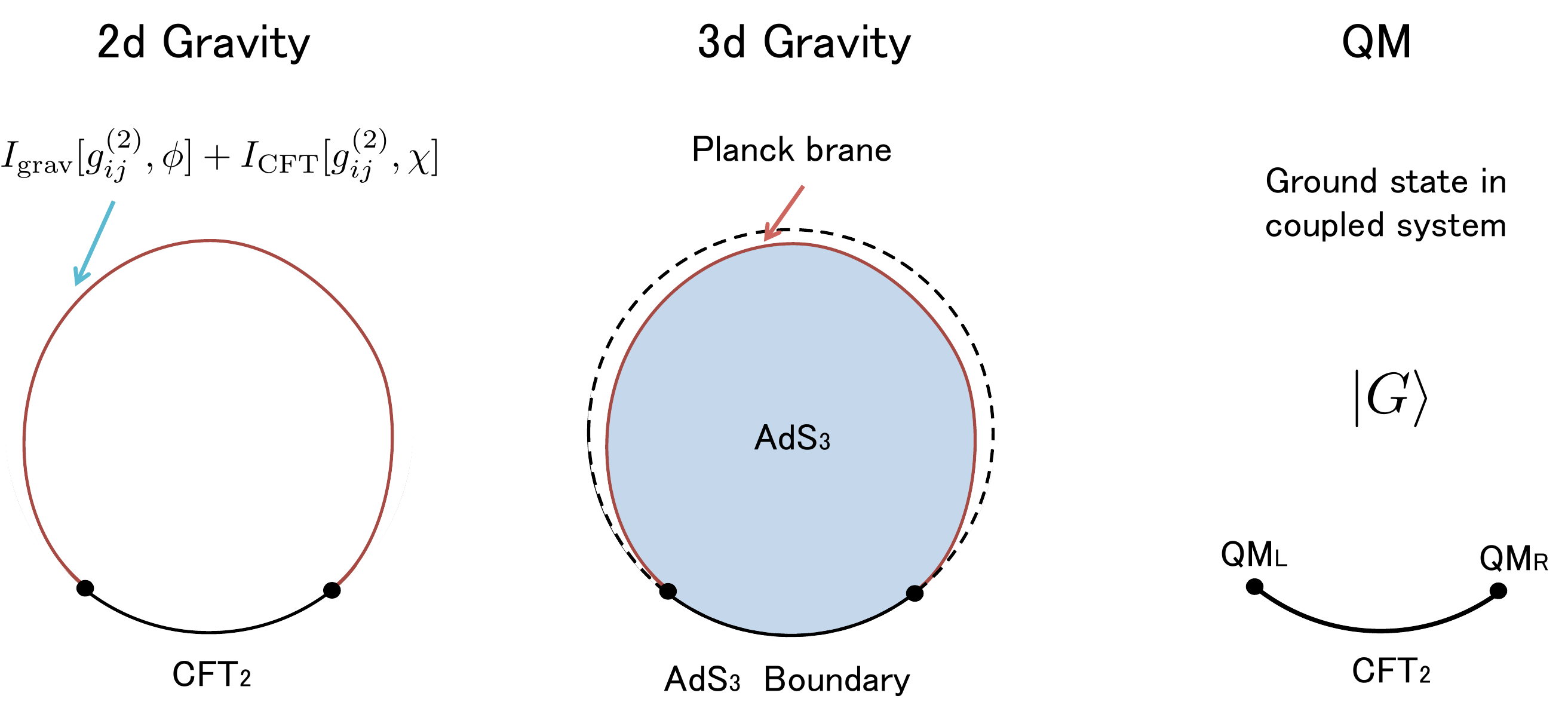}
\end{center}
\caption{3 descriptions of traversable wormholes coupled through CFT on a interval.
{\bf Left: } Traversable wormholes which are supported by negative null energy of the bulk matter.
We have CFTs on a cylinder which are partially sitting on a region with JT gravity.
{\bf Middle:} 3d gravity description when the bulk matters have their holographic duals.
The wormhole is now directly connected to the flat space region through the 3d direction.
{\bf  Right:} Full quantum mechanical description. The ground state of CFTs on a interval coupled to quantum mechanics on boundaries.
} 
\label{fig:wormholestate1}
\end{figure}
\begin{figure}[ht]
\begin{center}
\includegraphics[width=7cm]{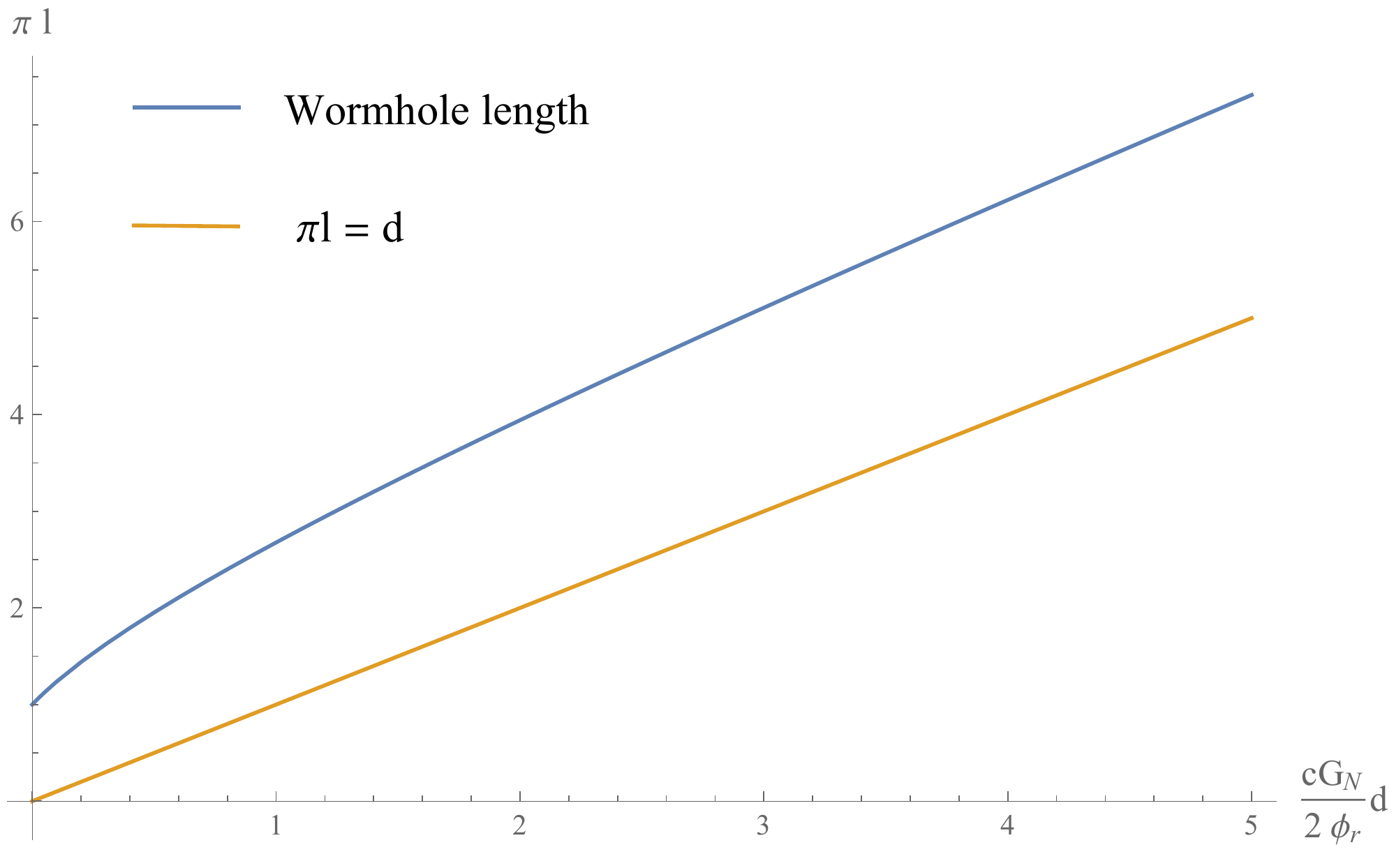}
\end{center}
\caption{The plot of the wormhole length $\pi \ell$.
} 
\label{fig:wormholelength1}
\end{figure}

We are interested in the ground state of this two coupled model.
The ground state is given by traversable wormholes, which is illustrated in figure \ref{fig:wormholestate1}.
Now we consider the traversable wormhole solutions with the finite "bath length" $d$ where the CFT$_2$ sitting on.
Here the finite bath length means that the ratio $d/\ell = a$ is a finite constant. 
In this case, we first put CFT on a cylinder with the length $\pi(1+a)$, which is a flat space.
Then, we obtain the stress tensor expectation value 
\be
\braket{T^{\text{cyl}}_{tt}} = \braket{T^{\text{cyl}}_{\sigma\sigma}} = -\f{c}{6\pi} \f{1}{(1+a)^2}.
\ee
The notations and the derivation from conformal anomaly are summarized in appendix \ref{sec:2dCFTformula}. In terms of null energy, they become 
\be
\braket{T^{\text{cyl}}_{y^+y^+}} = \braket{T^{\text{cyl}}_{y^-y^-}} = -\f{c}{12\pi} \f{1}{(1+a)^2}.
\ee
Next, we introduce the AdS$_2$ metric in the region $\sigma \in [0,\pi]$.
Then, the null energy in the AdS$_2$ region is 
\be
\braket{T^{\text{AdS}_2}_{y^+y^+}} = \braket{T^{\text{AdS}_2}_{y^-y^-}} = \f{c}{48\pi} -\f{c}{12\pi}\f{1}{(1+a)^2} = -\f{c}{48\pi} \f{(a+3)(1-a)}{(1+a)^2}.
\ee
The first term comes from the Weyl anomaly, see appendix \ref{sec:2dCFTformula}.
We can rewrite this as 
\be
\braket{T^{\text{AdS}_2}_{y^+y^+}} = \braket{T^{\text{AdS}_2}_{y^-y^-}} = -\f{c_{\text{eff}}(a)}{16\pi} . \label{eq:2coupledNullEnergyAdS}
\ee
Here we defined an "effective central charge"
\be
c_{\text{eff}}(a) = c\f{(3+a)(1-a)}{3(1+a)^2},
\ee
which is introduced in the wormhole regime by conformal matters.
This null energy is also directly calculated using the holographic stress energy tensor \cite{Balasubramanian:1999re} when the bulk matter fields are holographic.
Then, 
\ba
\phi(\sigma) &= \f{c_{\text{eff}}(a)}{8\pi} \bigg[ \f{\f{\pi}{2}- t' \sigma}{\tan (t' \sigma)} + 1\bigg] + \f{c}{24\pi} \notag \\
&= \f{2 \bar{\phi}_r}{\pi  \ell } \bigg[ \f{\f{\pi}{2}-  \f{\sigma}{\ell}}{\tan  \f{\sigma}{\ell}} + 1\bigg]+ \f{c}{24\pi} 
= \f{ \bar{\phi}_r}{\pi  \ell } \bigg[ \f{ \pi \ell - 2\sigma}{ \ell \tan  \f{\sigma}{\ell}} + 1\bigg]+ \f{c}{24\pi}
\ea
The inverse of wormhole length is given by $t' =1/\ell \equiv \f{c_{\text{eff}}(a)}{16 \bar{\phi}_r}$.
If we restore the $8\pi G_N$ which we put to be $1$ above, we obtain
\be
t' = \f{ c_{\text{eff}}(a)\pi G_N}{2\bar{\phi}_r}  = \f{c \pi G_N}{2\bar{\phi}_r} \f{(3+a)(1-a)}{3(1+a)^2}.
\ee
\be
M = -\f{c_{\text{eff}}(a)^2 \pi G_N}{32\bar{\phi_r}} = -\f{c^2 \pi G_N}{32\bar{\phi_r}} \f{(3+a)^2(1-a)^2}{9(1+a)^4}.
\ee
The wormhole length is 
\be
\ell = \f{2\bar{\phi}_r}{c \pi G_N} \f{3(1+a)^2}{(3+a)(1-a)}. \label{eq:2coupledWormholeLengthGen}
\ee
and the bath (region without gravity) length is 
\be
d = \pi \ell a =\f{2\bar{\phi}_r}{c G_N} \f{3(1+a)^2}{(3+a)(1-a)} a. \label{eq:bathlength1}
\ee
The last equation determines\footnote{The parameter $a$ is not a fixed parameter but a parameter which is determined from the saddle point equation of quantum gravity i.e. JT gravity in this setup.
This is because the wormhole length $\ell$ is a dynamical parameter rather than a fixed parameter.
The length of the bath region $d$ is a parameter of the theory because CFT in the bath region there are no dynamical gravity.} the parameter $a$ as a function of $d$.
As a function of $d$, $a$ becomes 
\be
a = F\Big(\f{ cG_N }{2 \bar{\phi}_r}d\Big)  
\ee
where $F(X)$ is the root of 
\ba
3a^3 + (X+6)a^2 + (2X +3) a -3 = 0,
\ea
as a polynomial of $a$.
The wormhole length is plotted in figure.\ref{fig:wormholelength1}.

For large $d \gg \f{2\bar{\phi}_r}{c G_N}$, $a \approx 1$.
This means $\pi \ell \approx d$.
It takes almost the same time to traverse wormholes with the time to go through the outside.
The energy in the bath region is 
\be
T_{tt}\cdot d = -\f{c}{6\pi} \f{t'^2  d}{(1+a)^2}  = -\f{c}{6 \pi} \f{ d}{( \ell + \ell \cdot a)^2} = -\f{c}{6} \f{\pi d}{(\pi l+d)^2}.
\ee
The energy density itself is 
\be
T_{tt} =  -\f{c}{6\pi} \f{t'^2  }{(1+a)^2} = -\f{c}{6} \f{\pi }{(\pi l+d)^2}.
\ee

\subsubsection{Variational ansatz}
We approximate the geometry by the $t = 0$ slice of the eternal black holes (= thermo field double states).
The wormhole length is $\ell = \f{1}{2\pi T_H}$.
The energy par a black hole is 
\be
M_{BH} = \f{\pi \bar{\phi}_r}{4 G_N} T_H^2 = \f{ \bar{\phi}_r}{16 \pi G_N} \f{1}{\ell^2}. 
\ee
The quantum matter contribution is 
\be
E_{mat} = \f{c}{24  \ell} -\f{c \pi}{6 (\pi \ell + d)}.
\ee
The total variational energy is 
\be
E(\ell) =2 M_{BH} + E_{mat} = \f{ \bar{\phi}_r}{8 \pi G_N} \f{1}{\ell^2} + \f{c}{24 \ell} -\f{c \pi}{6 (\pi \ell + d)} .
\ee
We minimize this energy 
\be
\f{\partial E}{\partial \ell} = -\f{ \bar{\phi}_r}{4 \pi G_N} \f{1}{\ell^3} - \f{c}{24  \ell ^2 }  + \f{c \pi^2}{6 (\pi \ell + d)^2}= 0.
\ee
By setting $\pi \ell a = d$, this equation becomes 
\be
d = \f{\pi \bar{\phi}_r}{2 G_N} 12 \f{(1+a)^2}{(3+a)(1-a)} a.
\ee
This equation is precisely the same with the former one (\ref{eq:bathlength1}) that determines the length of the wormhole as a function of $d$.
Then $\ell$ becomes 
\be
\ell = \f{2\bar{\phi}_r}{c G_N} \f{3(1+a)^2}{(3+a)(1-a)}.
\ee
This again reproduces the former result.
The approximation by the eternal black holes (TFD state) is a good description.

\subsubsection{Entanglement entropy for region with a boundary}
In this subsection, we consider entanglement entropy when we divide the system to two part at point $x = b$ in no gravity region in a full quantum mechanical description in figure \ref{fig:wormholestate1}.
In other words, we consider entanglement entropy for an interval of  $\pmb{[0,b]}$  in the notation of \cite{Almheiri:2019yqk}.
Here the bold notation means that we consider the interval in the full quantum mechanical description, not in the 2d gravity description.
In particular, we should notice that in 2d gravity picture we could not be able  to divide the spacetime, which is a circle, to two parts just only specifying one entangling surface $x = b$\footnote{This is related to the factorization problem in \cite{Harlow:2018tqv}.
Without assuming that {\it single} JT gravity factorizes to two systems, the concept to divide the system two to parts by only specifying entangling surface $x=b$ does not exist.}.
Correspondingly, the configuration without islands \cite{Almheiri:2019hni,Almheiri:2019yqk} does not exist in 2d gravity description in this case.

\begin{figure}[ht]
\begin{center}
\includegraphics[width=12cm]{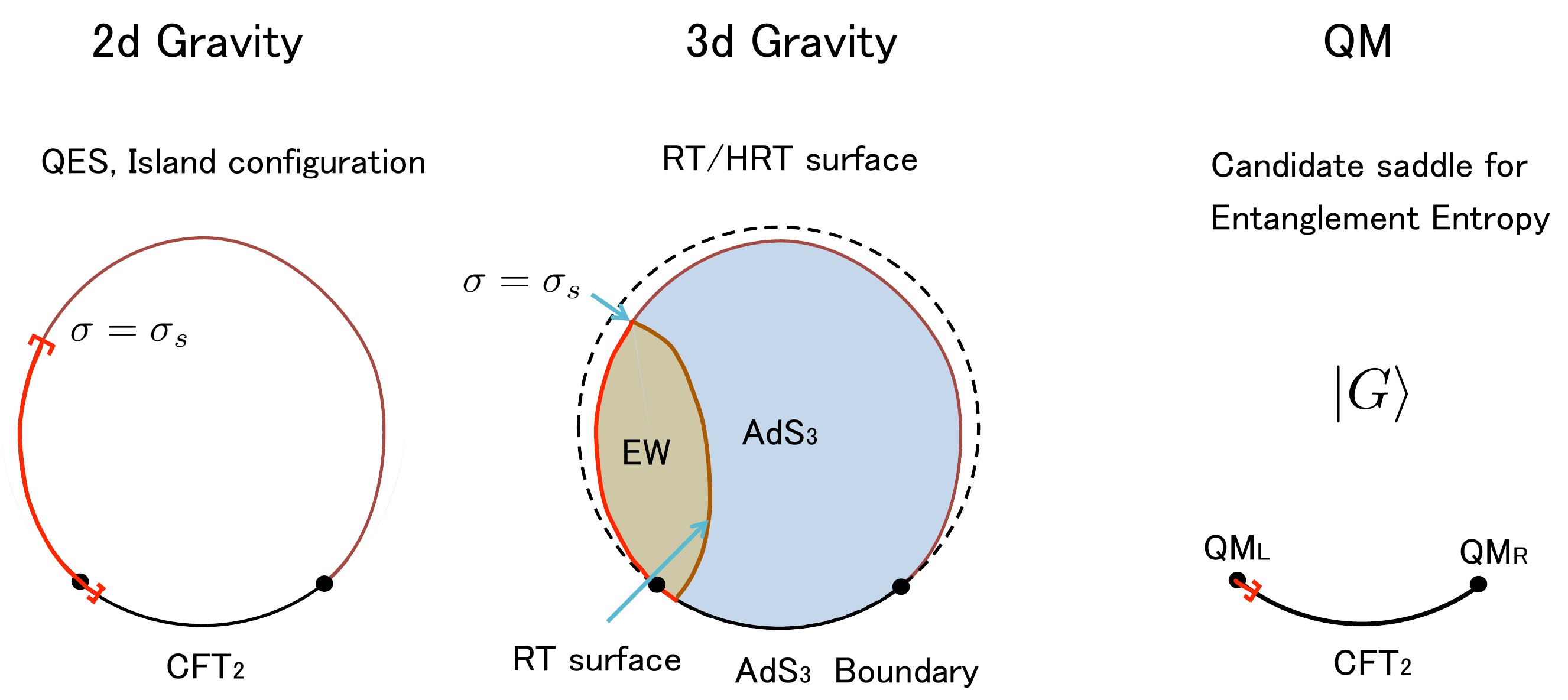}
\end{center}
\caption{We describe the three descriptions for the quantum extremal surface which contains the dual of a JT + matter system.
{\bf Left:} The JT gravity plus matter CFT description. We are computing quantum extremal surfaces.
{\bf Middle:} We use the 3d gravity description for matter CFT assuming that the matters are holographic.
  The shaded region (which is denoted by EW) is the entanglement wedge, which is surrounded by the entangling surface and the RT/HRT surfaces.
{\bf Right:} The full quantum mechanical description. We calculate entanglement entropy of CFT using the large $N$ expansion.
  This is another candidate saddle of the full answer.
} 
\label{fig:2coupledJTdefectIsland}
\end{figure}

The general form of the generalized entropy functional is given by \cite{Almheiri:2019psf,Almheiri:2019hni} 
\be
S(A) =  \text{Min}_{\mathcal{I}_g}\text{Ext}_{\mathcal{I}_g}\Bigg[ S_{eff}(A \cup \mathcal{I}_g) + \f{\text{Area}[\partial \mathcal{I}_g]}{4G_N}\Bigg].
\ee
Here $\mathcal{I}_g$ is a set of codimension one regions in the gravitating systems.
In our setup, generalized entropy functional for the $\pmb{[0,b]}$ is 
\ba
S_{\text{gen}}(\sigma) &= S_0 + 2\pi \phi(\sigma) + \f{c}{6} \log \Bigg (\Big(\f{\pi \ell + d}{\pi}\Big)^2 \f{\sin^2 \f{\pi(\sigma + b)}{\pi \ell + d}}{ \epsilon_B\ell \sin \f{\sigma}{\ell}}  \Bigg) \notag \\
&= S_0 + \f{ 2\bar{\phi}_r}{ \ell } \Big[ \f{ \pi \ell - 2\sigma}{ \ell \tan  \f{\sigma}{\ell}} + 1\Big] +  \f{c}{6} \log \Bigg ( \Big(\f{\pi \ell + d}{\pi}\Big)^2  \f{\sin^2 \f{\pi(\sigma + b)}{\pi \ell + d}}{ \epsilon_B \ell \sin \f{\sigma}{\ell}} \Bigg).
\ea
The first term is the topological contribution from the Einstein-Hilbert term of the JT gravity.
$\epsilon_B$ is a UV cutoff in no gravity regions.
The second term is the RT surface contribution.
We choose the normalization of $8\pi G_N = 1$ here and in this normalization $\f{\phi(\sigma)}{4G_N} = 2\pi \phi (\sigma)$. 
The third term is entanglement entropy of matter fields.
Note that $\f{\bar{\phi}_r}{\ell}\propto c$ and both term can compete.
 Entanglement entropy is given by the saddle point of the generalized entropy functional:
\be
\partial_{\sigma} S_{\text{gen}}(\sigma) = 0.
\ee
This becomes 
\be
\partial_{\sigma} S_{\text{gen}}(\sigma)  
=
-\f{c \ell + 24 \bar{\phi}_r}{6 \ell^2 \tan \f{\sigma}{\ell}}
 -\f{2 \bar{\phi}_r(\ell \pi - 2\sigma)}{\ell^3 \sin^2(\f{\sigma}{\ell})}
 + \f{\pi c}{3} \f{1}{(\pi \ell + d ) \tan \f{\pi(b+\sigma)}{\pi \ell + d}}.
\ee
The quantum extremal surface is given by the saddle point of the generalized entropy:
\be
S_{ext} =  S_{gen}(\sigma_s). \label{eq:QESsingle1}
\ee
The 2d gravity, 3d gravity and full quantum mechanical description for $S_{ent}(\sigma_s)$ are shown in figure \ref{fig:2coupledJTdefectIsland}.

\begin{figure}[ht]
\begin{center}
\includegraphics[width=5.5cm]{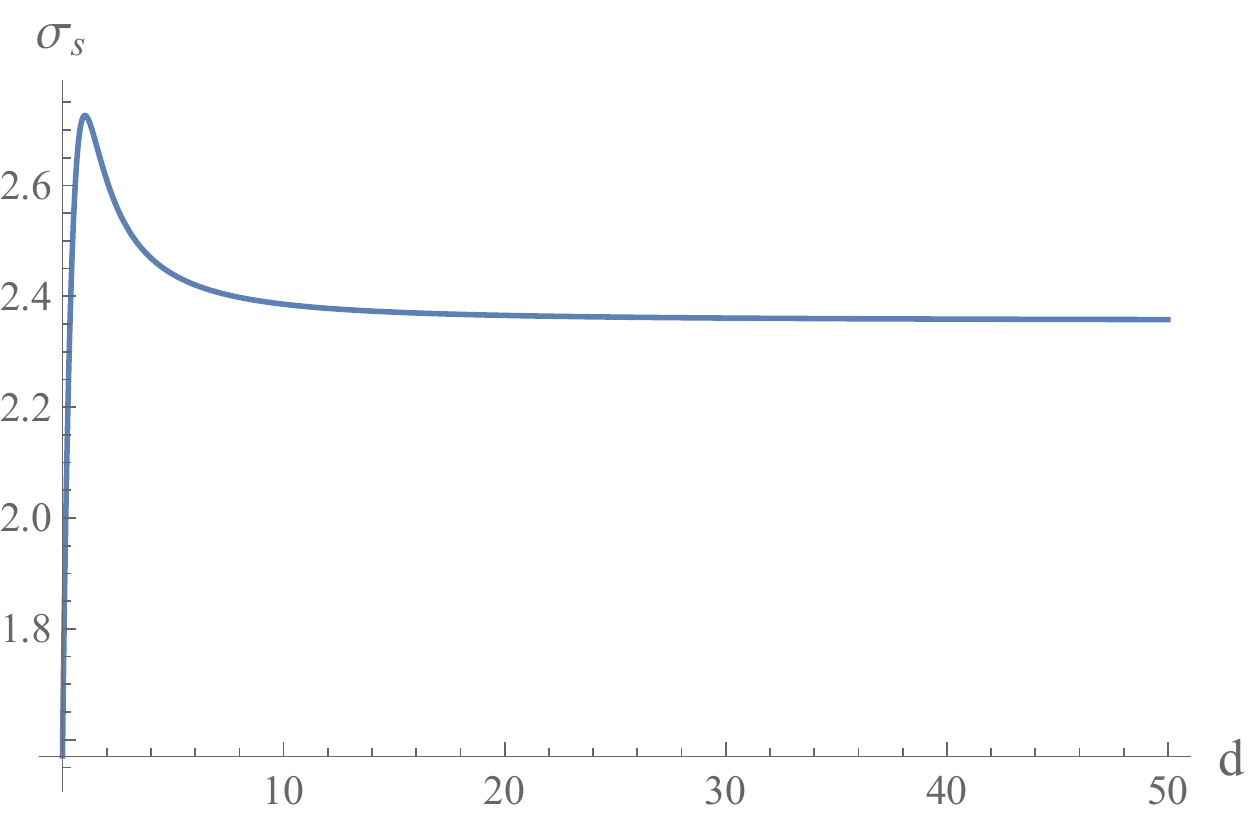} \qquad
\includegraphics[width=5.5cm]{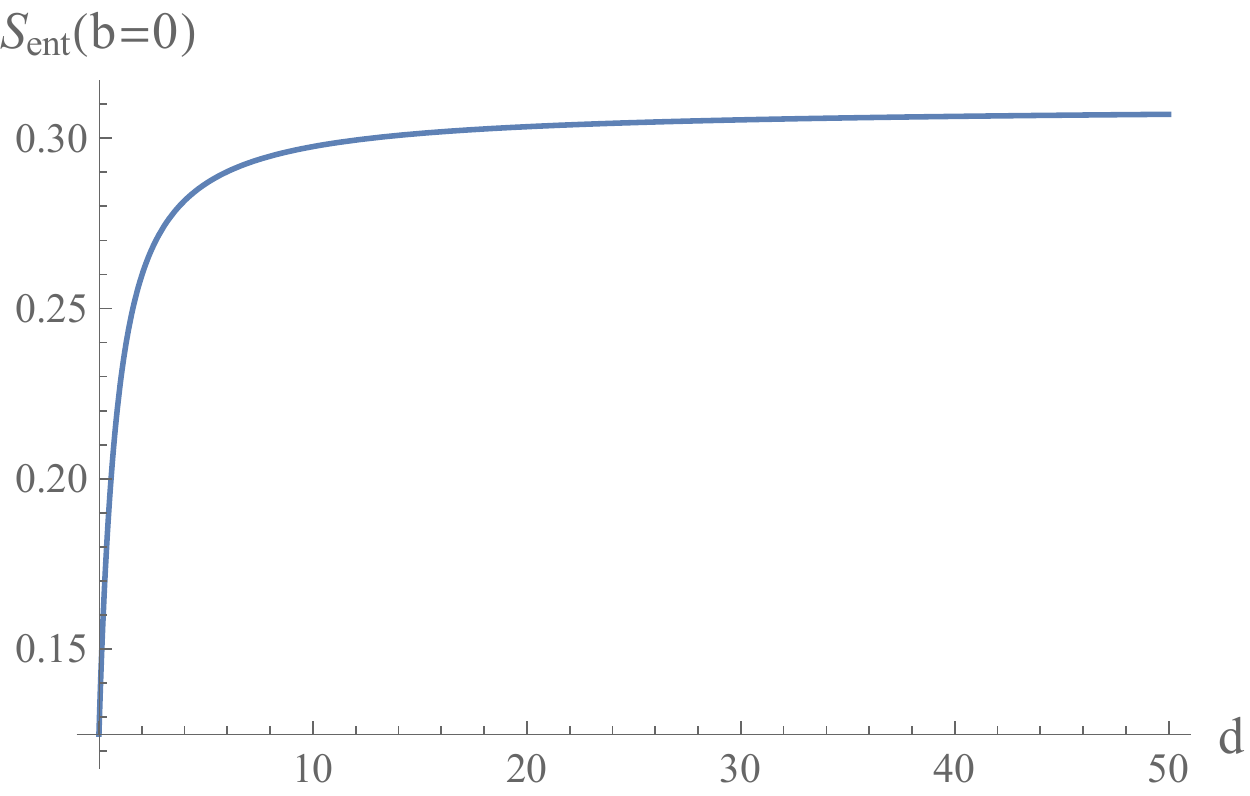}
\end{center}
\caption{The plots of the location of the quantum extremal surface and entanglement entropy for $b=0$ as function of $d$.
We set  $\ell = \f{16 \bar{\phi}_r}{c} = 1$, and $c = 1$. 
 {\bf Left}: The plot of the quantum extremal surface $\sigma_s$.
{\bf Right}: 
The plot of entanglement entropy.
We also omitted the extremal entropy part $S_0$.
We also subtract the divergence $\f{c}{6} \log \f{1}{\epsilon_B}$.
In this choice, the initial value is $ S_0 + \f{c}{8}+ \f{c}{6} \log \f{16 \bar{\phi}_r}{c } = \f{1}{8}$
Finally entanglement entropy approaches to $S_0 + \f{c}{6} + \f{c}{6} \log \f{12\pi \bar{\phi}_r}{ c} = \f{1}{6} + \f{1}{6} \log \f{3\pi}{4} \approx 0.309$. } 
\label{fig:2coupledEntanglementWboundary}
\end{figure}

For general $d$, we can calculate entanglement entropy \eqref{eq:QESsingle1} numerically.
We show the numerical solutions for $S_{ent}$ and the position of the quantum extremal surface $\sigma_s$ in figure \ref{fig:2coupledEntanglementWboundary}. 
Note that entropy \eqref{eq:QESsingle1} is a function of $d$ and $b$ because the wormhole length $\ell$ is also determined as a function of $d$ in \eqref{eq:2coupledWormholeLengthGen} .

When $d = 0$ i.e. we can ignore the length of the bath region, the saddle point is $\sigma_s = \f{\pi}{2} \ell$ with $\ell = \f{16 \bar{\phi}_r }{c}$, which is the middle of the AdS$_2$.
Entanglement entropy  for $b=0$ case becomes 
\ba
S_{gen}(\pi \ell/2) &=  S_0 + \f{2 \bar{\phi}_r}{\ell} + \f{c}{6} \log \f{\ell}{ \epsilon_B} \notag \\
&= S_0 + \f{c}{8}+ \f{c}{6} \log \f{16 \bar{\phi}_r}{c  \epsilon_B}
\ea

When $\sigma \ll \ell$ and $b \ll \ell$, we can approximate $\tan \f{\sigma}{\ell} \approx  \f{\sigma}{\ell}$, $\sin \f{\sigma}{\ell} \approx \f{\sigma}{ \ell}$ and  $\sin \f{\pi(\sigma + b)}{\pi \ell + d} \approx \f{\pi(\sigma + b)}{\pi \ell + d}$.
Then, the generalized entanglement entropy before the minimization is 
\be
S_{\text{gen}}(\sigma)  \approx S_0 + \f{2 \pi \bar{\phi}_r}{\sigma} + \f{c}{6} \log \f{(\sigma+b)^2}{\epsilon_B \sigma},
\ee 
which is exactly the same with the $0$ temperature generalized entropy.
The saddle point equation becomes 
\be
\f{\partial S_{\text{gen}}(\sigma)}{\partial \sigma} = -\f{2 \pi \bar{\phi}_r}{ \sigma^2 } + \f{c}{6} \bigg[ \f{2}{(\sigma  +b)} - \f{1}{\sigma}\bigg] = 0.
\ee
The solution is given by
\be
\sigma_s = \f{1}{2c} (bc + 12\pi \bar{\phi}_r + \s{b^2c^2 + 72 \pi \bar{\phi}_r bc + 144 \pi^2 \bar{\phi}_r^2}).
\ee
By introducing $\tilde{\sigma}_s = \f{c}{12\pi \bar{\phi}_r}\sigma_s $ and $\tilde{b} = \f{c }{12 \pi \bar{\phi}_r}b$, we  can rewrite 
\be
\tilde{\sigma}_s = \f{1}{2} (1 + \tilde{b} + \s{1 + 6 \tilde{b} + \tilde{b}^2}).
\ee
For $b = 0$, $\tilde{\sigma}_s = 1$.
In other words, in $b \to 0$ limit we obtain 
\be
\sigma_s = \f{12 \pi \bar{\phi}_r}{c}.
\ee
Entanglement entropy becomes 
\be
S_{gen}(\sigma_s) = S_0 + \f{c}{6} + \f{c}{6} \log \f{12\pi \bar{\phi}_r}{\epsilon_B c}.
\ee

\subsection{Entanglement entropy for regions without gravity}
\begin{figure}[ht]
\begin{center}
\includegraphics[width=12cm]{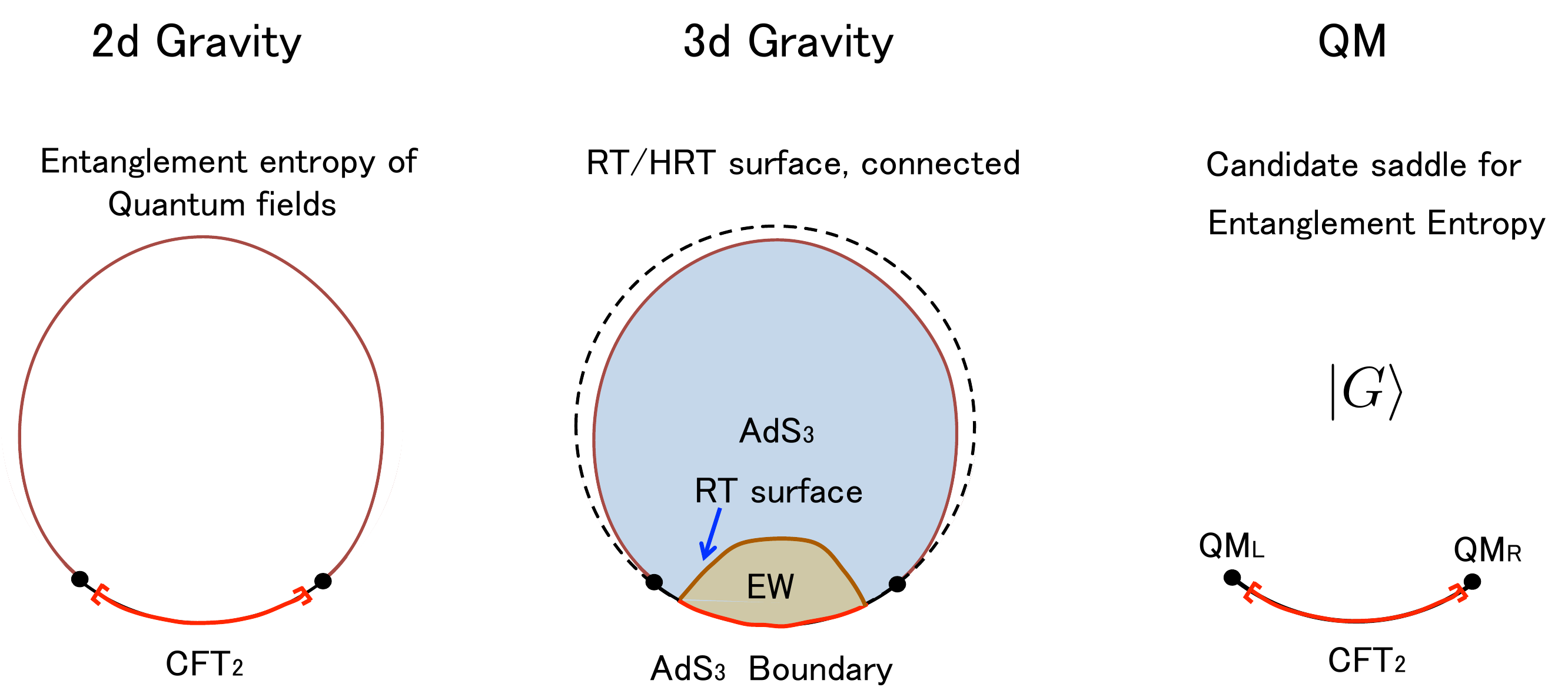}
\end{center}
\caption{We describe the three descriptions for the quantum extremal surface which does not contain islands.
{\bf Left:} The JT gravity plus matter CFT description. We are computing entanglement entropy of matter fields.
 {\bf Middle:} We use the 3d gravity description for matter CFT assuming that the matters are holographic.
  We are computing the lengths of the RT/HRT surfaces.
  The shaded region is the entanglement wedge, which is surrounded by the entangling surface and the RT/HRT surfaces.
  {\bf Right:} The full quantum mechanical description. We calculate entanglement entropy of CFT using the large $N$ expansion.
  This is a candidate saddle of the full answer.
} 
\label{fig:2coupledJTnoIsland}
\end{figure}
\begin{figure}[ht]
\begin{center}
\includegraphics[width=12cm]{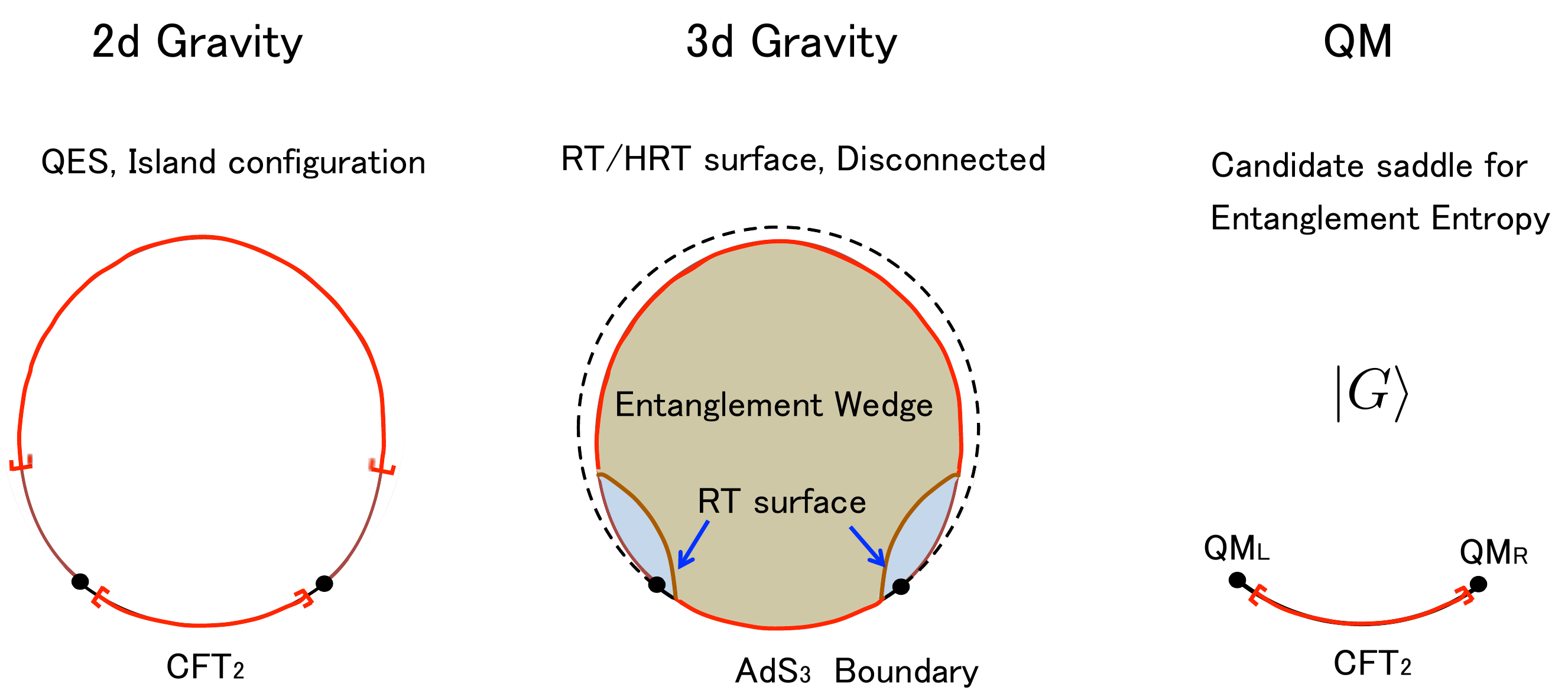}
\end{center}
\caption{We describe the three descriptions for the quantum extremal surface which contains islands.
{\bf Left:} The JT gravity plus matter CFT description. We are computing quantum extremal surfaces.
{\bf Right:} We use the 3d gravity description for matter CFT assuming that the matters are holographic.
  We are computing the lengths of the RT/HRT surfaces that end on the JT brane.
  The shaded region is the entanglement wedge, which is surrounded by the entangling surface and the RT/HRT surfaces.
{\bf Right:} The full quantum mechanical description. We calculate entanglement entropy of CFT using the large $N$ expansion.
  This is another candidate saddle of the full answer.
} 
\label{fig:2coupledJTIsland}
\end{figure}

Next we calculate  entropy of  $\pmb{[0_ + ,d  - 0]}$, i.e. entanglement entropy between bath and the two JT + matter systems.
In this case, there are at least two candidate of the quantum extremal surfaces .
The first one is  entanglement entropy of the interval $[0,d]$ on a circle of the total length $\pi \ell + d$.
This is usual entanglement entropy for a single interval and is evaluated as
\be
S_{\text{no-Island}} = \f{c}{3} \log \Big[\f{ \pi \ell + d}{\pi \epsilon _B} \sin \Big( \pi \f{d }{\pi \ell + d} \Big)  \Big].
\ee
The second one is a quantum extremal surface where we have the island.
Assuming the matter fields are holographic and use the Ryu-Takayanagi formula, this is the twice of the entropy of the entropy between single JT gravity + matter system we calculated in (\ref{eq:QESsingle1}):
\be
S_{\text{Island}} = 2 S_{gen}(\sigma_s).
\ee
Entanglement entropy of the region $\pmb{[0_ + ,d  - 0]}$ is given by the minimal value of these candidates:
\be
S_{ent} = \text{min} \{S_{\text{no-Island}} ,S_{\text{Island}}\}.
\ee
Assuming that $S_0$ is sufficiently large, for small $d$ the no-island saddle dominates.
$S_{\text{no-Island}}$ grows when we increase the length $d$.
When $d$ is extremely large, we expect that the island saddle dominates.
We can estimate the point where the dominant saddle exchanges.
For large $d$, the length of the bath and the wormhole are almost the same.
Therefore, we can evaluate the no-island saddle as 
\be
S_{\text{no-Island}} \approx \f{c}{3} \log  \f{d}{ \epsilon_B}
\ee
On the other hand, the island saddle is almost constant when $d \approx \pi \ell$ is large:
\be
S_{\text{Island}} = 2 S_{gen}(\sigma_s) = 2S_0 + \f{c}{3} + \f{c}{3} \log \f{12\pi \bar{\phi}_r}{\epsilon_B c}.
\ee
When these two are equal,
\be
S_{\text{Island}} = S_{\text{no-Island}} \to \f{c}{3} \log  \f{d}{ \epsilon_B} = 2S_0 + \f{c}{3} + \f{c}{3} \log \f{12\pi \bar{\phi}_r}{c \epsilon_B}.
\ee
This happens when
\be
\f{c}{12\pi \bar{\phi}_r}d = \exp \Big(\f{6S_0}{c}+ \f{3}{4}\Big) .
\ee
or 
\be
 d = \f{12\pi \bar{\phi}_r}{c} e^{\f{6S_0}{c}+ \f{3}{4}} = \f{3\pi}{4} \ell_0 e^{\f{6S_0}{c}+ \f{3}{4}}.
\ee
Here $\ell_0 = \f{16 \bar{\phi}_r}{c}$ is the wormhole length when the bath length is $0$, which is the minimal wormhole length of this traversable wormhole setup.
Therefore, when the wormhole is exponentially large with respect to the ratio $\f{6S_0}{c}$, the island saddle dominates.
For 4d traversable wormholes, these parameters become 
\be
S_0 = \f{\pi r_e^2}{G_N} = \f{\pi q^2}{g^2}, \qquad c = q.
\ee
Therefore, the ratio 
\be
\f{S_0}{c} = \f{\pi q}{g^2},
\ee
is huge when $q$ is large.
This means that it becomes hard to develop the island for large $q$ i.e. the magnetic charge and the mass of the original black hole are large.
Semiclassical limit is good only when $q$ is large, so it seems to be hard to develop islands.
On the other hand, if we use the SYK values, these parameter becomes \footnote{Precisely speaking, when we define bulk fermions mass from the SYK fermion operator dimensions through the usual AdS/CFT dictionally, the bulk fermions are not massless. }
\be
S_0 = N s_0, \qquad c = \f{N}{2}.
\ee 
In the SYK model $s_0$ is given by $ s_0 = \f{1}{2}\log 2 - \f{\pi^2 }{4q^2}$ in the large $q$ expansion \cite{Maldacena:2016hyu} and $s_0 \approx 0.2324 \approx \f{1}{2} \log (1.592)$ for $q = 4$ \cite{Cotler:2016fpe}.
Therefore the ratio is
\be
\f{S_0}{c} = 2s_0,
\ee
which is not big and does not depend on $N$.
In this type of theory it is easy to develop the island because $e^{\f{S_0}{c}}$ is order one quantity (w.r.t order $N$ counting) and $\ell_0 e^{\f{6S_0}{c}+ \f{3}{4}}$ is not so big.

\subsection{JT + several 2d holographic matters \label{sec:JTmanyHolographic} }
In the 4d traversable wormhole/magnetically charged blackholes setup\cite{Maldacena:2018gjk, Maldacena:2020skw}, we obtain large number of free 2d fermions because of the Landau degeneracy.
Similarly, we can consider to couple $2d$ several  holographic matters where the bulk fields have their own $3d$ gravity dual.
In other words, there are $N_H$ holographic matters that are coupled to JT gravity.
We label these holographic matters by $i$ and denote their Newton constants by $G_{N,(i)}^{(3)}$, AdS radius by $l_{AdS_3}^{(i)}$ and the metric by $g_{(i)\mu\nu}^{(3)}$.
Brown-Henneaux central charge is given by $c^i = \f{l_{AdS_3}^{(i)}}{8 \pi G_{N,(i)}^{(3)}}$.
For holographic CFTs, the "quantum" stress energy tensor expectation value is \cite{Balasubramanian:1999re} 
\be
T^{mat_i}_{\mu\nu} = \lim_{\epsilon\to 0} \f{1}{8 \pi G_{N,(i)}^{(3)}} \Big[K_{(i)\mu \nu} - K_{(i)} h_{(i)\mu\nu}^{(2)} - \f{1}{l_{AdS_3}^{(i)}}h  _{(i)\mu\nu}^{(2)}\Big] \Bigg|_{z = \epsilon}.
\ee
Here, $h_{(i)\mu\nu}^{(2)}$ is the induced metric on the UV cutoff surface $z= \epsilon$ and is expanded as 
\be
h^{(2)}_{(i)\mu\nu} = \f{g_{\mu\nu}^{(2)}}{\epsilon^2} + \cdots
\ee
where the 2d metric $g_{\mu\nu}^{(2)}$ is the metric in JT gravity, which will be set to be that of AdS$_2$ after imposing the equation of motion for JT gravity.
Note that we impose the same asymptotic boundary conditions for all the $N_H$ holographic matters because all the holographic CFTs are coupled to the same 2d gravity.
Now, the equation of motion for JT gravity becomes 
\be
-\f{1}{8 \pi G_N^{(2)}}(\nabla_{\mu} \nabla_{\nu} \phi  - g_{\mu \nu}^{(2)} \nabla ^2 \phi + g_{\mu \nu}^{(2)} \phi) =  \sum_{i = 1}^{N_H} \f{1}{8 \pi G_{N,(i)}^{(3)}} \Big[K_{(i)\mu \nu} - K_{(i)} h_{(i)\mu\nu}^{(2)} - \f{1}{l_{AdS_3}^{(i)}}h^{(2)}  _{(i)\mu\nu}\Big] \Bigg|_{z = \epsilon}  .
\ee
Note that all the holographic matters shares the same boundary condition $h^{(2)}_{(i)\mu\nu} = \f{g_{\mu\nu}^{(2)}}{\epsilon^2} + \cdots$, which represent the coupling to the same $2d$ JT gravity.
$K_{(i)\mu \nu}$ is the extrinsic curvature of the UV cutoff surface in the $i$-th holographic CFT.

These can be thought of as "Randall Sundrum junctions", since several 3d geometries are connected through the same JT gravity.
In particular, we can consider the special case where all the holographic matters are the same theory.
In this case, they have the same central charges and we obtain 
\be
-\f{1}{8 \pi G_N^{(2)}}(\nabla_{\mu} \nabla_{\nu} \phi  - g_{\mu \nu}^{(2)} \nabla ^2 \phi + g_{\mu \nu}^{(2)} \phi) = \f{1}{8 \pi G_{N}^{(3)}} \sum_{i = 1}^{N_H}  \Big[K_{(i)\mu \nu} - K_{(i)} h_{(i)\mu\nu}^{(2)} - \f{1}{l_{AdS_3}^{(i)}}h^{(2)}  _{(i)\mu\nu}\Big] \Bigg|_{z = \epsilon}
\ee

When we have two holographic matters, the setup is similar to the defect CFT model \cite{Erdmenger:2015xpq, Simidzija:2020ukv}.
In this case, JT brane exists between two holographic matters and looks like a defect between them.
We can consider the parameter regime where $N_H$ is of order $1/G_N^{(2)}$ so that we can balance the JT classical fields and "highly quantum" holographic matter effects.
In other words, each $c_i$ is $\mathcal{O}(1)$ in  the $1/G_N^{(2)}$ expansion but is still sufficiently large $c_i \gg 1$ so that we can use large $c$ limit for these holographic CFTs.

This JT $+$ several holographic matters model is convenient when we try to change the boundary of holographic CFT$_2$ gradually since we can associate different boundary conditions for different holographic matters.
Similar situation can arise from the 4d holographic CFT with $U(1)$ symmetry and strong magnetic fields \cite{Maldacena:2020sxe,DHoker:2009mmn}.
In this situation the dual $5d$ geometry flows from AdS$_5$ to AdS$_3 \times$ $\mathbb{R}^2$.
The magnetic flux $q$ which is vertical to $\mathbb{R}^2$ plays the role of $N_H$ there.

This JT gravity + many holographic matters setup is convenient for our purpose in latter sections.

\subsection{Traversable wormholes with Partial couplings \label{sec:PartialCouplings}}
\begin{figure}[ht]
\begin{center}
\includegraphics[width=12cm]{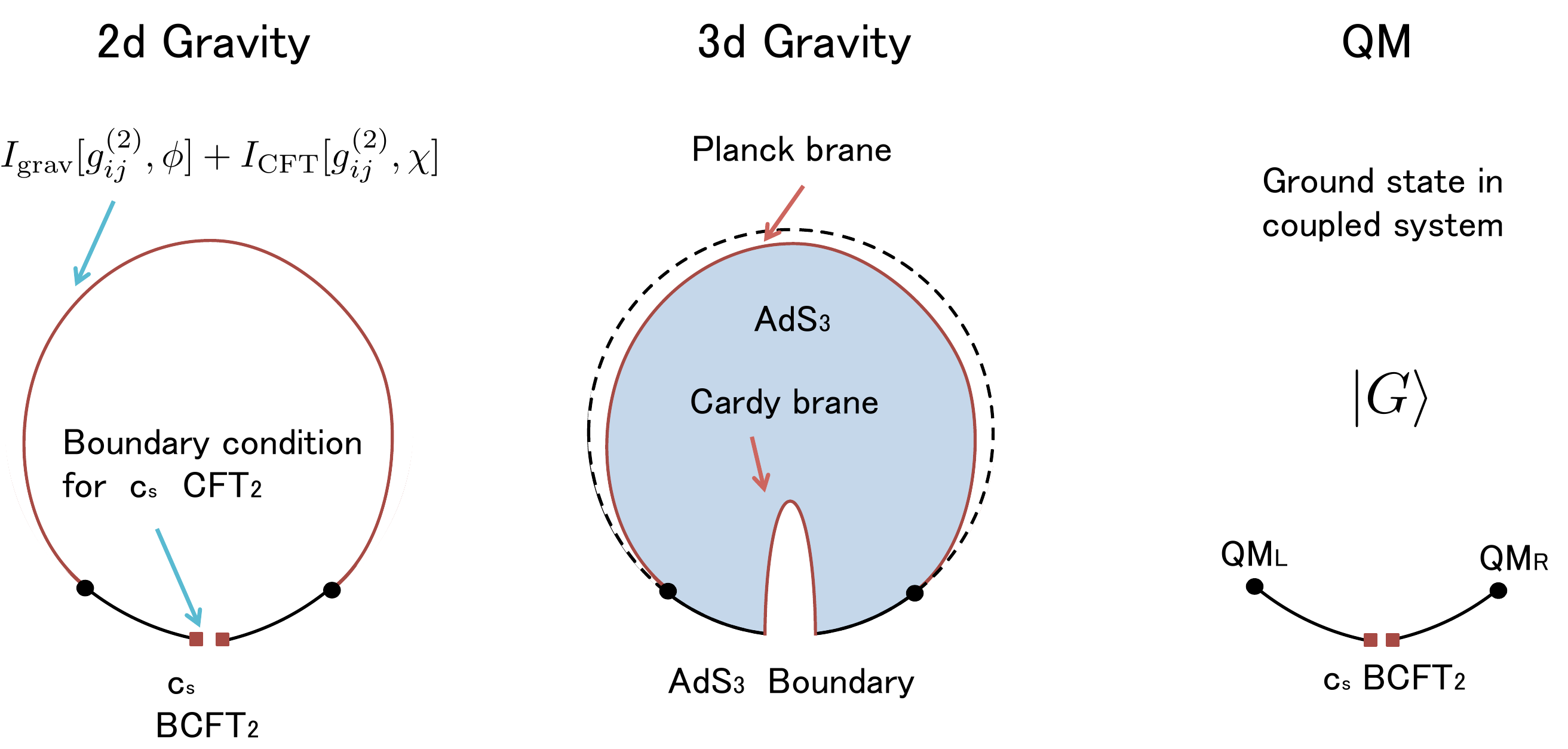}
\end{center}
\caption{The configuration for CFT with the central charge $c_s$ with boundary conditions. 
We suppress the $c_j$ CFT without boundaries that are treated in the same manner with the former case.
{\bf Left:} The JT gravity $+$ matter CFT description coupled to BCFT in the no gravity region.
{\bf Middle:} We take the 3d gravity description assuming that the matter CFT is holographic.
The 3d spacetime now has a Cardy brane, which is dual of boundary condition in AdS/BCFT setup.
{\bf Right:} The full quantum mechanical description. We couple BCFT to quantum mechanics. 
Again note that we suppress the $c_j$ CFT in this picture, which directly connects two quantum mechanics.
} 
\label{fig:2coupledBCFT}
\end{figure}

Until now we assume that we introduce the periodic boundary conditions for all the matter fields through the CFT in the region on which we do not have dynamical gravity.
Here we introduce the periodic boundary condition only for $c_j < c$ matter fields and we introduce a boundary condition in the no gravity region for $c_s =c -c_j $ fields that keeps a half of conformal symmetry \cite{CARDY1984514} \footnote{Here $j$ of $c_j$ stands for "join", which means that we are connecting in the no gravity region.
$s$ of $c_s$ stands for "split", which means that we split the direct coupling by introducing boundaries.}.
The way to introduce boundaries are shown in figure.\ref{fig:2coupledBCFT}.
These introduction of boundaries cut the direct couplings \cite{Shimaji:2018czt, Caputa:2019avh}.
In the coupled SYK models with "partial couplings" , we can still have similar traversable wormholes because of the all to all coupling nature \cite{Gao:2019nyj} and we expect the same thing for JT gravity.
We assume that the boundary conditions are the same in both side of the boundary so that the CFT achieve the minimal energy on a finite strip.
We deal with $c_{s}$ as a continuous parameter.
This is valid when the CFTs are collections of free CFTs.
In a model with holographic matters, we consider the model in section \ref{sec:JTmanyHolographic} and then we can treat $c_j$ as a continuous parameter.
For simplicity, we put these boundary conditions at $x = d/2$ on the strip $[0,d]$.
In this case , the ground state energy for $c_s$ fields on a flat finite strip is 
\be
T_{++} = T_{--} = -\f{c_s}{48 \pi} \f{1}{(1+a)^2}.
\ee
If we couple the matter to AdS$_2$ for $ x \in [\pi a/2 , \pi(1 + a/2)]$ region, the energy in the AdS$_2$ region is 
\be
\braket{T_{y^+y^+}^{\text{AdS}_2}} = \f{c_s}{48 \pi} - \f{c_s}{48 \pi} \f{1}{(1+a)^2} = \f{c_s}{48\pi }\f{a(a+2)}{(a+1)^2}. \label{eq:BCFTbulkEnergy}
\ee
Therefore, if we introduce the boundary conditions for fields outside the wormhole, these matters only introduce the positive energy in the AdS$_2$ region.
The negative energy we have in total is 
\ba
\braket{T_{y^+y^+}^{\text{AdS}_2}} &= -\f{c_j}{48\pi} \f{(1-a)(3+a)}{(a+1)^2} + \f{c_s}{48\pi} \f{a(a+2)}{(a+1)^2} \notag \\
&=-\f{1}{48\pi} \f{c_j(1-a)(3+a) - c_sa(a+2)}{(a+1)^2} \notag \\
&= -\f{1}{48\pi} \f{-c a^2 - 2ca +3 c_j}{(a+1)^2}  =  -\f{c_{\text{eff}}(a)}{16\pi} .
\ea
where $c_{\text{eff}}(a) = \f{-c a^2 - 2ca +3 c_j}{3(a+1)^2} $ plays a role of "effective central charges" for making traversable wormholes.
Therefore the effective central charge decreases and the wormhole length becomes longer and longer.
In particular for $d \ll \ell$, $c_s$ do not give any contribution and the wormhole length is determined as $\ell = \f{16 \bar{\phi}_r}{c_j}$.

The wormhole length $\ell $ is 
\be
\ell = \f{2\bar{\phi}_r}{c_{\text{eff}}(a) G_N} = \f{2\bar{\phi}_r}{ G_N} \f{3(1+a)^2}{3c_j -2 ca - ca^2} . \label{eq:WHlengthPartial}
\ee
and the length of the bath ( region without gravity ) in terms of parameter $a$ is 
\be
d = \pi \ell a = \f{2\bar{\phi}_r}{ G_N} \f{3(1+a)^2}{3c_j -2 ca - ca^2} .
\ee
This fix the parameter $a$ and the wormhole length $\ell$ in terms of $d$.
For $\f{cG_N}{2 \phi_r} d \gg 1$, the parameter $a$ becomes
\be
a = \f{d}{\pi \ell} \approx \s{1 + \f{3 c_j}{c}} -1 .
\ee
Since this is smaller than $1$, the wormhole length $\pi \ell$ is larger than $d$, which is expected from the achronal average null energy condition.
We should also notice that the traversable wormhole solution do not exist when  $c_s = c$ (or equivalently $c_j =0$) because we do not couple two sides of AdS$_2$ and  do not introduce  negative null energy in the bulk any more. 
Therefore it is not possible to construct the traversable wormhole solutions.
Actually, even when $c_a$ is the order one and not  the order of $1/G_N^{(2)}$, the wormhole length $\ell$ becomes very large and we can not ignore the quantum gravity effect so that we can not trust the classical treatment of the wormholes.

Entanglement entropy of the no gravity region is also calculated in the same manner.
Here we assume that CFTs are given by holographic CFTs, and BCFTs are modeled by the AdS/BCFT setup \cite{Takayanagi:2011zk,Fujita:2011fp} with tensionless Cardy brane.
Here we consider entanglement entropy for the interval $\pmb{ [0_ + ,d  - 0]}$, which means entanglement between CFT on no gravity region and quantum mechanics QM$_L$ and QM$_R$ in the full quantum mechanical description.
In the configuration without islands, entanglement entropy of quantum fields are given by\footnote{Entanglement entropy of the region $A = [0,l]$ on a strip $[0,L]$ is given by \cite{Calabrese:2009qy}
\be
S_A = \f{c}{6} \log \Big(\f{2L}{\epsilon} \sin \f{l}{L}\Big) + \log g + c_1'.
\ee
where $\log g$ is the boundary entropy \cite{PhysRevLett.67.161} and $c_1'$ is the non universal constant without the presence of the boundaries.} 
\ba
S_{\text{no-Island}} &= \f{c_j}{3}\log \Big( \f{\pi \ell + d}{\epsilon_B \pi} \sin \f{ \pi d}{\pi \ell + d} \Big) + \f{c_s}{3}\log \Big( \f{2 (\pi \ell + d)}{\epsilon_B \pi} \sin \f{ \pi d}{(\pi \ell + d)} \Big) +2  \log g_s \notag \\
&\approx  \f{c_j}{3}\log \f{d}{\epsilon_B} + \f{c_s}{3}\log \f{2d}{\epsilon_B} + 2\log g_s = \f{c}{3} \log \f{d}{\epsilon_B} + \f{c_s}{3}\log 2  + 2\log g_s  \qquad (\text{for}\ \ \pi \ell \gg d) .
\ea 
On the other hand, in the island configuration the generalized entropy functional is
\ba
S(\sigma_a,\sigma_b) &= 2S_0 + 2\pi \phi(\sigma_a) + 2\pi \phi(\sigma_b) + \f{c_j}{6} \log \Big[ \Big(\f{\pi \ell + d}{\pi} \Big)^2 \f{\sin^2 \f{ \pi \sigma_a}{\pi \ell + d}}{\epsilon_B  \ell \sin \f{\sigma_a}{\ell}} \Big] + \f{c_j}{6} \log \Big[ \Big(\f{\pi \ell + d}{\pi} \Big)^2 \f{\sin^2 \f{\pi \sigma_b}{\pi \ell + d}}{\epsilon_B \ell \sin \f{\sigma_b}{\ell}} \Big] \notag \\
&\ +  \f{c_s}{6} \log \Big[ \Big(\f{2(\pi \ell + d)}{\pi} \Big)^2 \f{\sin^2 \f{ \pi \sigma_a}{2(\pi \ell + d)}}{\epsilon_B \ell \sin \f{\sigma_a}{\ell}} \Big] + \f{c_s}{6} \log \Big[ \Big(\f{2(\pi \ell + d)}{\pi} \Big)^2 \f{\sin^2 \f{ \pi \sigma_b}{2(\pi \ell + d)}}{\epsilon_B \ell \sin \f{\sigma_b}{\ell}} \Big]
\ea 
Assuming $\pi \ell \gg  d, \sigma_a,\sigma_b$ , this entropy functional reduces to the extremal black hole one\footnote{Without the assumption that the BCFT is  replaced by the AdS/BCFT model, we will get entanglement entropy for an interval with the presence of the boundary, which is not universal.}
\ba
S(\sigma_a,\sigma_b) &\approx  2S_0 + 2\pi \phi(\sigma_a) + 2\pi \phi(\sigma_b) + \f{c_j}{6} \log \f{\sigma_a}{\epsilon_B} +  \f{c_j}{6} \log \f{\sigma_b}{\epsilon_B} 
\ + \f{c_s}{6} \log \f{\sigma_a}{\epsilon_B} +  \f{c_s}{6} \log \f{\sigma_b}{\epsilon_B} \notag \\
&= 2S_0 + 2\pi \phi(\sigma_a) + 2\pi \phi(\sigma_b) + \f{c}{3} \log  \f{\sigma_a}{\epsilon_B} + \f{c}{3} \log \f{\sigma_b}{\epsilon_B} .
\ea 
So the entanglement entropy only gets the boundary entropy contribution in the no island configuration, but except that the structure is basically the same with the former case without boundaries.

\section{4 coupled JT gravity \label{sec:FourCoupledJT}}
\begin{figure}[ht]
\begin{center}
\includegraphics[width=14cm]{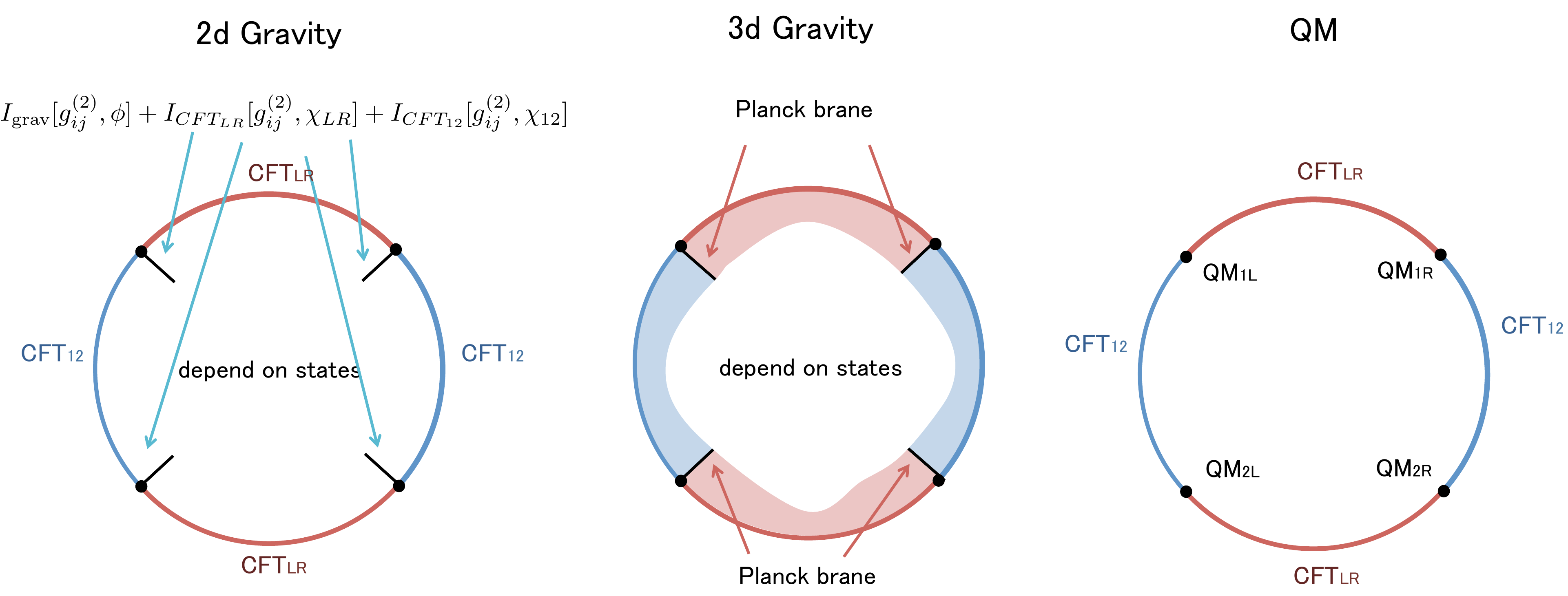}
\end{center}
\caption{The descriptions of the theory we consider in the 4 coupled JT gravities. 
} 
\label{fig:4coupledJTsetup}
\end{figure}

In this section, we consider the four coupled JT gravity coupled to conformal matters.
The setup is a generalization of the  two coupled JT gravity we discussed above and similar to the four coupled SYK models we studied in section \ref{sec:FourCoupledSYK}.
First, we assume that the matter fields consists of at least two decoupled CFTs: CFT$_{LR}$ with central charge $c_{LR}$ and CFT$_{12}$ with central charge $c_{12}$.
In other words, the JT gravity + matter theory is given by the action 
\be
S =I_{\text{grav}}[g_{ij}^{(2)},\phi] + I_{CFT_{LR}}[g_{ij}^{(2)},\chi_{LR}] + I_{CFT_{12}}[g_{ij}^{(2)},\chi_{12}],
\ee
where $I_{\text{grav}}[g_{ij}^{(2)},\phi]$ is the JT gravity action \eqref{eq:JTaction},  the $\chi_{LR}$ abstractly represents the fields for CFT$_{LR}$ and  the $\chi_{12}$ abstractly represents the fields for CFT$_{12}$. 
Then, we also prepare two copies of CFT$_{LR}$ and CFT$_{12}$  on lengths $d_{LR}$ and $d_{12}$ without gravity.
We couple JT + matter system through these CFTs on no gravity region in a way that is shown in figure \ref{fig:4coupledJTsetup}.

\subsection{CFT on 2 traversable wormholes}

\begin{figure}[ht]
\begin{center}
\includegraphics[width=14cm]{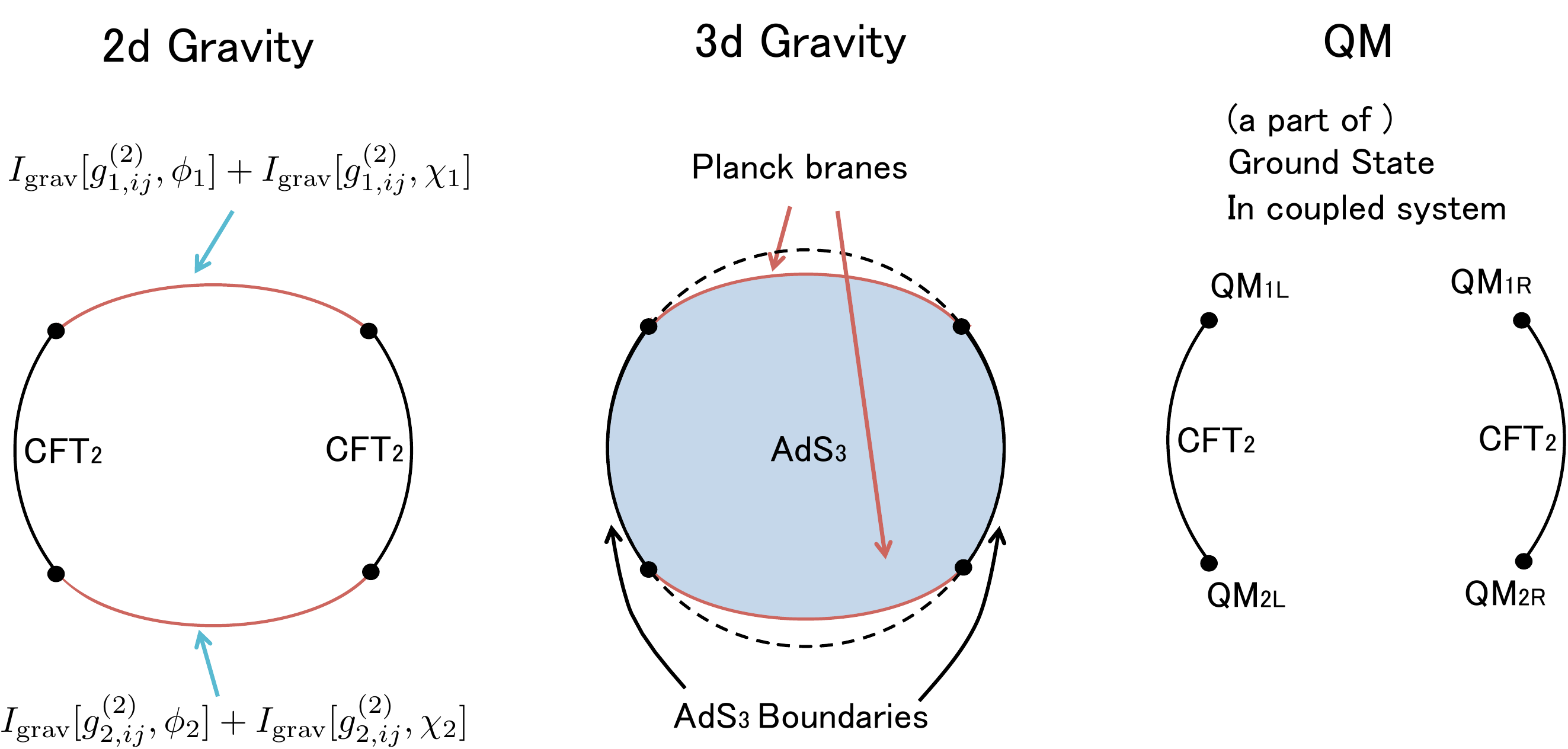}
\end{center}
\caption{The configuration for CFT sitting on two traversable wormholes. 
We suppress the $c_{LR}$ CFT that are treated in the same manner with the former case.
} 
\label{fig:CFTonTwoWHs}
\end{figure}

To construct the wormhole solutions in four coupled JT gravities, first we consider CFT on a cylinder of a length $L = 2\pi(1 + b)$.
Then, the expectation value of the stress tensor is 
\be
\braket{T^{\text{cyl}}_{tt}} = \braket{T^{\text{cyl}}_{\sigma\sigma}} = -\f{c}{24\pi} \f{1}{(1+b)^2}.
\ee
In terms of null energy, they become 
\be
\braket{T^{\text{cyl}}_{y^+y^+}} = \braket{T^{\text{cyl}}_{y^-y^-}} = -\f{c}{48\pi} \f{1}{(1+b)^2}.
\ee
We couple the region $\sigma \in [0,\pi]$ to AdS$_2$ metric.
Furthermore, we put another JT gravity on the region $\sigma \in [\pi(1+b),\pi(2+b)]$.
Therefore, this CFT is sitting on two different wormholes!
Then, the null energy in each AdS$_2$ region is
\be
\braket{T^{\text{AdS}_2}_{y^+y^+}} = \braket{T^{\text{AdS}_2}_{y^-y^-}} = \f{c}{48\pi} -\f{c}{48\pi}\f{1}{(1+b)^2} = \f{c}{48\pi} \f{b(b+2)}{(1+b)^2}. \label{eq:CFTtwoWHenergy}
\ee 
First of all, this coupling between two side actually do not introduce the negative energy in the bulk.
Rather, it is introducing a positive energy in the 2$d$ bulk.
This contribution takes the same form with (\ref{eq:BCFTbulkEnergy}), in which we introduce the boundary conditions for CFT in the no gravity region \footnote{Actually two situations are related by the folding trick.}.

Let us understand this when this CFT on 2 traversable wormholes is holographic.
One interesting thing is that now two traversable wormholes are directly connected  through 3$d$ spacetime which is dual to the entangled state of bulk  matter.
From 3$d$ gravity point of view, this configuration is actually a traversable wormhole in 3$d$ since two asymptotic boundaries are connected in the interior as depicted in the figure \ref{fig:CFTonTwoWHs}.
The positive energy is interpreted as a cost to make this 3$d$ wormhole traversable.
This positive energy in the bulk makes wormholes longer.
This behavior is also similar to the SYK interaction $\mu_{12}(H_{LL}^{12} + H_{RR}^{12})$ in section \ref{sec:4SYKPhaseDiagram}, which causes the decrease of the energy gap in L-R wormhole phase rather than increasing the energy gap.


Taking $b \to \infty $, we obtain the CFT on infinite line with coupling to AdS$_2$ on the region $\sigma \in [0,\pi]$:
\be
\braket{T^{\text{AdS}_2}_{y^+y^+}} = \braket{T^{\text{AdS}_2}_{y^-y^-}} = \f{c}{48\pi} .
\ee 
This induces a positive null energy on AdS$_2$ region.

The length in flat space region on which CFT lives is given by
\be
d_{12} = \pi \ell b.
\ee
When we take $d_{12}=0$, then the 3d gravity geometry looks like that in wedge holography \cite{Akal:2020wfl}.



\subsection{Wormhole solutions in 4 coupled JT gravities \label{sec:wormhole4coupled}}
\subsubsection{variational anzatz}
Here we first consider the variational approximation 
We approximate the geometry by the $t = 0$ slice of the eternal black holes (= thermo field double states).
The wormhole length is $\ell = \f{1}{2\pi T_H}$.
The energy par a black hole is 
\be
M_{BH} = \f{\pi \bar{\phi}_r}{4 G_N} T_H^2 = \f{ \bar{\phi}_r}{16 \pi G_N} \f{1}{\ell^2}. 
\ee
The quantum matter contribution is 
\be
E_{mat} = 2 \f{c_{LR} + c_{12}}{24 \ell} - 2\f{c_{LR} \pi}{6 (\pi \ell + d_{LR})} - \f{c_{12} \pi}{6 \cdot 2 (\pi \ell + d_{12})}.
\ee
The total variational energy is 
\be
E(\ell) = 4 M_{BH} + E_{mat} = \f{ \bar{\phi}_r}{4 \pi G_N} \f{1}{\ell^2} +  \f{c_{LR} + c_{12}}{12 \ell} - \f{c_{LR} \pi}{3 (\pi \ell + d_{LR})} - \f{c_{12} \pi}{12 (\pi \ell + d_{12})}.
\ee
We minimize this energy 
\be
\f{\partial E}{\partial \ell} = -\f{ \bar{\phi}_r}{2 \pi G_N} \f{1}{\ell^3} -  \f{c_{LR}+ c_{12}}{12 \ell^2} + \f{c_{LR} \pi^2}{3 (\pi \ell + d_{LR})^2} + \f{c_{12}\pi^2}{12 (\pi \ell + d_{12})^2}= 0. \label{eq:variational4coupled}
\ee
$\ell$ is then determined by solving the equation \eqref{eq:variational4coupled}.
We analyze this in some special cases.
First we consider the case of $d_{12} = 0$.
In this case, the energy from the $c_{12}$ CFT completely cancels.
Therefore, the length of the wormhole becomes the same with the case without $c_{12}$ CFT.
Second, we consider the case of $d_{LR} = 0$ and $d_{12} \to \infty$ \footnote{Here we mean $\f{d_{12}}{\bar{\phi}_r} \gg 1$. In this regime, the solution with $\pi \ell \approx d_{12}$ may not be trusted since the quantum gravity effect becomes important.}.
In this case, the matter energy becomes 
\be
E_{mat} = -\f{3c_{LR} - c_{12}}{12 \ell}. \label{eq:Energy12InfL}
\ee
In this regime, the wormhole length becomes that of single traversable wormhole with effective central charge $c_{LR} - \f{c_{12}}{3}$ for $c_{12}<3c_{LR}$ cases. 
On the other hand when $c_{12} > 3c_{LR}$ the matter energy no more introduce the negative energy in AdS$_2$ region and we can not make traversable wormholes.
This means that we can not take $d_{12}$ to be large while keeping the length $\ell$ to be finite.
There is a cost to entangle the $c_{12}$ fields through bulk and there is a limit to support such entanglement.
For $c_{12} > 3c_{LR}$, as we take $d_{12}$ to be large, $\ell$ also becomes large rather than approaches to a constant.

\subsubsection{solving the JT Gravity equation of motion directly}
We solve the equation of motion of the JT gravity \eqref{eq:dialtonEOM1} directly. 
Finally we found that the variational method gives the exact answer.
We assume the form 
\ba
ds^2 &= \f{-dt^2 + d\sigma^2}{ \ell ^2 \sin^2 \f{\sigma}{\ell}}, \notag \\
\phi(\sigma) &= \f{\bar{\phi}_r}{\pi \ell} \Big [\f{\pi \ell - 2\sigma}{\ell \tan \f{\sigma}{\ell}} + 1 \Big] + y.
\ea
Then, first we obtain 
\be
y = \f{c_{LR}+c_{12}}{24\pi}.
\ee
The null energy becomes 
\be
\braket{T_{x^+x^+}^{\text{AdS}_2}} = \f{c_{LR}}{48\pi} -\f{c_{LR}}{12 \pi (1+a)^2} + \f{c_{12}}{48\pi} -\f{c_{12}}{48 \pi (1+b)^2} = -\f{c_{\text{eff}}(a,b)}{16\pi}.
\ee
here we defined the effective central charge through
\be
c_{\text{eff}}(a,b) = \f{c_{LR}}{3} \Big( \f{4}{(1+a)^2}- 1 \Big) - \f{ c_{12}}{3} \Big( 1 - \f{1}{(1+b)^2} \Big)
\ee
Then, we can use the same equation \eqref{eq:WHlengthPartial} in a single traversable wormhole case.
The (inverse) length is determined as 
\be
t'  = \f{ c_{\text{eff}}(a,b)\pi G_N}{2\bar{\phi}_r}.
\ee
Using $t' = \ell^{-1}$, the definition of $c_{\text{eff}}(a,b)$ , $d_{LR} = \pi \ell a$ and $d_{12} = \pi \ell b$, we obtain
\be
\f{\bar{\phi}_r}{8 \pi G_N \ell^3 } = \f{c_{LR} + c_{12}}{48 \ell^2} - \f{c_{LR} \pi^2}{12(\pi \ell + d_{LR})^2 } - \f{c_{12} \pi^2}{12(\pi \ell + d_{12})^2 }. 
\ee
This is the same equation with (\ref{eq:variational4coupled}) that we obtained from the variational approximation.

\subsubsection{ Entanglement entropy of wormhole solutions}
 We consider entanglement entropy in the wormhole solutions in 4 coupled JT gravities.
 Because of symmetry, we can focus on the L-R wormhole solution without loss of generality.
 We first consider entanglement entropy between L and R.
 Similarly to the single traversable wormhole case, there only be island configuration because only specifying the entangling surface on the region without gravity does not divide the system to two parts.
 The generalized entropy functional becomes 
\ba
 & S_{gen}(\sigma_1,\sigma_2) \notag \\
 & = 2 S_0 + 2\pi \phi(\sigma_1) + 2\pi \phi(\sigma_2) + \f{c_{LR}}{6} \log \Bigg(\Big(\f{\pi \ell + d_{LR}}{\pi} \Big)^2 \f{\sin ^2 \f{\pi (\sigma_1 + b_1)}{\pi \ell + d_{LR}}}{\epsilon_B \ell\sin \f{\sigma_1}{\ell} } \Bigg) \notag \\
 & \qquad +  \f{c_{LR}}{6} \log \Bigg(\Big(\f{\pi \ell + d_{LR}}{\pi} \Big)^2 \f{\sin ^2 \f{\pi (\sigma_2 + b_2)}{\pi \ell + d_{LR}}}{\epsilon_B \ell\sin \f{\sigma_2}{\ell} } \Bigg) + \f{c_{12}}{6} \log \Bigg(\Big(\f{2(\pi \ell + d_{12})}{\pi} \Big)^2 \f{\sin ^2 \f{\pi (\sigma_1 + \sigma_2 +  d_{12})}{\pi \ell + d_{12}}}{\ell^2 \sin \f{\sigma_1}{\ell}\sin \f{\sigma_2}{\ell} } \Bigg). \label{eq:LRwormholeGent}
\ea
Entanglement entropy is obtained by taking the extremal of the generalized entropy.
Generically the saddle point of \eqref{eq:LRwormholeGent} can be found numerically.
For a simple case of $b_1 = b_2 = d_{LR}/2$, the extremal point is given by $\sigma_1 = \sigma_2 = \f{\pi \ell}{2}$.
In this case, entanglement entropy between L and R system is 
\be
S_{LR} = 2S_0 + 4 \f{\bar{\phi_r} }{ \ell} + \f{c_{LR}}{3} \log \bigg(\f{\ell}{\epsilon_B} \Big(\f{\pi \ell + d_{LR}}{\pi \ell}\Big)^2\bigg) + \f{c_{12}}{3} \log \bigg(\f{2 (\pi \ell + d_{12})}{\pi \ell}\bigg)
\ee
where $b_1,b_2$ are the distance from the AdS$_2$ asymptotic boundary in no gravity region and $\sigma_1,\sigma_2$ are the positions of quantum extremal surfaces.

\begin{figure}[ht]
\begin{center}
\includegraphics[width=9cm]{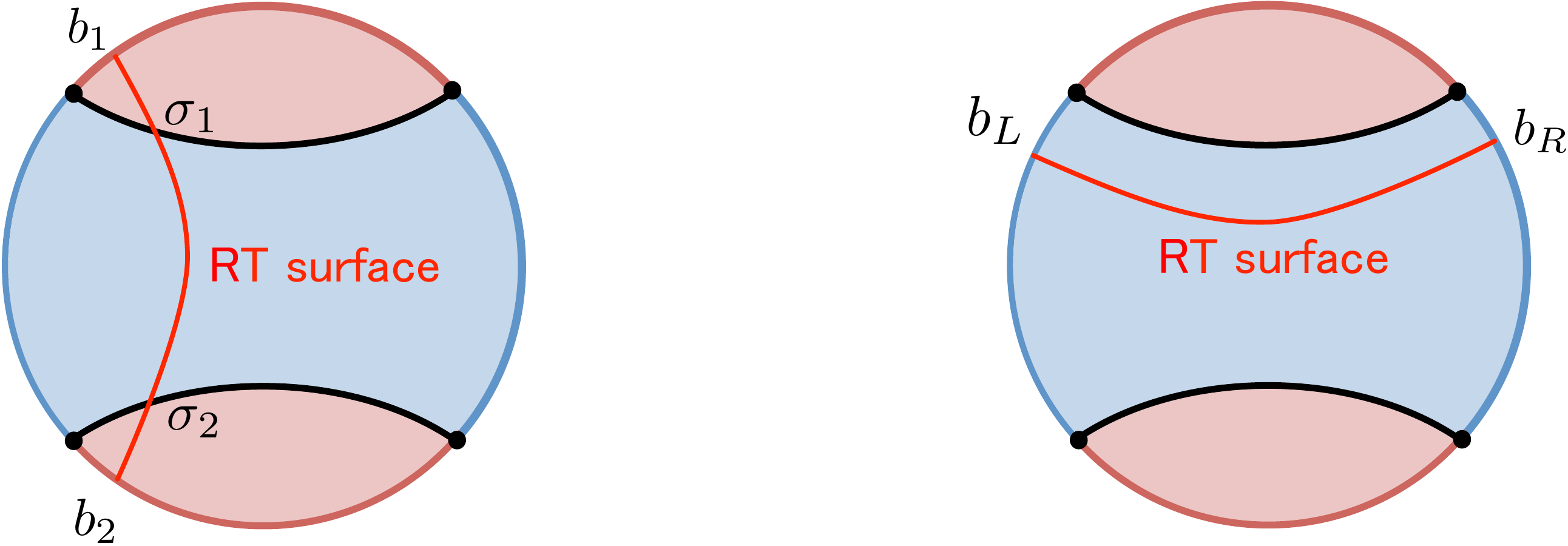}
\end{center}
\caption{ Entanglement entropy in L-R wormhole solutions with bulk holographic matters.
{\bf Left:} Entanglement between L and R system.
{\bf Right:} Entanglement between 1 and 2 system.
} 
\label{fig:4coupledJTconfig1}
\end{figure}

Next we consider entanglment entropy between 1 and 2 system.
In this case, we can consider the configuration without islands.
Entanglement entropy is simply given by the matter entanglement entropy
\be
S_{12} = \f{c_{12}}{3} \log \bigg( \Big ( \f{2(\pi \ell + d_{12})}{\pi \epsilon_B} \Big) \sin \f{\pi(b_L + b_R + \pi \ell)}{2 (\pi \ell + d_{12})} \bigg).
\ee

Therefore, in L-R wormhole solution entanglement entropy between L and R is much larger than that between 1 and 2 because of $2S_0$ contribution.
We can think of entanglement entropy as an order parameter, which is an analog of the spin operators in the four coupled SYK models.

\subsection{Symmetric solution}
We can find a solution which becomes symmetric under the change of 1-2 and L-R direction when $c_{LR} = c_{12}$.
We call this solution symmetric solution. 
The solution is simply given by the 4 extremal black holes 
\be
\phi(z) = \phi_0 + \f{\bar{\phi}_r}{z}. \label{eq:PoincareDilaton}
\ee
and all the CFTs are in the vacuum state on infinite lines.
There energy stress tensor vanishes $\braket{T^{mat}_{\mu\nu}} = 0$ everywhere, so the set of the dilaton profile and the quantum stress tensor satisfies the equation of motion. 
This solution exists for any $c_{LR}$ and $c_{12}$ they are not symmetric under the exchange of 1-2 and L-R direction when $c_{LR}  \neq c_{12}$.
The energy of this solution is $0$, which is bigger than the wormhole states.


\begin{figure}[ht]
\begin{center}
\includegraphics[width=10cm]{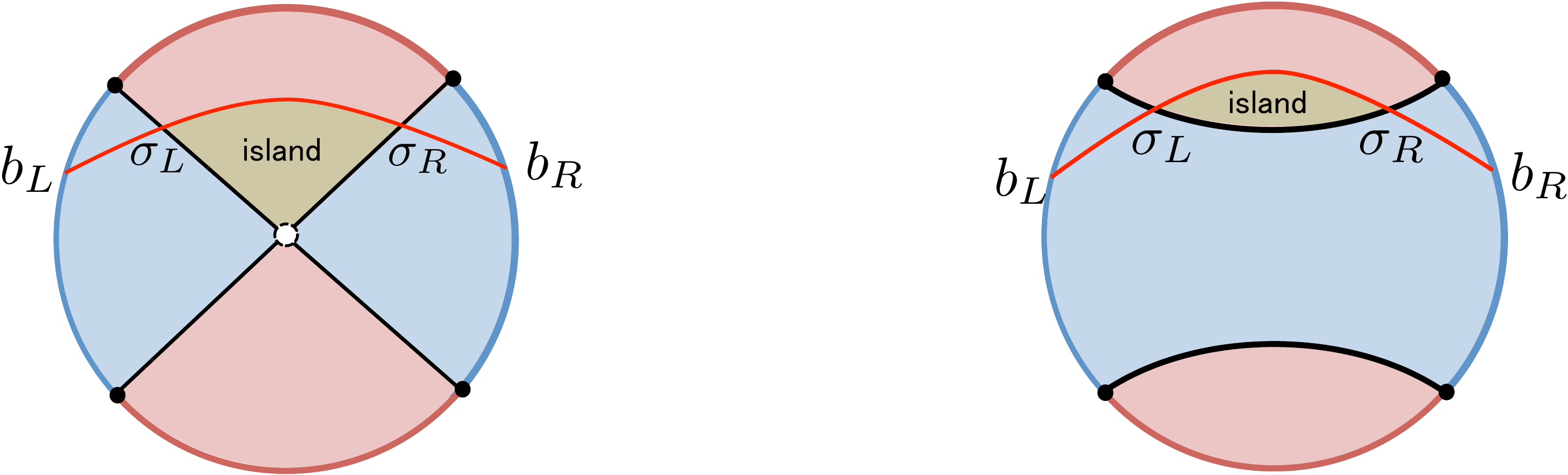}
\end{center}
\caption{The configurations for RT surface when the matters are holographic.
{\bf Left :} The symmetric solution. The dashed circle at the center represent the "infinity" of this geometry. {\bf Right: }Island phase in the L-R wormhole solution. } 
\label{fig:IslandSymmetric}
\end{figure}

We consider entanglement entropy in this solution.
In this solution, the island always dominate since the configurations without islands diverge.
The entropy functional is 
\be
S(\sigma_L,\sigma_R) = 2S_0 + \f{ 2\pi \bar{\phi}_r}{\sigma_L} + \f{2\pi\bar{\phi}_r}{\sigma_R} +\f{c_{12}}{6} \log \f{(\sigma_L +b_L)^2}{\epsilon_B\sigma_L}+\f{c_{12}}{6} \log \f{(\sigma_R +b_R)^2}{\epsilon_B\sigma_R} + \f{c_{LR}}{6}\log \f{(d_{LR} + \sigma_L + \sigma_R)^2}{\sigma_L \sigma_R}. \label{eq:genEntFuncSym}
\ee
where $b_L,b_R$ are the distances from the points on which asymptotic AdS$_2$ boundaries are located.
$\sigma_L, \sigma_R$ are the positions of quantum extremal surfaces. 
Assuming $d_{LR}$ is small ($d_{LR}  \ll \f{\bar{\phi}_r}{c}$), $b_L = b_R \equiv b$  and  $\sigma _L = \sigma_R \equiv \sigma$, the entropy functional becomes
\be
S(\sigma,\sigma) =  2\Big[ S_0 + \f{ 4\pi \bar{\phi}_r}{\sigma}  + \f{c_{LR}}{6} \log \f{(\sigma + b)^2}{\epsilon_B \sigma} \Big] + \f{c_{12}}{3} \log 2.
\ee
Note that the matter contribution in the L-R coupling only gives a constant in this parameter regime. 
Thus, we obtain the same solution with the zero temperature case for small $b$
\be
\sigma_s = \f{12  \pi\bar{\phi}_r}{c_{LR}}.
\ee
Entanglement entropy is given by 
\be
S(\sigma_s,\sigma_s) = 2S_0 + \f{c_{LR}}{3} + \f{c_{LR}}{3}\log\f{12 \pi \bar{\phi}}{\epsilon_B c_{LR}} + \f{c_{12}}{3} \log 2.
\ee
In particular in doubly holographic models, a portion of 3$d$ spacetime is   contained in the entanglement wedge of the right side.
This is a {\it 3d island} in this state.
In a doubly holographic setup, the JT branes are directly connected to other branes through the 3d direction, though the branes themselves are disconnected.
This is still enough to allow the state to have order $2S_0$ entropy for both of L-R direction and 1-2 direction.

Let us compare with the possible island phase in entanglement entropy between system 1 and 2.
We can consider the island phase in figure \ref{fig:IslandSymmetric}. 
This involves the computation of two interval entanglement entropy, which is not universal.
For simplicity, we consider the case that the bulk matter fields are holographic.
Then, the generalized entropy functional is given by
\ba
S(\sigma_L,\sigma_R) &= 2S_0 + \f{\phi(\sigma_L)}{4 G_N}+ \f{\phi(\sigma_R)}{4 G_N} + \f{c_{LR}}{6} \log \bigg(\Big( \f{\pi \ell + d_{LR}}{\pi} \Big)^2 \f{\sin^2 \f{\pi (\sigma_L + \sigma_R + d_{LR} )}{\pi \ell + d_{LR}}}{\ell^2 \sin \f{\sigma_L}{\ell}\sin \f{\sigma_R}{\ell}}\bigg) \notag \\
 & \qquad + \f{c_{12}}{6} \log \bigg(\Big(\f{2(\pi \ell + d_{12})}{\pi}\Big)^4  \ \f{\sin^2 \f{\pi (\sigma_L + b_L)}{2(\pi \ell+ d_{12})}\sin^2 \f{\pi (\sigma_R + b_R)}{2(\pi \ell + d_{12})}}{\epsilon_B^2  \ell^2  \sin \f{\sigma_L}{\ell} \sin \f{\sigma_R}{\ell}} \bigg)  \label{eq:4coupled12Island}
\ea
For $\ell \gg d_{12}, d_{LR}$ and assuming $\sigma_L, \sigma_R \ll \ell$ ,  the generalized entropy functional \eqref{eq:4coupled12Island} reduces to that for symmetric solution \eqref{eq:genEntFuncSym}.
Therefore, entanglement entropy for symmetric solution is basically the same with the entanglement entropy for long wormhole.

\subsection{$\mathbb{Z}_2$ symmetry breaking at special point}

\begin{figure}[ht]
\begin{center}
\includegraphics[width=16cm]{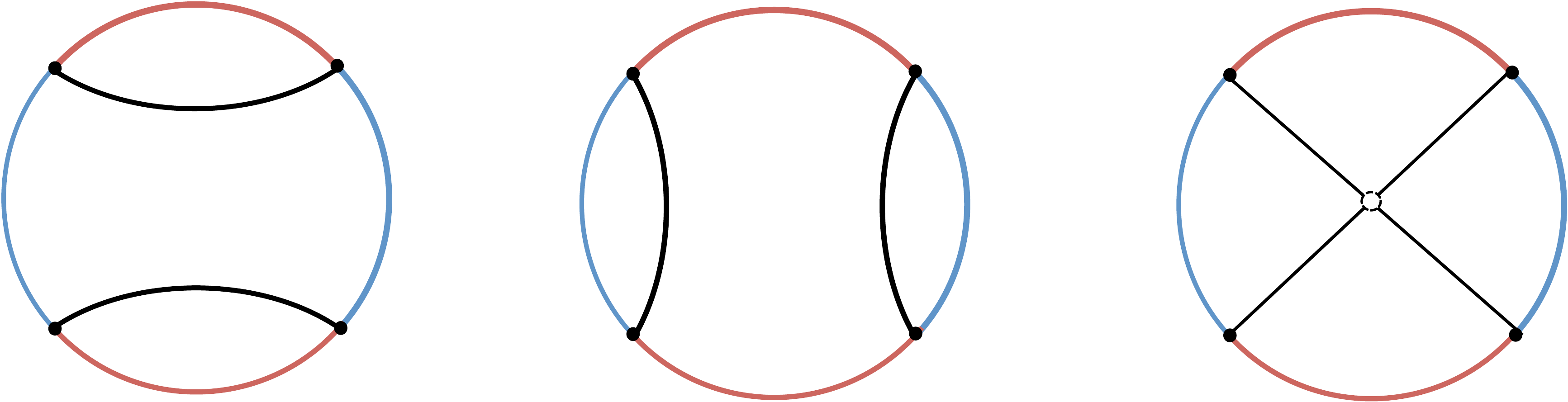}
\end{center}
\caption{ Configuration of the 4 coupled JT gravities with conformal matters. 
{\bf Left: }Traversable wormholes in L-R direction.
{\bf Middle:} Traversable wormholes in 1-2 direction.
{\bf Right:}A symmetric saddle of 4 coupled JT gravity + 2d CFT baths. 
The dashed circle at the center represent the "infinity" of this geometry.
} 
\label{fig:4coupledJTconfig1}
\end{figure}

\begin{figure}[ht]
\begin{center}
\includegraphics[width=15cm]{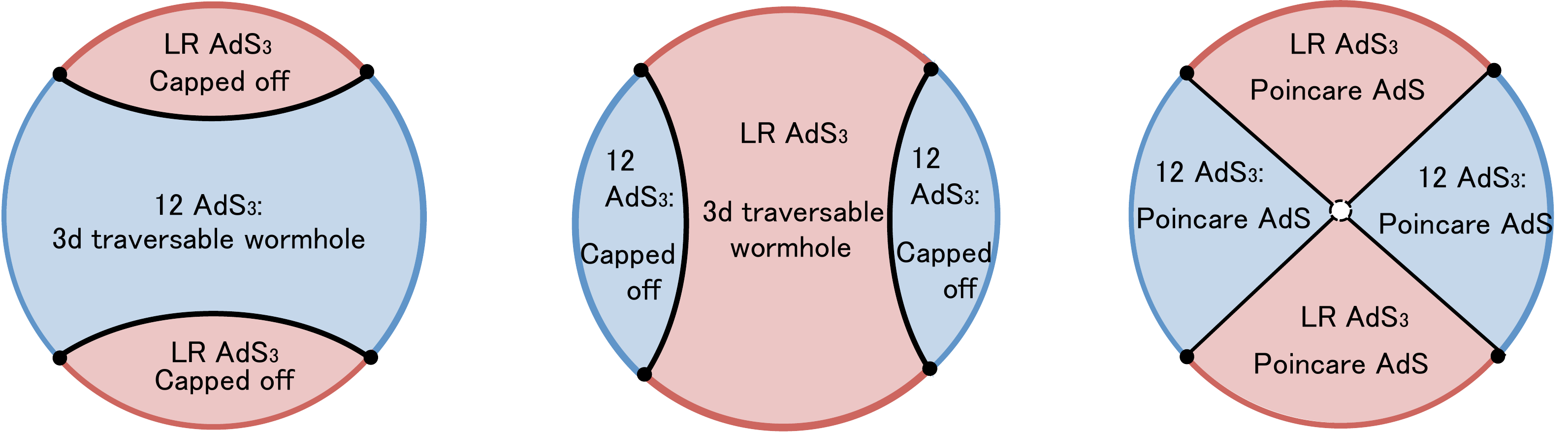}
\end{center}
\caption{ Configuration of the 4 coupled JT gravities with holographic matters. 
{\bf Left: }Traversable wormholes in L-R direction. The L-R CFT is capped off whereas the 1-2 CFT makes a 3d traversable wormhole.
{\bf Middle:} Traversable wormholes in 1-2 direction.
In this case L-R CFT makes a 3d traversable wormhole.
{\bf Right:}A symmetric saddle of 4 coupled JT gravity + 2d CFT baths. 
Branes are disconnected but they are connected through 3d directions. 
} 
\label{fig:4coupledJTconfigHolographic1}
\end{figure}

Suppose we have the same fields in L-R and 1-2 direction i.e, the same matter contents and especially the same central charge $c_{LR} = c_{12}$ and the same distance $d_{LR} = d_{12}$.
Then the system is $\mathbb{Z}_2$ invariant, which is an analog of  the $\mathbb{Z}_2$ symmetry \eqref{eq:z2symmetrySYK} in 4 coupled SYK model,  because the asymptotic boundary condition in 2d/3d gravity prescription is symmetric.
The configurations we found are shown in figure \ref{fig:4coupledJTconfig1} in 2d gravity description and in figure \ref{fig:4coupledJTconfigHolographic1} when the matter fields are holographic.
We considered the wormhole solution which connect left and right systems in section \ref{sec:wormhole4coupled}.
We can exchange the role of L-R direction and 1-2 direction and obtain the 1-2 wormhole solution.
Because the energy in the wormhole solution is negative whereas the energy in symmetric solution is $0$, the wormhole solutions  are always dominant saddles.
Since each wormhole state is not invariant under $\mathbb{Z}_2$ symmetry that exchange L-R $ \leftrightarrow$ 1-2 direction, this symmetry is broken by each wormhole configuration.

We can use entanglement entropy as an order parameter.
The difference of entanglement entropy in L-R wormhole phase is 
\be
S_{LR} - S_{12} = 2 S_0 + 4 \f{\bar{\phi}_r}{\ell} + \f{2c}{3} \log  \f{\pi \ell + d}{\pi \ell}.
\ee
On the other hand, entanglement entropy in symmetric solution is 
\be
S_{LR} -S_{12} = 0.
\ee
Therefore, the difference of entanglement entropy quantitatively characterizes the pattern of symmetry breaking.
 
 \subsection{Partial coupling in 4 coupled JT gravities and phase transition \label{sec:PartialCoupling4}}
Here we study what happens when we decrease the L-R coupling starting from the L-R wormhole saddle.
To reduce the coupling between L and R, we introduce boundaries for L-R CFTs as we did in section \ref{sec:PartialCouplings}.
Here we separate the central charge  as $c_{LR} = c_{LR}^j + c_{LR}^s$ and introduce the boundary conditions for $c_{LR}^s$ CFTs whereas we continue to impose the transparent boundary conditions for $c_{LR}^j$ CFT in the no gravity region.
We imagine that the boundary conditions are described by BCFT and moreover the boundary conditions are the same (or more precisely the CPT conjugate) between 1 and 2 systems.

We already calculated these BCFT contribution in \eqref{eq:BCFTbulkEnergy}.
Using this, the energy which we should minimize in L-R wormhole phase is 
\be
E(\ell) = \f{\bar{\phi}_r}{4\pi G_N} \f{1}{\ell^2} + \f{c_{LR} + c_{12}}{12 \ell} - \f{c_{LR}^j\pi}{3(\pi\ell + d_{LR})} - \f{c_{LR}^s \pi}{12 (\pi \ell + d_{LR})}- \f{c_{12}}{12 (\pi \ell + d_{12})}. \label{PartialC4LRenergy}
\ee
Note that when $d_{12} = d_{LR}$, the  CFT with split boundary condition with central charge $c_{LR}^s$  gives the same contribution with  CFT with central charge $c_{12}$ that sits on two traversable wormholes as we observed in  \eqref{eq:CFTtwoWHenergy}.
In this case, introducing boundaries have the same effect with replacing $(c_{LR}, c_{12}) \to (c_{LR}^j, c_{12}+ c_{LR}^s  )$.

On the other hand, in 1-2 wormhole phase the energy which we should minimize is 
\ba
E(\ell) &= \f{\bar{\phi}_r}{4\pi G_N} \f{1}{\ell^2} + \f{c_{LR} + c_{12}}{12 \ell} - \f{c_{12}\pi}{3(\pi\ell + d_{12})} - \f{c_{LR}^j \pi}{12 (\pi \ell + d_{LR})}- \f{c_{LR}^s}{12 (\pi \ell + d_{LR})} \notag \\
&=  \f{\bar{\phi}_r}{4\pi G_N} \f{1}{\ell^2} + \f{c_{LR} + c_{12}}{12 \ell} - \f{c_{12}\pi}{3(\pi\ell + d_{12})} - \f{c_{LR} \pi}{12 (\pi \ell + d_{LR})}. \label{PartialC412energy}
\ea
Therefore, in this phase the energy function is not changed.

We can also have symmetric solutions. This is because BCFT on a half line with single boundary has vanishing stress tensor $\braket{T^{mat}_{\mu\nu}} = 0$.
This solution is still given by the dilaton profile \eqref{eq:PoincareDilaton}. 
The ADM energy for symmetric solutions remains $0$.

\begin{figure}[ht]
\begin{center}
\includegraphics[width=7.5cm]{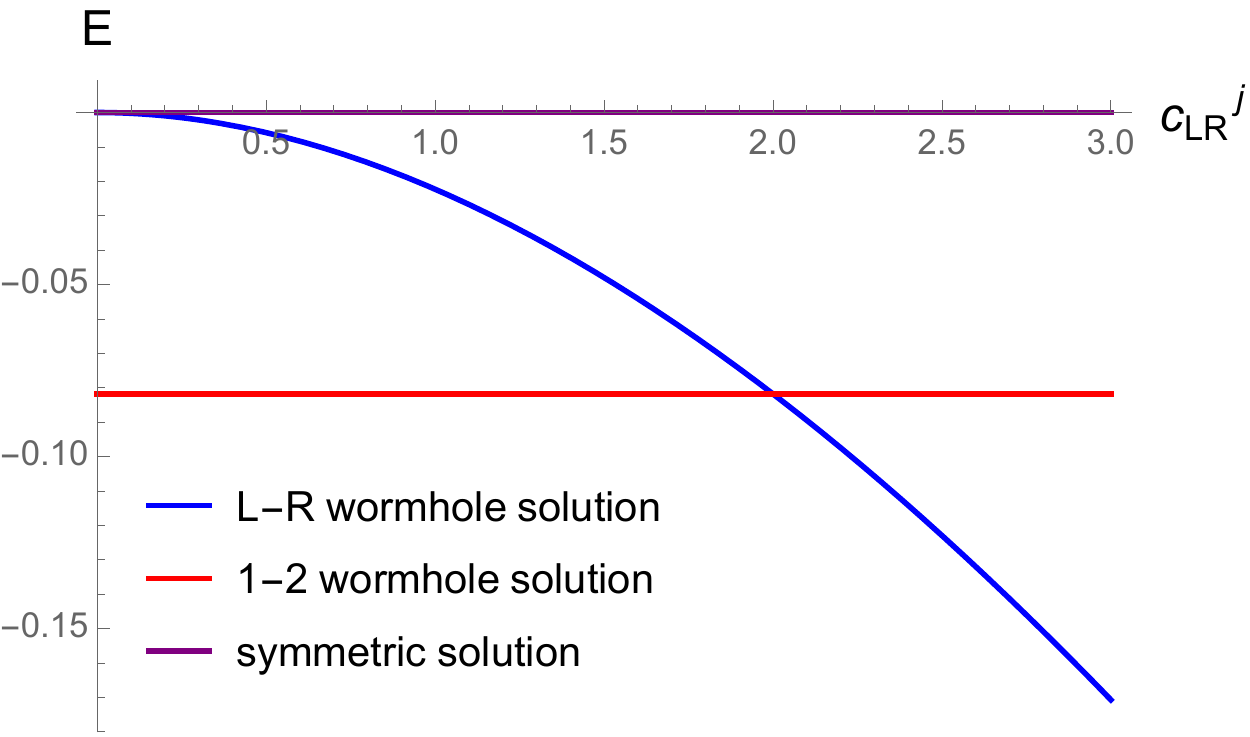}
\end{center}
\caption{The plot of the energy in L-R wormhole, in 1-2 wormhole and in the symmetric solution.
The energy in L-R wormhole phase is obtained by minimizing \eqref{PartialC4LRenergy} whereas The energy in L-R wormhole phase is obtained by minimizing \eqref{PartialC412energy}.
The parameters are taken to be $\f{\bar{\phi}_r}{8 \pi G_N} = 1$, $d_{LR} = 1$, $d_{12} = 1$, $c_{LR} = 3$ and $c_{12} =2$.   } 
\label{fig:SplittingEnergyPlot}
\end{figure}

For $d_{LR},d_{12}\ll \ell$ regime, the energy  as a function of $c_{LR}^j$  on each solutions is
\be
E = 
\begin{cases}
\displaystyle
 -\f{(c_{LR}^j)^2 \pi G_N}{ 16 \bar{\phi}_r} \qquad \text{for L-R wormhole solution} \\
\displaystyle -\f{(c_{12})^2 \pi G_N}{ 16 \bar{\phi}_r} \qquad \text{for 1-2 wormhole solution} \\
 0 \qquad  \text{for symmetric solution} 
\end{cases}
\ee
For generic parameters $d_{12}, d_{LR}$, we show a plot in figure \ref{fig:SplittingEnergyPlot}.
The phase diagram is very similar with that in the 4 coupled SYK in figure \ref{fig:Mu300B1000}, though the energy for 1-2 wormhole  and the symmetric solution do not meet.

\subsection{Changing central charge $c_{LR}$ and $c_{12}$ keeping $c_{LR} + c_{12}$ fixed \label{sec:Reconnecting4}}
Here we consider to change the central charge $c_{LR}$ and $c_{12}$ while we are keeping $c_{LR} + c_{12} \equiv c_{tot}$ to be fixed.
What we are imagining is to cut some part of $c_{12}$ CFT and reconnect in L-R direction.
We can also do the opposite i.e. cutting some part of $c_{LR}$ and reconnect in $12$ direction.
We can gradually change these central charges assuming that $c_{LR}$ consists from decoupled many CFTs like free fermions or several holographic matters introduced in \ref{sec:JTmanyHolographic} so that we can gradually change these central charges.

Let us consider the simple case of $d_{LR} = d_{12} \ll \ell$.
The energy in the L-R wormhole saddle is 
\ba
E= 2(M_L + M_R) = -2\f{\bar{\phi}_rt'^2}{16 \pi G_N} &= -4\f{\bar{\phi}_r}{16 \pi G_N} \Big ( \f{c_{LR}\pi G_N}{2\bar{\phi}_r}\Big)^2 \notag \\
&= -\f{c_{LR}^2 \pi G_N}{ 16 \bar{\phi}_r}.
\ea
The energy in the 1-2 wormhole saddle is  
\ba
E= 2(M_1 + M_2) = -2\f{\bar{\phi}_rt'^2}{16 \pi G_N} &= -\f{c_{12}^2 \pi G_N}{ 16 \bar{\phi}_r} = -\f{(c_{tot} - c_{LR})^2 \pi G_N}{ 16 \bar{\phi}_r}.
\ea
Therefore for $d_{LR},d_{12}\ll \ell$ regime, the energy as a function of $c_{LR}$ on each solutions is
\be
E = 
\begin{cases}
\displaystyle
 -\f{c_{LR}^2 \pi G_N}{ 16 \bar{\phi}_r}.
 \qquad \text{for L-R wormhole solution} \\
\displaystyle -\f{(c_{tot} - c_{LR})^2 \pi G_N}{ 16 \bar{\phi}_r} \qquad \text{for 1-2 wormhole solution} \\
 0 \qquad  \text{for symmetric solution} 
\end{cases}
\ee

When we increase the central charge $c_{12}$ from 0 to some value greater than $c_{LR}$, there is a phase transition.
The symmetric saddle have zero energy and the wormhole saddles always have smaller energy.

\begin{figure}[ht]
\begin{center}
\includegraphics[width=7.5cm]{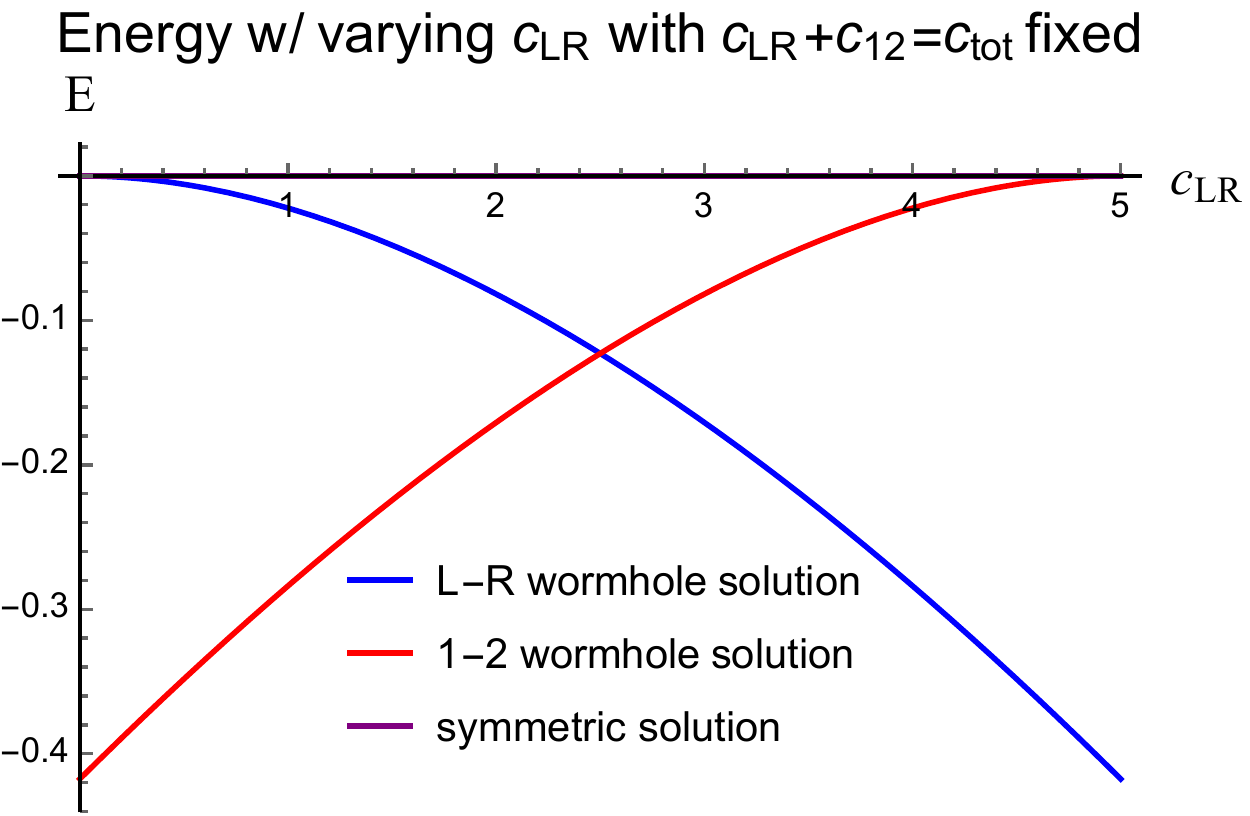}
\end{center}
\caption{The plot of the energy in L-R wormhole, in 1-2 wormhole and in the symmetric solution.
The parameters are taken to be $\f{\bar{\phi}_r}{8 \pi G_N} = 1$, $d_{LR} = 1$, $d_{12} = 1$, $c_{tot} = 5$.   } 
\label{fig:ReconnectingEnergyPlot}
\end{figure}
For generic parameters $d_{LR},d_{12}$, we plot a example in figure \ref{fig:ReconnectingEnergyPlot}.
The plot is very similar to that in the 4 coupled SYK model in figure \ref{fig:MuTot300B1000}.
The difference is that the symmetric saddle point

\subsection{JT gravity with both of holographic matters and Free CFTs \label{sec:partialDHolo}}
Since our construction of the solutions only depends on the central charges of matters, we can consider several pattern in our models if we do not care about the $\mathbb{Z}_2$ symmetry at the $c_{LR} = c_{12}$ point.
Especially we can consider 

1. LR CFT: Free CFT, 12 CFT: Free CFT

2. LR CFT: Holographic, 12 CFT: Holographic

3. LR CFT: Holographic, 12 CFT: Free CFT

\noindent
The case $1$ is basically the two dimensional picture of our setup whereas the case $2$ is the three dimensional picture in which both of L-R and 1-2 CFTs have holographic dual.
We can also consider where the L-R CFT has holographic dual whereas the 1-2 CFT is a bunch of free fields.
One can interpret that the free CFTs are  end-of-the world brane degrees of freedoms \cite{Penington:2019kki,Marolf:2020xie,Balasubramanian:2020hfs}.
From the perspective of CFTs on flat space, JT gravity + Holographic matter + free CFT theory gives an interface between holographic CFTs and free CFTs.
If these interfaces exist, they give a way to embed the holographic states into the free CFT Hilbert spaces.
Actually the four coupled model serves a way to such an embedding as we will see later. 

Note that the JT gravity coupled to order $1/{G_N}$ fermions can arise from the 4d magnetically charged near extremal black holes with a single 4d fermion \cite{Maldacena:2018gjk}.
Similarly, the JT gravity coupled to "order $1/{G_N}$" number of  holographic CFTs can arise from the 4d black holes  \cite{Maldacena:2020sxe} in the Randall-Sundrum  I\hspace{-.1em}I model \cite{Randall:1999vf}.
A model with a dark photon with couplings to both of 4d fermions and holographic CFT will give rise to a JT gravity both with 2d free fermions and 2d holographic CFTs.
In this case, originally we have 4d gravity + 4d fermions + 4d holographic CFT with UV cutoff, which is equivalent to the brane world model where the 4d fermions and 4d gravity live on the planck brane of 5d gravity and especially the 4d fermions are EOW brane degrees of freedom.
From these 4d theory we obtain 2d nearly AdS$_2$ gravity + 2d free fermions + 2d holographic CFTs near horizons in near extremal magnetically charged black holes.

\begin{figure}[ht]
\begin{center}
\includegraphics[width=10cm]{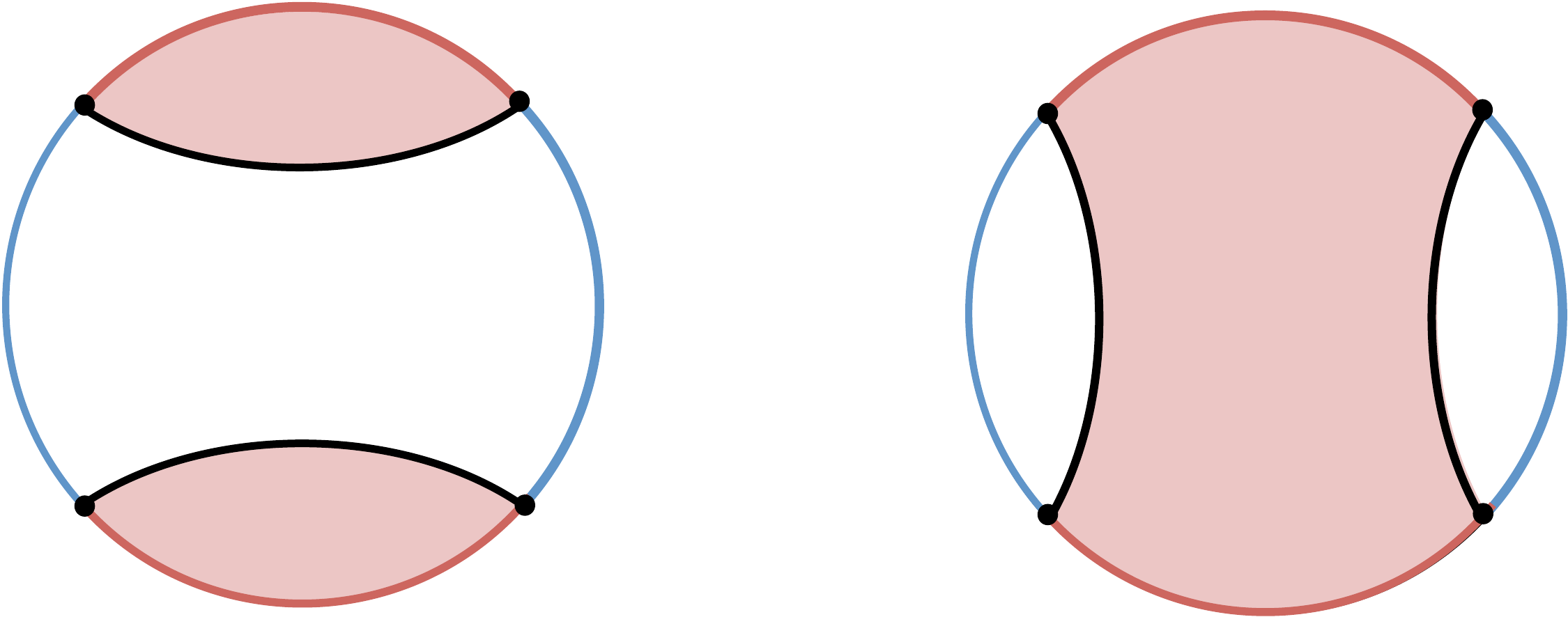}
\end{center}
\caption{ Configuration of the 4 coupled JT gravities with L-R holographic matters and 1-2 free CFTs. 
{\bf Left: }Traversable wormholes in L-R direction. The L-R CFT is capped off. 
In the middle there are "nothing" but the matter fields on the branes are entangled.
{\bf Right:} Traversable wormholes in 1-2 direction.
In this case L-R CFT makes a 3d traversable wormhole.
} 
\label{fig:4coupledJTconfigPartial1}
\end{figure}

\subsection{Wick rotation and Bra-ket wormholes }
\subsubsection{State preparation interpretation of the 4 coupled JT gravities}
It is interesting to consider the Wick rotation in the fully quantum mechanical description.
Here we first work in the Euclidean signature so the Wick rotation do nothing but we change the interpretations of the Euclidean geometry by exchanging the role of the space and the Euclidean time.

\begin{figure}[ht]
\begin{minipage}{0.49\hsize}
\begin{center}
\includegraphics[width=5.5cm]{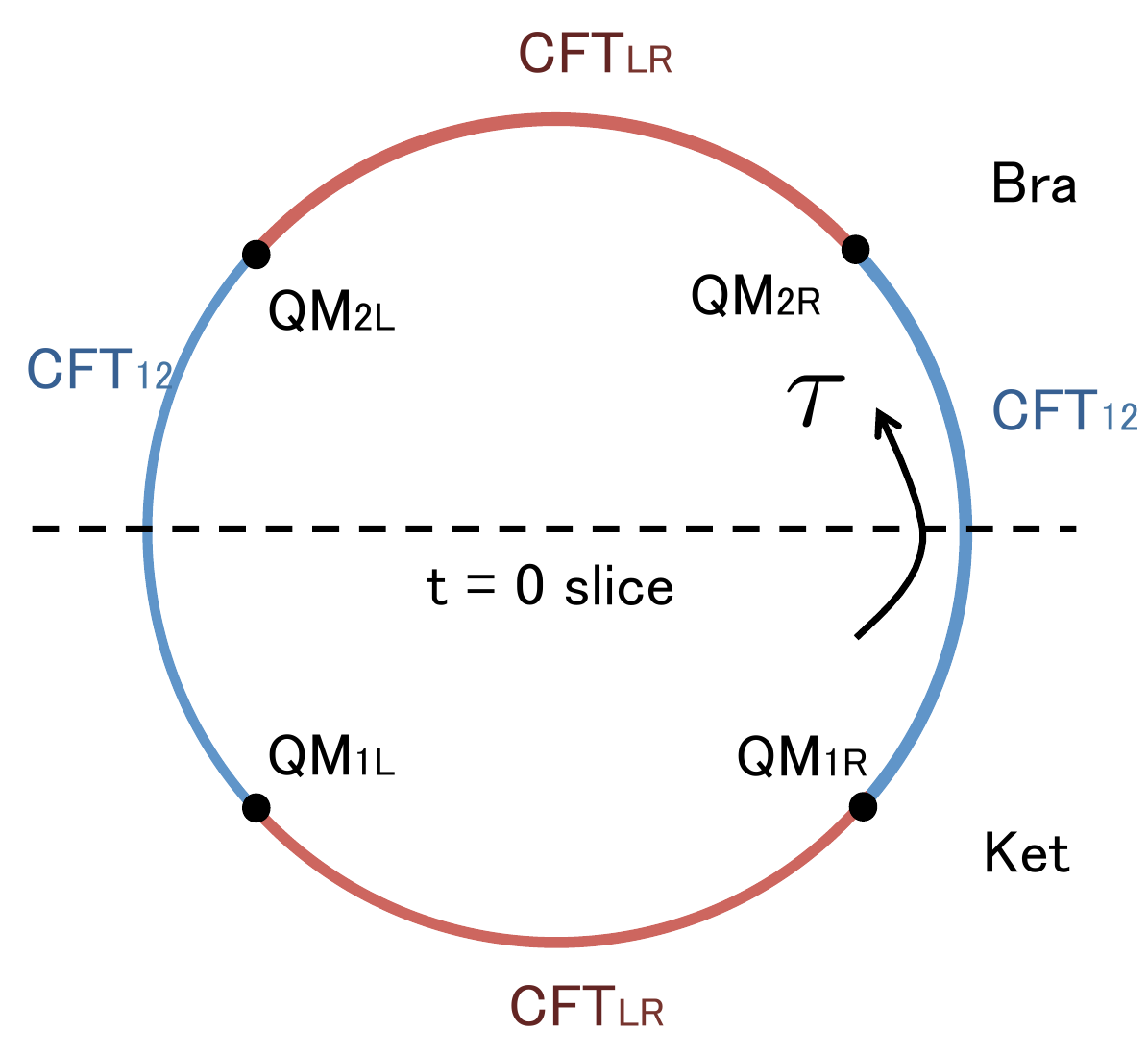}
\end{center}
\end{minipage}
\begin{minipage}{0.49\hsize}
\begin{center}
\includegraphics[width=6cm]{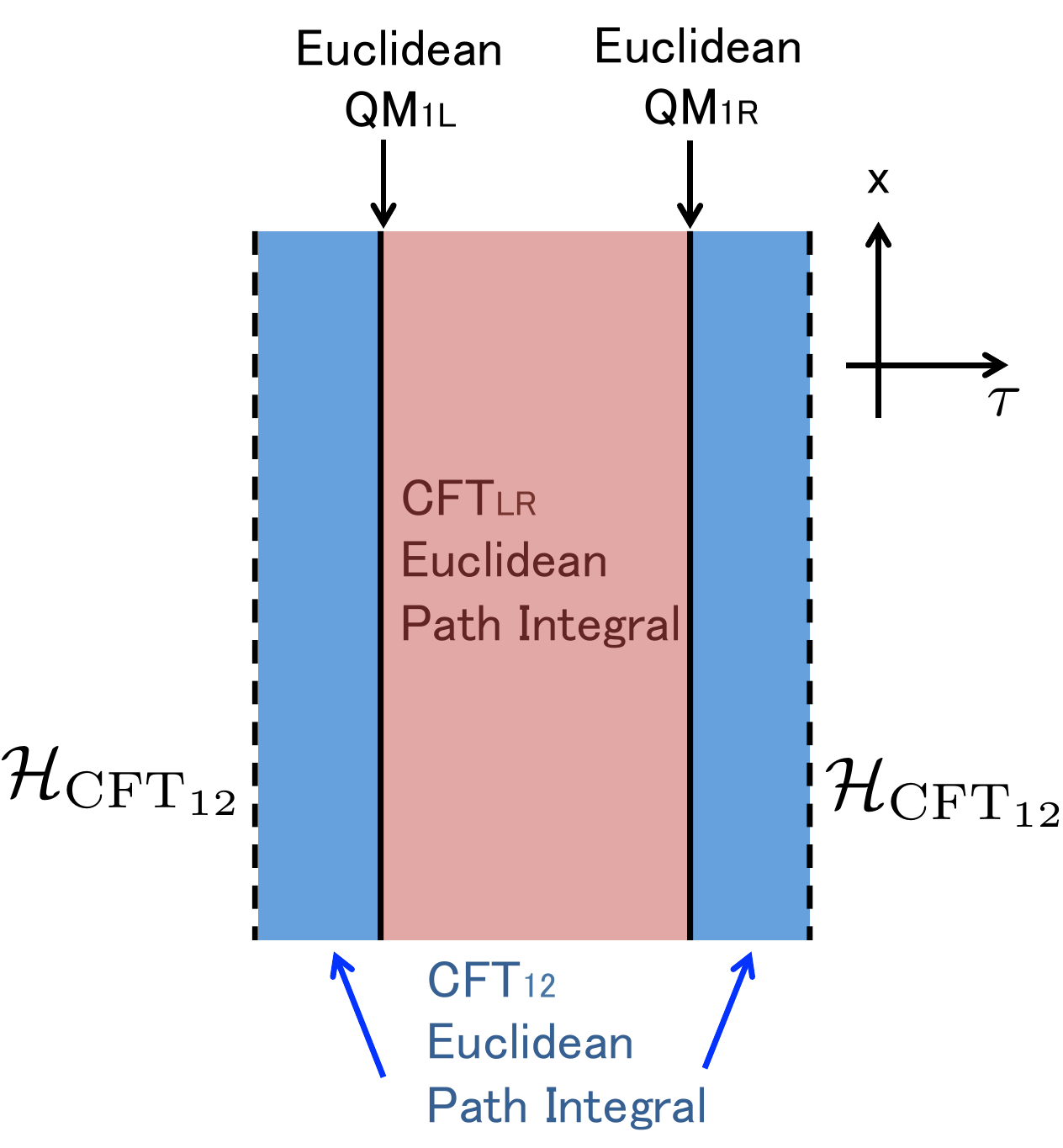}
\end{center}
\end{minipage}
\caption{The four coupled JT gravities in the state preparation interpretation in the full quantum mechanical description.
{\bf Left:} Here we suppress the spacial direction rather than the (Euclidean) time direction. The configurations are reflection symmetric and we take a Hilbert space at $t=0$ slice, which is the fixed slice under the reflection.
{\bf Right:}We include the spacial direction.
 The picture of the Euclidean path integral for the ket part.
} 
\label{fig:4coupledJTconfigPartial1}
\end{figure}

Since these configurations are (Euclidean) time reflection symmetric, these geometry can be understood as an Euclidean path integral for a state  preparation, see Figure \ref{fig:4coupledJTconfigPartial1}.
For example, we can take the time slice in the middle of the 1-2 CFT path integrals (the dotted line in figure \ref{fig:4coupledJTconfigPartial1}).
Then, these Euclidean path integral give states $\ket{\Psi}$ for the copies of Hilbert space of CFTs with central charge $c_{12}$: $\ket{\Psi_{12}} \in \mathcal{H}_{\text{CFT}_{12}} \otimes \mathcal{H}_{\text{CFT}_{12}}$.
The CFT$_{LR}$ and the domain walls play the role to characterize the entanglement structure between two copies, but their Hilbert spaces do not appear finally in the $t = 0$ time slice in this interpretation. 
In this interpretation, we can think that the label $1$ is for ket and the label $2$ for bra. 

Now, we consider the wormhole configurations in this state preparation interpretation.
The wormholes are now connecting the "bra" and "ket" path integral.
Therefore, the wormholes are interpreted as braket wormhole \cite{Chen:2020tes}, which is a kind of Euclidean wormhole.

As we observed, we have several saddles so we take the saddle point that minimizes the partition function.
The configurations are given by exchanging the role of Euclidean time and the spacial direction in wormhole/symmetric saddles: the 2d gravity prescriptions are given by figure.\ref{fig:4coupledJTconfig1} and the 3d gravity prescriptions are given by figure.\ref{fig:4coupledJTconfigHolographic1}.
The evaluation is completely the same of the partition function, which is given by $Z \approx e^{- L E_0 }$ where $E_0$ is the energy  evaluated in the wormhole solutions and $L$ is the length of the spacial direction\footnote{ This is the energy in the traversable wormhole picture.
In the traversable wormhole picture, $L$ plays the role of the inverse temperature $\beta$.}, which we take to be very large.
Then, the dominant saddle point is given by the wormhole configuration that has minimum energy.
The symmetric saddle can not be dominant since this have larger energy than wormhole saddles.
Therefore, for $d_{12},d_{LR} \ll \ell$ case,  we obtain 
\ba
&c_{12} > c_{LR}: \text{Bra-ket wormhole phase}, \\ \notag 
&c_{12} < c_{LR}: \text{No Bra-ket wormhole phase} .\notag
\ea
and the phase transition happens at $c_{12} = c_{LR}$.

\subsubsection{Projections on braket wormhole states}

It is interesting to consider the transition in section \ref{sec:Reconnecting4} and \ref{sec:PartialCoupling4} in the state preparation interpretation.
First let us consider the partial couplings in section \ref{sec:PartialCoupling4}.
In that context, we introduced  boundary conditions for $c_{12}^s$ fields.
After changing the role of time and space direction, these boundary conditions become spacelike.
These spacelike boundaries are called boundary states \cite{Cardy:1989ir,Onogi:1988qk,Ishibashi:1988kg}.
Since the boundary conditions we introduce are local, the boundary states $\ket{B}$ are schematically represented as 
\be
\ket{B} = \prod_{x} \ket{\psi_x}.  
\ee
In our context, we fix the future boundary condition for $c_{12}^s$ CFTs. 
These are interpreted as local projection of quantum states \cite{Numasawa:2016emc} onto boundary states. 
The projection operator $P = \ket{B}\bra{B}$ is then schematically written as 
\be
P = \prod_{x} \ket{\psi_x} \bra{\psi_x}.
\ee
The introduction of the spacelike boundaries for $c_{12}^s$ CFTs is interpreted as 
\be
\ket{\Psi_{12}} \rightarrow  (P_{c_{12}^s} \otimes \mathbb{I}_{c_{12}^j})^L\otimes (P_{c_{12}^s} \otimes \mathbb{I}_{c_{12}^j})^R \ket{\Psi_{12}}  = \ket{B_{c_{12}^s}^L,B_{c_{12}^s}^R}\braket{B_{c_{12}^s}^L,B_{c_{12}^s}^R|\Psi_{12}}.
\ee
Here $ P_{c_{12}^s} = \ket{B_{c_{12}^s}}\bra{B_{c_{12}^s}}$ is the projection operator onto the boundary state $\ket{{B_{c_{12}^s}}}$ and $\braket{B_{c_{12}^s}^L,B_{c_{12}^s}^R|\Psi_{12}} \in \mathcal{H}_{CFT_{{12}^j}}\otimes \mathcal{H}_{CFT_{{12}^j}}$ where $\mathcal{H}_{CFT_{{12}^j}}\otimes \mathcal{H}_{CFT_{{12}^j}}$ is  the Hilbert space for remaining CFT with the central charge $c_{12}^j$ after projections.
In this context, the partition function for traversable wormholes with partial couplings evaluates the (unnormalized) norm $|\braket{B_{c_{12}^s}^L,B_{c_{12}^s}^R|\Psi_{12}}|^2$.
The interpretation of the phase transition in \ref{sec:PartialCoupling4} becomes as follows.
Initially we do not specify the final state and we have braket wormholes. 
Then, we select the final state for $c_{12}^s$ CFT by introducing the boundaries.
Finally, at $c_{LR} = c_{12}^j$ there is a phase transition and beyond that point braket wormholes dissappear!  

Dividing by the normalization $\braket{\Psi_{12}|\Psi_{12}}$, we can compute the two point function of projection operators:
\be
\braket{P_{c_{12}^s}^L P_{c_{12}^s}^R} = \f{\braket{\Psi_{12}|P_{c_{12}^s}^L P_{c_{12}^s}^R|\Psi_{12}}}{\braket{\Psi_{12}|\Psi_{12}}}.
\ee  
This corresponds to the ratio of the partition functions: 
\be
\braket{P_{c_{12}^s}^L P_{c_{12}^s}^R} = e^{-L(E_0(c_{12}^j) - E_0(c_{12}))}. \label{eq:TwoPtProjection}
\ee
where $E_0(c_{12}^j)$  is the energy as a function of  $c_{12}^j$, an example of which is given in figure  \ref{fig:SplittingEnergyPlot}.
The behavior of the two point function \eqref{eq:TwoPtProjection} is shown the left panel of figure \ref{fig:ExpValue}.
 
\begin{figure}
\begin{center}
\includegraphics[width=12cm]{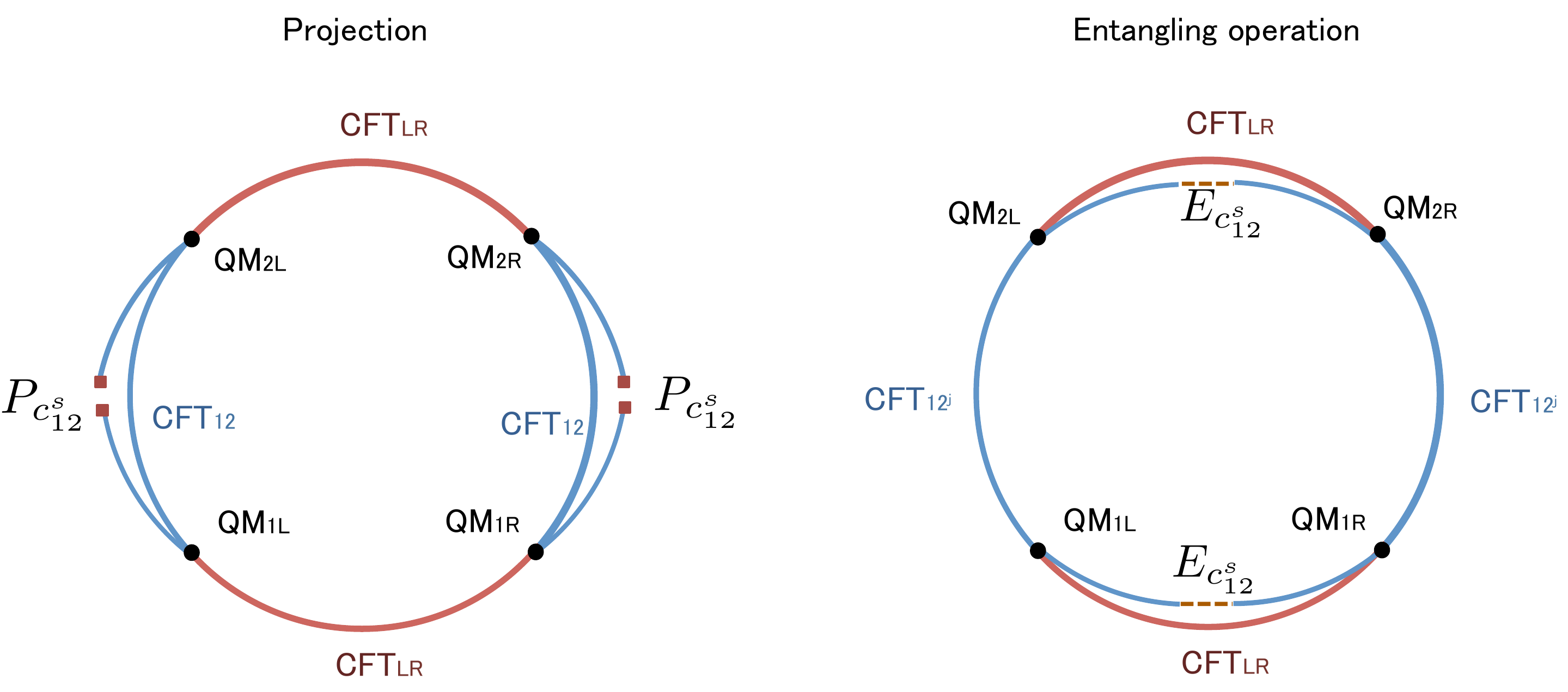}
\end{center}
\caption{The schematic picture for operations for the states $\ket{\Psi_{12}}$ which is prepared by the Euclidean path integral.
{\bf Left:} The projection of $\ket{\Psi_{12}}$. We introduce spacelike boundaries, which is denoted  by the solid squares with $P_{c_{12}^s}$,  for CFT$_{{12}^s}$ with the central $c_{12}^s$ and take the trace for the remaining CFT$_{{12}^j}$.
{\bf Right:} The partial entangling of $\ket{\Psi_{12}}$. We introduce spacelike transparent boundary conditions, which is denoted  by the dashed lines with $E_{c_{12}^s}$, for     CFT$_{{12}^s}$ with the central $c_{12}^s$ and take the trace for the remaining CFT$_{{12}^j}$.
} 
\label{fig:ProjectionEntangling}
\end{figure}

\subsubsection{Entangling operations on braket wormhole states}

\begin{figure}
\begin{center}
\includegraphics[width=7cm]{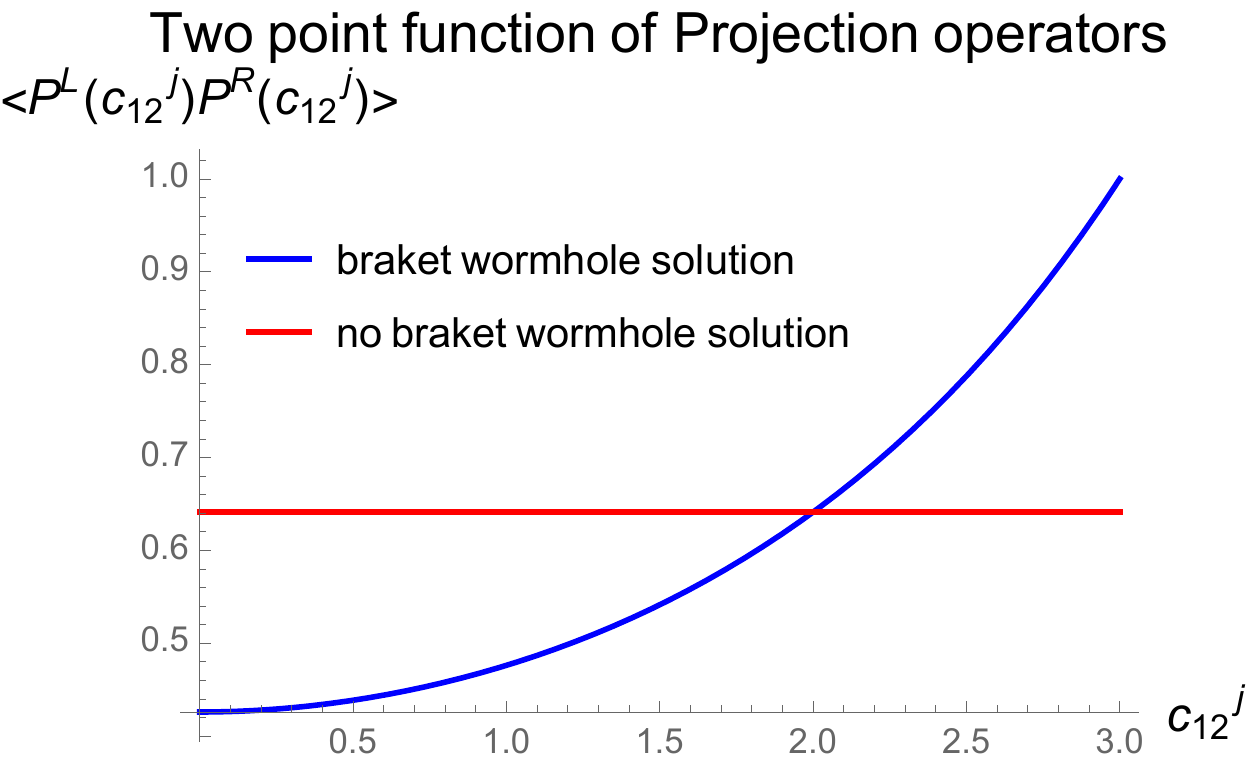}
\includegraphics[width=6.7cm]{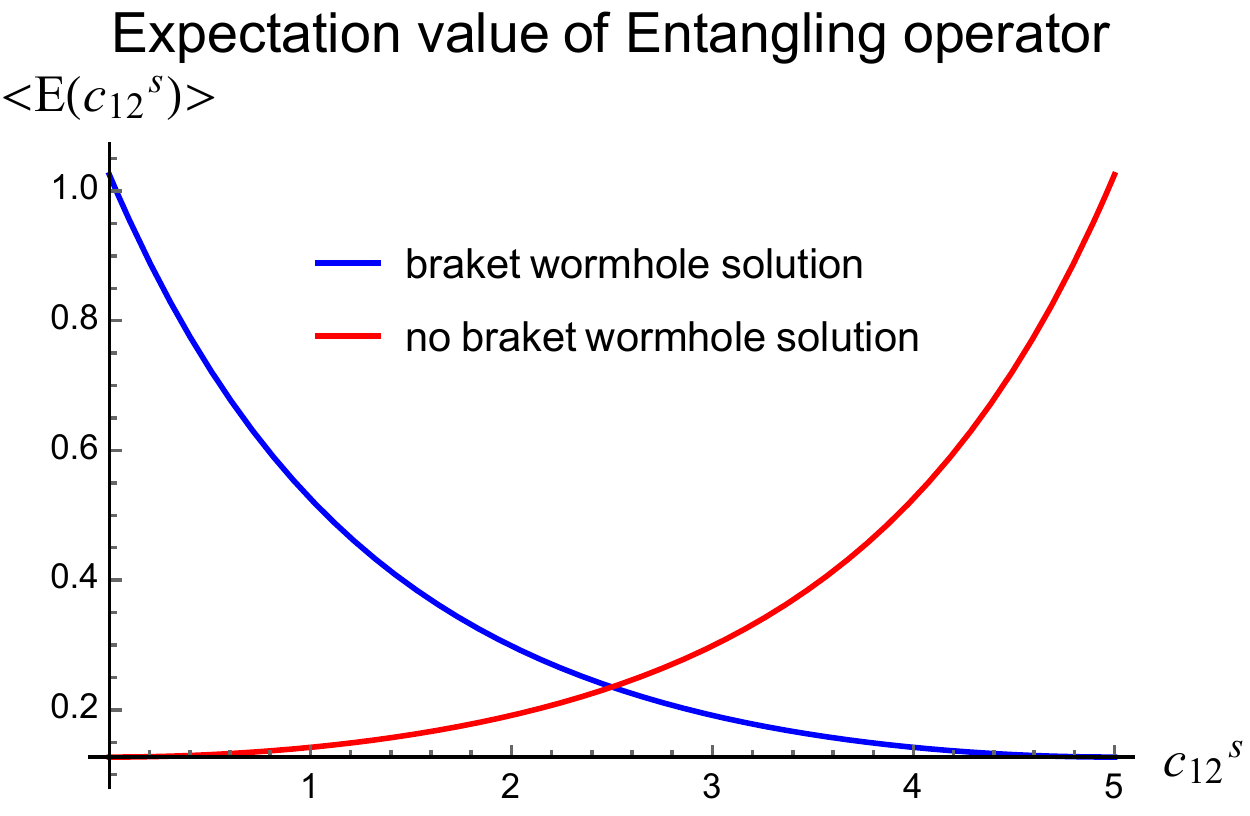}
\end{center}
\caption{The plot of projection operator two point functions and the expectation value of entangling operators.
The parameters are taken to be $\f{\bar{\phi}_r}{8 \pi G_N} = 1$, $d_{LR} = d_{12} = 1$ and $L = 5$.
We choose the maximal one in two saddles.
{\bf Left:} The projection operator two point functions $ \braket{P_{c_{12}^j}^L P_{c_{12}^j}^R}$ as a function of $c_{12}^j$. 
{\bf Right:} The expectation value of an entangling  operator $ \braket{E_{c_{12}^s}}$ as a function of $c_{12}^s$. 
Here we assume that initially $c_{LR} = 0$ and $c_{12} = c_{tot}$.
} 
\label{fig:ExpValue}
\end{figure}

Next, we consider the case in section \ref{sec:Reconnecting4} where we consider to first split some of CFT$_{12}$, say CFT$_{12}^s$ with  central charge $c_{12}^s$, and then reconnect them in L-R direction.
This amounts to changing $(c_{LR},c_{12}) \to (c_{LR} + c_{12}^s, c_{12}-c_{12}^s)$.
Here we attaching the spacelike perfectly transparent condition for $c_{12}^s$ CFTs.
These gluing conditions are interpreted as attaching local entanglement \cite{Numasawa:2016emc} between two sides. 
In other words, we partially project the state onto maximally entangled state, which is schematically written as 
\be
\ket{I}_{LR} = \prod_{x}\Big( \sum_{n_x}\ket{n_x}_L\ket{n_x}_R\Big).  
\ee
The projection operator onto maximally entangled state are then schematically written as 
\be
E =\ket{I}_{LR}\bra{I}_{LR} = \prod_{x}\Big( \sum_{n_x}\ket{n_x}_L\ket{n_x}_R\Big)\Big( \sum_{m_x}\bra{m_x}_L\bra{m_x}_R\Big) .
\ee
After this entangling procedure, the state becomes
\be
\ket{\Psi_{12}}  \rightarrow  E_{c_{12}^s} \otimes \mathbb{I}_{c_{12}^j}^L  \otimes \mathbb{I}_{c_{12}^j}^R \ket{\Psi_{12}} = \ket{I_{c_{12}^s}}\braket{I_{c_{12}^s}|\Psi_{12}}.
\ee
Here $ E_{c_{12}^s} = \ket{I_{c_{12}^s}}\bra{I_{c_{12}^s}}$ is the projection operator onto the maximally entangled state $\ket{{I_{c_{12}^s}}} \in \mathcal{H}_{CFT_{{12}^s}}\otimes \mathcal{H}_{CFT_{{12}^s}}$ and $\braket{I_{c_{12}^s}|\Psi_{12}} \in \mathcal{H}_{CFT_{{12}^j}}\otimes \mathcal{H}_{CFT_{{12}^j}}$ where $\mathcal{H}_{CFT_{{12}^j}}\otimes \mathcal{H}_{CFT_{{12}^j}}$ is  the Hilbert space for remaining CFT with the central charge $c_{12}^j = c_{12} - c_{12}^s$ after projections.
In this context, the partition function for traversable wormholes with partial couplings evaluates the (unnormalized) norm $|\braket{I_{c_{12}^s}|\Psi_{12}}|^2$.
The interpretation of the phase transition in \ref{sec:Reconnecting4} becomes as follows, which is similar to the boundary state cases.
Initially we do not specify the final state and we have braket wormholes. 
Then, we select the final state for $c_{12}^s$ CFT by introducing entanglement between two side.
Finally, at $c_{LR} = c_{12}^j$ there is a phase transition and beyond that point braket wormholes disappear\footnote{Using central charges as  variables is similar to the toy model of black hole evaporation in \cite{Akers:2019nfi}}.

Therefore, both of projections and entangling operation induce the disappearance of the bra-ket wormholes.
In other words, two pairs of cosmological spacetime annihilate in Euclidean regime.

Dividing by the normalization $\braket{\Psi_{12}|\Psi_{12}}$, we can compute the one point function of entangling operators:
\be
\braket{E_{c_{12}^s}} = \f{\braket{\Psi_{12}|E_{c_{12}^s}|\Psi_{12}}}{\braket{\Psi_{12}|\Psi_{12}}}.
\ee  
This again corresponds to the ratio of the partition functions: 
\be
\braket{E_{c_{12}^s}} = e^{-L(E_0(c_{12}^s) - E_0(0))}. \label{eq:OnePtEntangling}
\ee
where $E_0(c_{12}^s)$  is the energy as a function of  $c_{12}^s$, an example of which is given in figure  \ref{fig:ReconnectingEnergyPlot}.
Initially at $c_{12}^s = 0$, the entangling operator is the identity operator and the expectation value is $1$.
The behavior of the two point function \eqref{eq:OnePtEntangling} is shown the right panel of figure \ref{fig:ExpValue}.
Here we focus on the case where initially $c_{LR} = 0$.
In this case, the initial state are two decoupled braket wormholes, which is expected to be a product states $\ket{\Psi_{12}} = \ket{\mathcal{B}}_L\ket{\mathcal{B}}_R$.
Then at $c_{12}^s = c_{12}$  we do entangling operation for all the CFT$_{12}$.
The property of the maximally entangled state leads to $\bra{I}_{LR}(\ket{\mathcal{B}}_L\ket{\mathcal{B}}_R) = \braket{\mathcal{B}|\mathcal{B}}$.
Then, the expectation value of the entangling operator becomes 
\be
\braket{E_{c_{12}}} = \f{(\bra{\mathcal{B}}_L\bra{\mathcal{B}}_R)(\ket{I}_{LR} \bra{I}_{LR})(\ket{\mathcal{B}}_L\ket{\mathcal{B}}_R)}{(\bra{\mathcal{B}}_L\bra{\mathcal{B}}_R)(\ket{\mathcal{B}}_L\ket{\mathcal{B}}_R)} = 1.
\ee
Therefore, when we increase $c_{12}^s$ from $0$ to $c_{12}$, entangling operator expectation value decreases but finally should come back to $1$ at $c_{12}^s = c_{12}$.
The inclusion of the phase transition in  \eqref{fig:ExpValue} correctly reproduce this behavior.
On the other hand, if we always use  the braket wormhole solution, the entangling operator expectation becomes much smaller than we expect. 
We interpret this as a kind of information loss and the exchange of saddle is needed to reproduce the correct behavior for factorized state $\ket{\mathcal{B}}_L\ket{\mathcal{B}}_R$.

\subsubsection{Lorentzian continuation}
It is interesting to consider Lorentzian continuation of the state preparation interpretation.
The 2d metric and the dilaton profile is given by\footnote{We shift $\sigma \to \sigma + \f{\pi}{2}\ell$ so that the $t=0$ slice become $\sigma = 0$.}
\ba
ds^2 &= \f{-dt^2 + d\sigma^2}{\ell ^2 \cos^2 \f{\sigma}{\ell}}, \notag \\
\phi(\sigma) &=  \f{\bar{\phi}_r}{\pi \ell} \Big(1 + \f{\sigma}{\ell} \tan \f{\sigma}{\ell} \Big).
\ea
We analytically continue to $\sigma \to i t$, $t \to ix$. 
Then, the metric and the dilaton profile become
\ba
ds^2 &= \f{-dt^2 + dx^2}{\ell ^2 \cosh^2 \f{t}{\ell}}, \notag \\
\phi(t) &=  \f{\bar{\phi}_r}{\pi \ell} \Big(1 - \f{t}{\ell} \tanh \f{t}{\ell} \Big). \label{eq:FLRWuniv}
\ea
These are interpreted as a closed universe \cite{Chen:2020tes}.
The Lorentzian geometry \eqref{eq:FLRWuniv} corresponds to an Friedmann-Lemaitre-Robertson-Walker (FLRW) universe that emerges from a singularity at $\eta = - \infty$, reach the maximum size at $\eta = 0 $ and recollapses at $\eta = \infty$ \footnote{In three dimensions with negative cosmological constant, we can construct Euclidean wormhole solutions \cite{Maldacena:2004rf,VanRaamsdonk:2020tlr}, which becomes FLRW universe after analytic continuation to Lorentzian signature. 
The geometry here is the 2d analog of the relation between Euclidean wormholes and the closed universes.}. 
If the conformal matter is holographic, the original metric is given by that of the global coordinate:
\be
ds_{(3)}^2 = \l_{AdS_3}^2 (-\cosh ^2 \rho dt^2 + d\rho^2 + \sinh ^2 \rho  d\sigma^2).
\ee
After the analytic continuation $\sigma \to i t$ and $t \to ix$, the metric becomes 
\be
ds_{(3)}^2 = \l_{AdS_3}^2 (-\sinh ^2 \rho dt^2 + d\rho^2 + \cosh ^2 \rho d x^2).
\ee
This is the metric for BTZ black holes \cite{Hartman:2013qma}.
Let us consider the braket wormhole phase.
In  2d gravity description, there are two closed universes with CFT$_{LR}$ and CFT$_{12}$.
CFT$_{12}$ are entangled with the same CFT$_{12}$ on a Minkowski spacetime whereas CFT$_{LR}$ are entangled with the other closed universe.
In  3d gravity description when matter fields are holographic, matter quantum entanglement are geometrized in a way dipected in \ref{fig:LorentzianBraket}.
\begin{figure}[ht]
\begin{center}
\includegraphics[width=16cm]{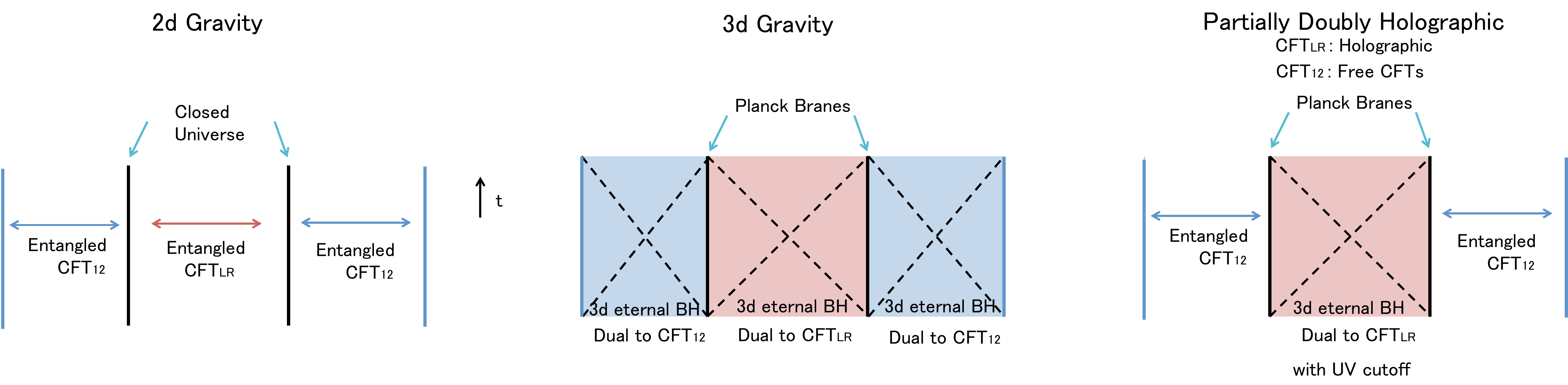}
\end{center}
\caption{{\bf Left :} The 2d gravity description of the state $\ket{\Psi_{12}}$ in 1-2 wormhole phase. {\bf Middle :} The 3d gravity description of the state $\ket{\Psi_{12}}$ in 1-2 wormhole phase. {\bf Right :} $\ket{\Psi_{12}}$ in 1-2 wormhole phase in partially doubly holographic setup.
CFT$_{LR}$ have holographic dual but CFT$_{12}$ is a collection of free fields, for example.
We can think of 3d eternal black holes with UV cutoff by planck branes, which are embedded in CFT$_{12}$ Hilbert spaces. } 
\label{fig:LorentzianBraket}
\end{figure}
The length of the circle which the CFT is living on gives the temperature of the CFT, or entanglement temperature of the thermo field double state in 2d gravity description.
Each temperature is given by
\be
\begin{cases}
\beta_{12} = \pi \ell + d_{12} \\
\beta _{LR} = 2(\pi \ell + d_{LR})
\end{cases}
\rightarrow
\begin{cases}
T_{12} = \f{1}{\pi \ell + d_{12}} \\
T_{LR} = \f{1}{2(\pi \ell + d_{LR})}
\end{cases}\label{eq:TempLR12}
\ee
where $\beta_{12} (\beta_{LR})$ is the inverse temperature for CFT$_{12}$ (CFT$_{LR}$) and $T_{12} (T_{LR})$ is the temperature.
In 3d gravity description, this is consistent that we obtain a black hole geometry. 

Now we consider entanglement entropy between two copies of CFT$_{12}$.
First it is easy to consider the 3d gravity description.
In this case, we can use the holographic entanglement entropy formula \cite{Ryu:2006bv,Ryu:2006ef,Hubeny:2007xt} for bulk matter CFTs.
We have 3 horizons in figure \ref{fig:LorentzianBraket}, which areas give candidate saddles.
The area of the horizon, or equivalently the thermal entropy for CFT is given by $S = \f{\pi c}{3 \beta }L$ \cite{Bloete:1986qm} where $c$ is the central charge, $L$ is the length of the spacial circle and $\beta$ is the inverse temperature. 
Therefore, holographic entanglement entropy between two sides are given by
\be
S_{ent} = \text{min} \Big\{\f{\pi c_{LR}}{3 \beta_{LR}}L ,\f{\pi c_{12}}{3 \beta_{12}}L \Big\}. \label{eq:EEforCFT12}
\ee
where the first one is the horizon area of L-R black hole and the second one is that of 1-2 black holes.
Because the bra-ket wormhole is dominated when $c_{LR} < c_{12}$.
Moreover, we found that $T_{LR} < T_{12}$ (here we focus on the case $d_{12} = d_{LR}$) in \eqref{eq:TempLR12}.
Therefore entanglement entropy between two side is 
\be
S_{ent} = \f{\pi c_{LR}}{3 \beta_{LR}}L. \label{eq:LRentanglementH}
\ee
In particular, when $c_{LR} = 0$ entanglement entropy becomes $0$.
This is consistent with the observation in \cite{Chen:2020tes}.
In this case, we have the product state $\ket{\Psi_{12}} = \ket{\mathcal{B}}\ket{\mathcal{B}}$ and each of closed universe is contained in the entanglement wedge of CFT on a flat space.
Though we only worked for holographic bulk matters cases, we expect this will be justified by replica wormhole argument \cite{Penington:2019kki,Almheiri:2019qdq} even for non holographic bulk matters.
We will not fully analyze this problem but discuss later in the discussion section.

It is interesting when CFT$_{12}$ is a collection of free CFTs and CFT$_{LR}$ is holographic.
Even in this case \eqref{eq:LRentanglementH} will be correct.
Then, we use holographic entanglement entropy  formula for states of free CFTs!
This is natural from the perspective of quantum error correction \cite{Harlow:2016vwg}.
Entanglement is a nature of state, not of the theory, and for the holographic code in free CFT we expect that we can use the holographic entanglement entropy formula.

\begin{figure}[ht]
\begin{center}
\includegraphics[width=8cm]{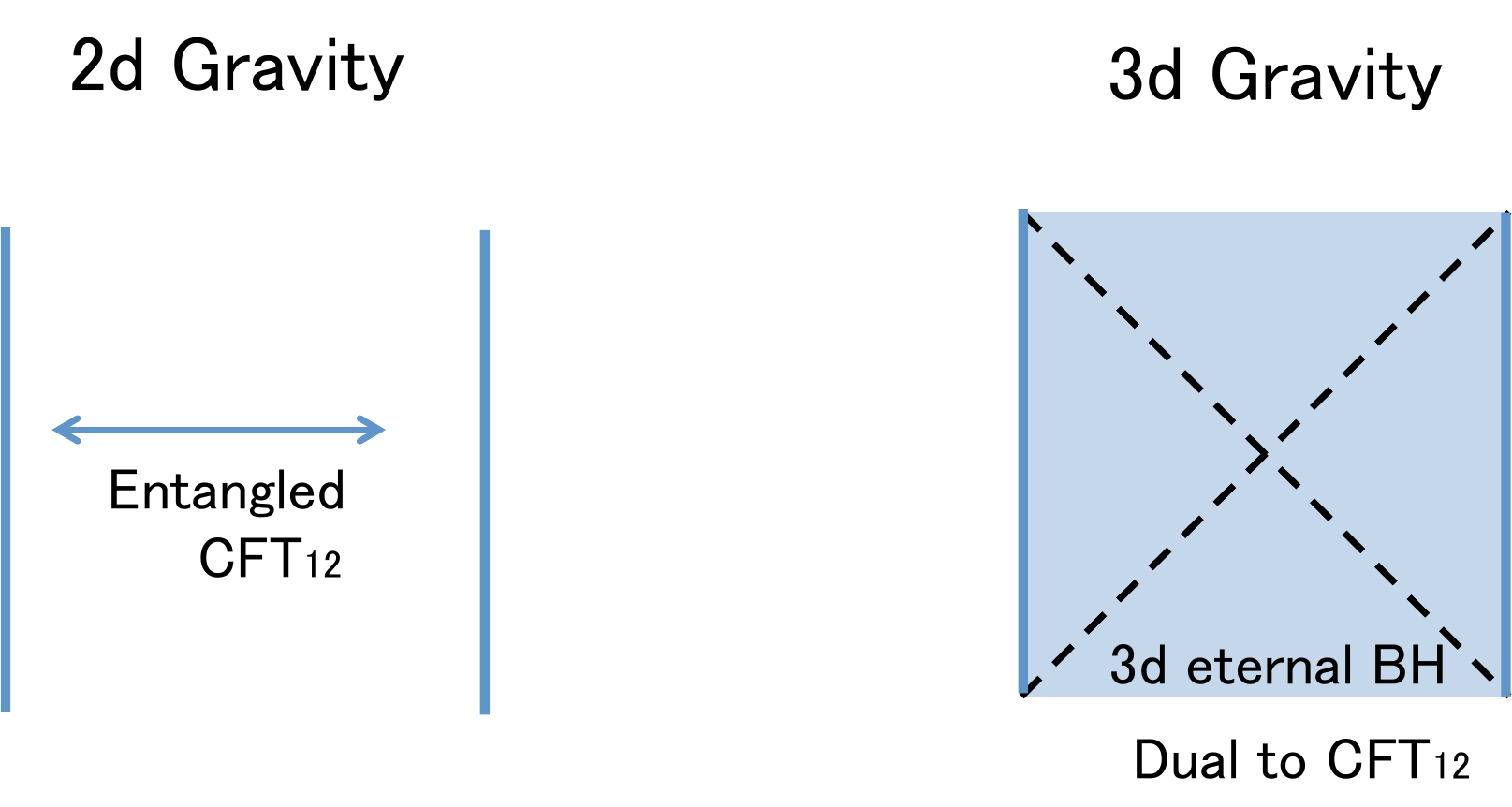}
\end{center}
\caption{The Lorentzian "geometry" in no bra-ket wormhole phase. In both gravity descriptions, JT gravities with conformal matters do not appears in Lorentzian signature but they only determine entanglement temperature.  {\bf Left :} The "2d gravity" description of the state $\ket{\Psi_{12}}$ in no bra-ket wormhole  (L-R wormhole) phase. The semiclassical state is an ordinary thermofield double state.
{\bf Right :} The 3d gravity description of the state $\ket{\Psi_{12}}$ in  no bra-ket wormhole (L-R wormhole) phase.
The semiclassical geometry is just an ordinary eternal black hole. } 
\label{fig:LorentzianBraket}
\end{figure}

In the no bra-ket wormhole phase, closed universes and CFT$_{LR}$ do not appear anymore in Lorentzian signature.
The only role of the JT gravity is to determine the entanglement temperature between two copies of CFT$_{12}$:
\be
\beta_{12} = 2(\pi \ell + d_{12})  \qquad 
\rightarrow  \qquad 
T_{12} = \f{1}{2(\pi \ell + d_{12})} .
\label{eq:TempNoBraket}
\ee
Here $\ell$ is the wormhole length in the Euclidean signature and it depends on $\bar{\phi}_r$.
Thought we have two JT gravity in Euclidean signature, they annihilate in the Euclidean regime and disappears in the Lorentzian signature.

\section{Discussion \label{sec:DiscussionSummary}}
\subsection{Summary}
Here we describe the summary of this paper.
In this paper, we construct models of four coupled SYK models and nearly AdS$_2$ gravities and study them.
In both of SYK models and JT gravities, these coupling can be thought of as coupling of two traversable wormholes.
This coupling introduce entanglement between two traversable wormholes.
These theories shows the first order phase transitions at the symmetric point both of them exhibit the $\mathbb{Z}_2$ symmetry breaking that exchanges the wormhole configurations.
These models also have symmetric solutions under $\mathbb{Z}_2$ symmetry, which have larger energy than wormhole configurations.
The direct interaction at the boundaries acts in a state dependent way.
In one wormhole phase they shorten the wormhole length when we introduce the interactions.
On the other hand, in the other wormhole phase the direct boundary interaction lengthen the wormhole length.

In the four coupled SYK models, we study the solutions and their properties numerically at large $N$ limit beyond the low energy description by the Schwartzian action.
Our tool is the Schwinger-Dyson equation.
Each of wormhole configurations disappears in some parameter regimes.
The order parameters  are given the (difference of) spin operators that consist from the fundamental fermions.
The first order phase transition only exists when the coupling of the four coupled interaction (mass terms) is small.
For larger couplings (mass terms), the phase transition disappears and the theory exhibits the crossover.
The decreasing energy gap, or the lengthening the wormhole length still exists even in the cross over regime.
The theory has a duality, which becomes symmetry at a special point.
Based on the action for  collective fields $G$ and $\Sigma$, we can evaluate the effective potential for the four coupled theory.
By explicitly studying this potential, the symmetric solution contains unstable direction.

We also studied the four coupled nearly AdS$_2$ gravities, or four coupled JT gravities.
Here we meant that the theory have four boundaries.
We assume that the JT gravities are coupled to at least two different CFTs.
Then, we couple these JT + CFTs through CFTs on flat space without dynamical gravity.
The quantum mechanical dual description looks like four 2d CFTs that are connected through domain walls where each domain wall supports quantum mechanics that is dual to JT gravity + CFTs.
We construct two wormhole solutions and one symmetric solution in this setup.
When we coupled two traversable wormholes, the matter fields on different wormholes are entangled.
If these matter fields are holographic, the dual 3d geometry directly connect two wormholes, which is the realization of the ER=EPR.
Furthermore,  this 3d geometry becomes a traversable wormhole from the perspective of CFTs on flat space without dynamical gravity.
When we increase entanglement between wormholes there is a phase transition where the wormhole configuration changes.
One of the interesting feature of this gravity model is that we can compute entanglement entropy. 
We study entanglement entropy which plays a role of order parameter.
In some configuration, we need to take into account island contribution, which leads to the wormhole entropy.

We also consider the wick rotation of four coupled JT gravities.
In this case the wormholes are interpreted as braket wormholes.
By considering the JT gravity coupled to both of free CFTs and holographic CFTs, these braket wormhole configuration serves a way to embed holographic states into free CFTs' Hilbert spaces.
The change of the coupling is interpreted as either of partial projection onto local product states or partial entangling operation.
The first order phase transition happens when we increase the number of projection or entangling operation.
The first order phase transition is now interpreted as a braket wormhole transition, which means a transition of entanglement structure in family of states prepared by Euclidean path integrals.


\subsection{Similarity with symmetry breaking in holographic QCD \label{eq:similarityHQCD}} 
In the four coupled JT gravities with holographic matter fields, the symmetry breaking is realized as an brane connection in 3d space time.
Actually, there is already a well known example of symmetry breaking by brane connection.
That is the holographic QCD model known as Sakai-Sugimoto-Witten model \cite{Sakai:2004cn,Witten:1998zw}.
In that model, the connection of  D8 branes and anti D8 branes (denoted by  $\overline{\text{D8}}$ branes), represent the breaking or restoration of the chiral symmetry.
At zero temperature, the D8 branes and $\overline{\text{D8}}$ branes are connected in the bulk, which means the chiral symmetry is broken.
On the other hand, at finite temperature and when the gluon sector is deconfined,  there are two configurations \cite{Aharony:2006da}.
One configuration is the same with that in the zero temperature, which means the chiral symmetry breaking.
The second configuration is the branes ending on the horizon in the black hole geometry.
In Euclidean signature, these branes are actually connected in the shrinking thermal circle in the bulk.
When we go to the Lorentzian signature, the branes of two sides in the thermo field double state is actually connected at the Einstein-Rosen (ER) bridge.

A difference is that in the holographic QCD case there are chiral symmetries that act on each defect in the dual picture and single brane connection breaks these symmetries whereas our four coupled model do not have such an symmetry acting on each defects.
This is more similar to the complex SYK symmetry breaking \cite{Kim:2019upg,Klebanov:2020kck} where each complex SYK cluster has a $U(1)$ symmetry \cite{Sachdev:2015efa} and the symmetry is spontaneously broken in coupled models \cite{Sahoo:2020unu}.
Another comment is that the symmetry breaking also plays an important role in quantum chaos in the SYK or JT gravity from semiclassical analysis \cite{Saad:2018bqo}.

\subsection{The connection to the bubble of wormholes}
The story of Sakai-Sugimoto-Witten model is related to the bubble of wormholes \cite{Fabinger:2000jd,Horava:2007hg}.
In type IIA string theory, when we put the theory on a $\mathbb{R}^{1,8} \times I$ and introduce the 8 D8-branes with an O8-plane on each boundary of an interval $I$, we obtain the type I' theory \cite{Horava:1995qa}.
In the strong coupling limit we obtain M-theory ending on the end of the world M9 brane that carries the $10d$ $E_8$ super Yang-Mills (SYM) \cite{Horava:1995qa,Horava:1996ma}, which is an "end of the world brane degrees of freedom".

The configurations above are supersymmetric and stable. However, when we flip a chirality of one of the boundaries, the supersymmetry is completely broken \cite{Fabinger:2000jd}.
An possible scenario when two branes are separated is that the spacetime hole is nucleated and spacetime completely disappears.
These are described by the $\mathbb{Z}_2$ orbifold of the bubble of nothing bounce solution \cite{Kim:2019upg, Witten:1981gj}.
The interesting thing is that these bubbles connect two boundaries and plays a role of wormhole from the end of the world M9 brane perspective.
After this bubble is nucleated, the $E_8$ SYM degrees of freedom is connected and we obtain a single $E_8$ degrees of freedom, which is similar to the D8 brane connections and the Higgs mechanism in the holographic QCD.
This is more similar with the situation in section \ref{sec:partialDHolo} where L-R CFT is holographic and 1-2 CFT is free CFTs that are interpreted as the EOWs brane degrees of freedom.
The first similarities are the point that the L-R CFT is separated to to pieces under the wormhole transition by nucleating the "bubble of nothing"  in the middle.
Second, we obtain a single free fields that are sit on two traversable wormholes, which is similar to obtaining a single $E_8$ SYM.
When the separation of the two boundaries are smaller than the string scale, we expect that the theory is tachyonic and perturbatively unstable. 
Actually, there is  a tachyonic string theory with single $E_8$ gauge degrees of freedom \cite{Horava:2007hg}, which is the counterpart of the chiral symmetry breaking in the holographic QCD.
This is analogous to the SYK symmetric unstable solution in this paper and it is interesting to study more how they are parallel.

\subsection{Interpretations of results in 4d traversable wormhole setup}
In section  \ref{sec:PartialCoupling4} and  \ref{sec:Reconnecting4}, we study the change of boundary conditions in no gravity region and study their effects.
It is interesting to consider the interpretation of these changes in 4d traversable wormhole setup in \cite{Maldacena:2018gjk, Maldacena:2020sxe}.

\begin{figure}[ht]
\begin{center}
\includegraphics[width=15cm]{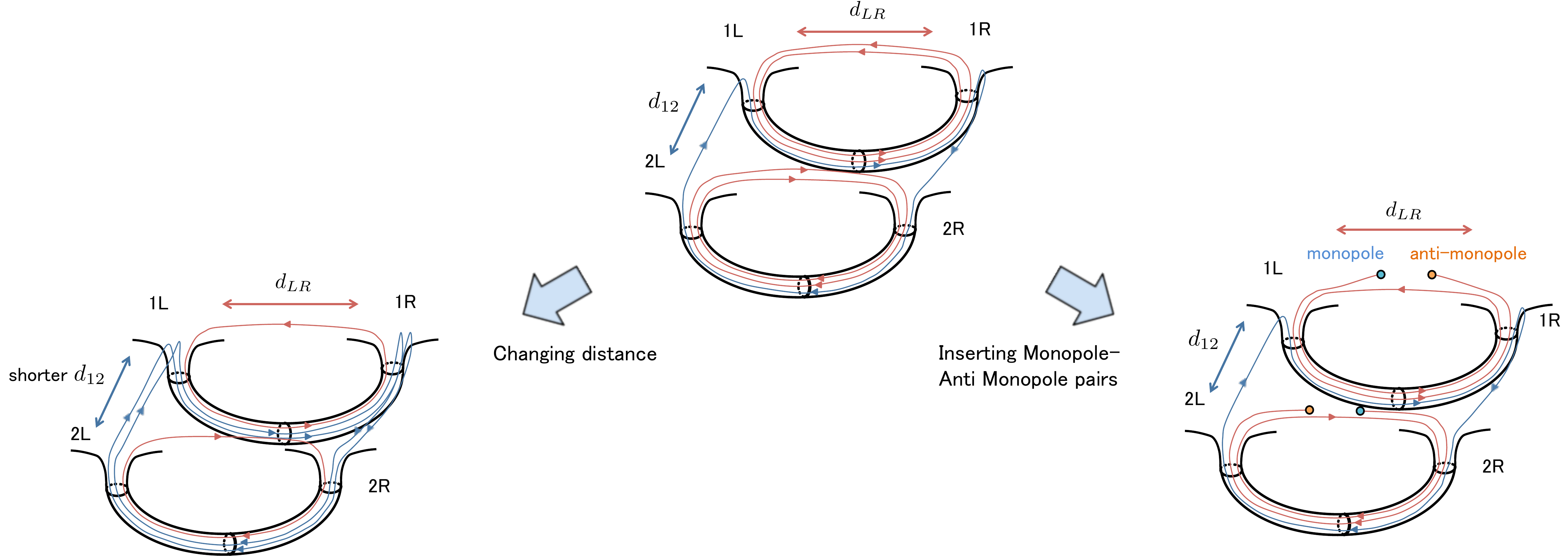}
\end{center}
\caption{The interpretation  of the setup \ref{sec:PartialCoupling4} and  \ref{sec:Reconnecting4} in 4d traversable wormholes.
Changing distance causes the change of $c_{12}$ with $c_{tot} =c_{12} + c_{LR}$, which is discussed in section  \ref{sec:Reconnecting4}.
Introducing monopole anti-monopole pairs outside the wormholes will introduce the boundaries for 2d fermions, which is discussed in section \ref{sec:PartialCoupling4}.  }
\label{fig:4dWormholeBdyChange}
\end{figure}

First, we consider the change of the distance between two traversable wormholes.
Then, this causes the change of magnetic field configurations.
Since each 2d free fermion propagates along the magnetic lines, this will cause the change of number of 2d fermions that are propagating along from L to R, or 1 to 2.
This changes the central charge $c_{LR}$ and $c_{12}$.
On the other hand, the total central charge $c_{tot} = c_{LR} + c_{12}$ will not be changed because these are total number of 2d free fermions propagating in the wormhole, which is determined by the magnetic flux that penetrates the sphere $S^2$.
Therefore, we can interpret that the change of boundary conditions in  \ref{sec:Reconnecting4} is modeling this change of magnetic field configurations and its effects.

Next, we consider to insert monopole anti-monopole pairs to cut the magnetic line outside the wormholes.
Then some of magnetic lines terminate on (anti-) monopoles and number of magnetic lines that are cut depends on the total monopole charges.
Therefore, fermions on such magnetic lines essentially lives on intervals with boundaries on the location of (anti-) monopoles.
4d fermions around monopoles or dyons and their connection to boundary CFT are discussed in \cite{Callan:1982au,Polchinski:1984uw,Maldacena:1995pq}.
We can interpret that the partial couplings by the introduction of boundaries outside the wormholes in section \ref{sec:PartialCoupling4} are modeling this insertion of monopole anti-monopole pairs.

In more realistic 4d setup, the things outside the wormholes are also described by the 4d physics whereas we consider the situation where we can do what we want to do outside the wormholes.
For example, the mouses of wormholes can attract with each other because they have opposite charges.
To avoid this, we can introduce a rotation to avoid attraction but it leads to a radiation and wormholes still have finite life times \cite{Maldacena:2018gjk}.
Because the transitions between wormholes involve topology changes, which are non perturbative effects and it may takes longer time than a life time.

\subsection{Generalization to many coupled SYK/JT gravities \label{sec:manycoupled}}

\begin{figure}[ht]
\begin{center}
\includegraphics[width=12cm]{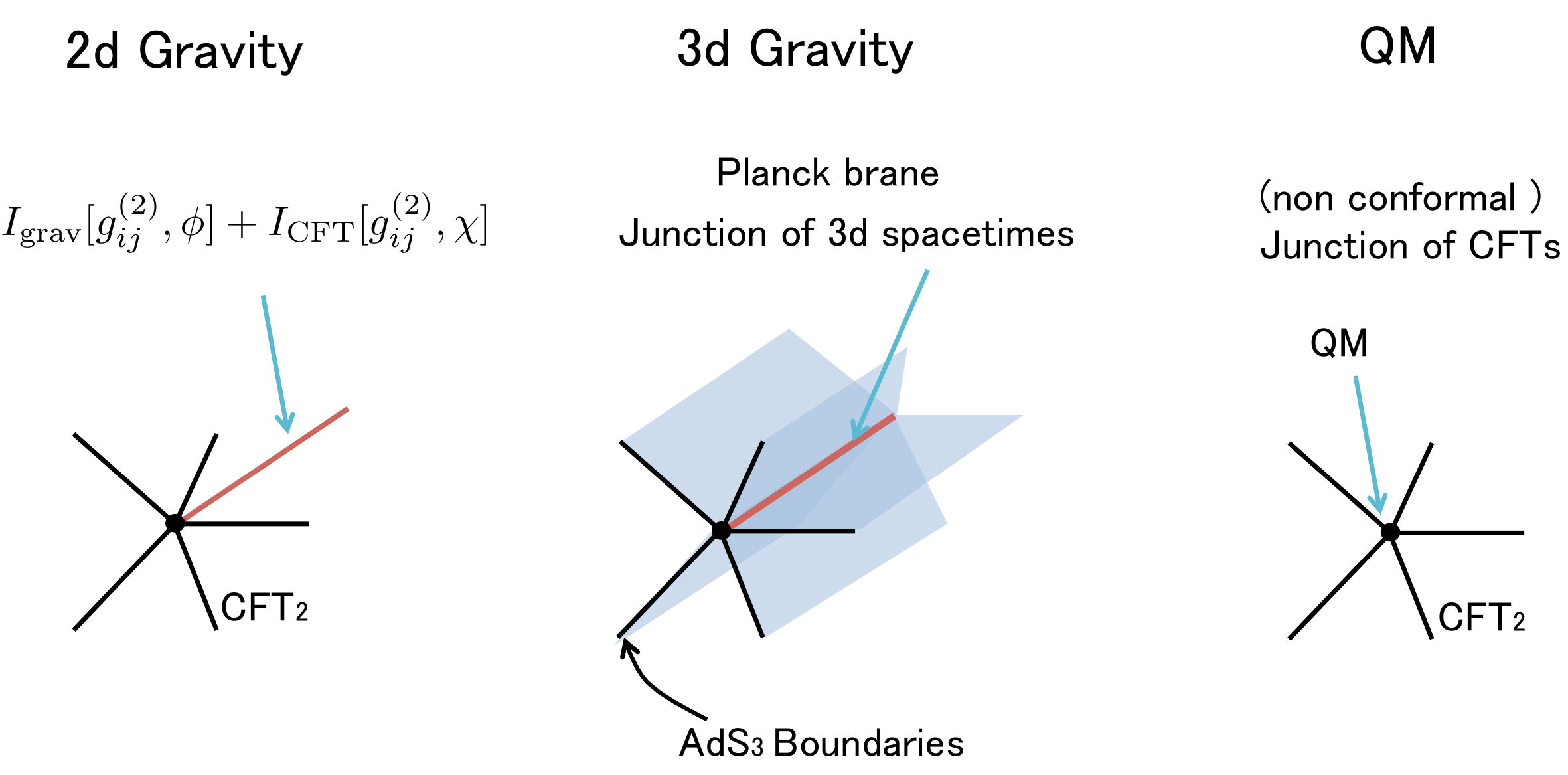}
\end{center}
\caption{ Three description of a junction of five CFTs. 
{\bf Left: }2d gravity description. The red line denotes the JT gravity + five CFTs.
There are also five CFTs on regions without dynamical gravity.
They are connected by the transparent boundary conditions at junction point.
{\bf Middle: }3d gravity description when the matter CFTs have holographic duals.
Five AdS$_3$ geometry are joined at an end of the world brane.
{\bf Right:} Full quantum mechanical description.
Five CFTs on half lines are joined at center on which a quantum mechanical degrees of freedom with gravity dual.
This quantum mechanical degrees of freedom is not conformal and we can think of the  theory as  non conformal junction CFTs.
} 
\label{fig:JunctionCFT}
\end{figure}

It is interesting to consider to couple more SYK/JT gravities.
In the SYK side, there are several works on the SYK chains \cite{Gu:2016oyy,Gu:2017njx,PhysRevLett.119.216601,Altland:2019lne,Jian:2017tzg}.
In a similar manner, we can consider $n$-coupled SYK models.
It is interesting to study this generalization in detail.

We can also consider more general couplings of JT gravities based on our construction of the four coupled JT gravities.
One generalization is to introduce many domain walls in the full quantum mechanical description.
Note that many D5 branes on black brane backgrounds and their connection to dimer states are discussed in \cite{Kachru:2009xf, Kachru:2010dk}.
Based on the relation between brane connections and wormhole configuration in section \ref{eq:similarityHQCD}, it is natural to expect that the states with many domain walls will have similar structure.


Another generalization is to consider the quantum mechanical domain walls with JT gravity dual to junctions of CFTs, and then couple different junctions.
For example, we can consider the junction of five CFTs as depicted in figure \ref{fig:JunctionCFT}.
Actually, JT gravity + $N$ free CFTs can be thought of as a junction of $N$ free CFTs.
In 4d traversable wormholes,  free fermions on flat space region without dynamical gravity effectively live on different lengths of intervals \cite{Maldacena:2018gjk}.
Therefore, it is more natural to think of them as a junction CFT.
In 3d gravity description, we also already encounters this type in section \ref{sec:JTmanyHolographic}.
The $N_H = 5$ setup there can be thought of as the junction of five holographic CFTs in figure \ref{fig:JunctionCFT}.
Based on these junctions, we can couple them. 
It is interesting study this type of couplings and their entanglement structures.

\subsection{Relation to Replica wormholes}
It is interesting to consider a relation to replica wormholes \cite{Penington:2019kki,Almheiri:2019qdq}.
Let us consider the replica method justification of the calculation of entanglement entropy in \eqref{eq:EEforCFT12} in a partially doubly holographic model.
In other words, CFT$_{LR}$ is holographic and CFT$_{12}$ is a collection of free fields.
We mainly consider the braket wormhole phase.

\begin{figure}[ht]
\begin{center}
\includegraphics[width=14cm]{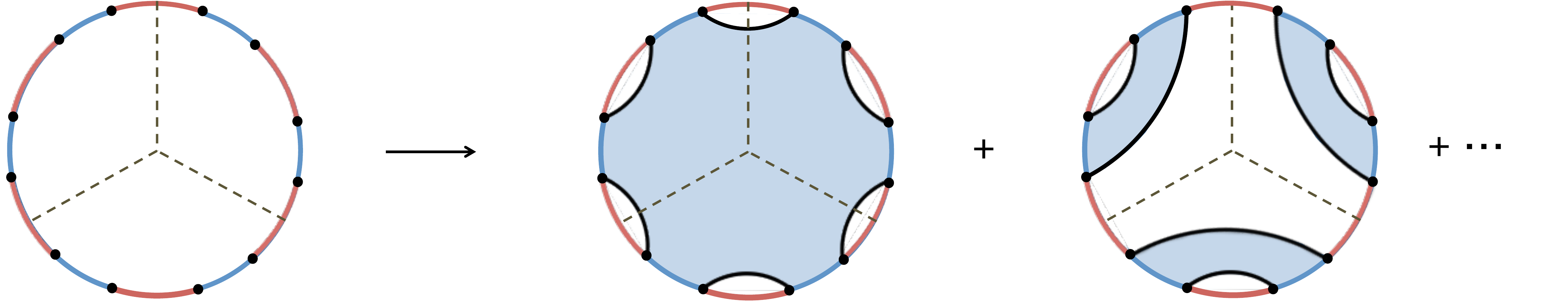}
\end{center}
\caption{ {\bf Left: } Quantum mechanical description of the R\'enyi entropy for the density matrix $\rho_{L} = \Tr _R \ket{\Psi_{12}}\bra{\Psi_{12}}$ with replica number $3$.
{\bf Middle:} Replica wormhole configuration with replica symmetry in 3d gravity description for CFT$_{LR}$.
{\bf Right:} Disconnected configuration with replica symmetry in 3d gravity description for CFT$_{LR}$.
\label{fig:Replica3}}
\end{figure}
We consider the $n$-th R\'enyi entropy.
In this case, we encounter the situation with $4 \times n = 4n$ domain walls in the Euclidean path integral.
This is an example of many coupled JT gravities that we discussed in section \ref{sec:manycoupled}.
Then we can evaluate the replica partition function using the JT gravity description.
We describe replica symmetric configurations in figure \ref{fig:Replica3}.
The first configuration in figure \ref{fig:Replica3} will be relevant to compute entanglement entropy between two closed universes, which gives the minimal one in the braket wormhole phase, see \eqref{eq:EEforCFT12}.
These configurations are replica wormholes for 3d gravities that are holographic duals of bulk matters.
The replicas of braket wormholes are connected in this configuration.
If we view these configurations as traversable wormholes, the 3d gravity parts are traversable wormholes with $2n$ boundaries.
Because CFT$_{LR}$ is entering on all 2d traversable wormholes, the matter stress tensor from CFT$_{LR}$ on each AdS$_2$ region is evaluated 
\be
\braket{T_{++}^{\text{AdS}_2}} = \f{c_{LR}}{48  \pi \ell ^2 } - \f{c_{LR}}{48 }\f{\pi}{n^2 (\pi \ell + d_{LR})^2}.
\ee
Therefore as we increase the replica number $n$, the energy from CFT$_{LR}$ increases and it approaches to $\f{c_{LR}}{48 \pi \ell^2}$.
However as we observed around \eqref{eq:Energy12InfL}, even in this limit the finite $\ell$ solution exists because in the braket wormhole phase the central charges satisfy $c_{LR} < c_{12}$ whereas the condition to have finite $\ell$ solution is $c_{LR} < 3 c_{12}$\footnote{Here the role of $c_{LR}$ and $c_{12}$ are exchanged compared to the notation around \eqref{eq:Energy12InfL}. }.
Therefore we expect that the replica wormhole solutions always exists for any R\'enyi index $n$.  
These configurations are interpreted as both of replica wormholes and traversable wormholes, which gives a connection between them.

On the other hand, the second configuration in figure \ref{fig:Replica3} will be relevant to compute entanglement entropy between a closed universe and CFT on a flat space without dynamical gravity.
Though we expect that there are solutions with finite wormhole length $\ell$ for $n \approx 1$, it is not apparent whether there are saddles with these configurations because CFT$_{12}$ no more introduces negative energy for generic replica number $n$.

We also expect that there are also replica non symmetric solutions generically.
We skip the full analysis of other saddles as a future problem.


\section*{Acknowledgements}
We thank Vijay Barasubramanian, Tomonori Ugajin, Gabor Sarosi, Arjun Kar, Simon Ross, Tomoki Nosaka,  Tadashi Takayanagi, Netta Engerhardt, Daniel Harlow and Shinsei Ryu for useful discussions.
T.Numasawa was supported by the Simons Foundation through the ``It from Qubit" Collaboration T.Numasawa is supported by a JSPS postdoctoral fellowship for research abroad .
Part of the numerical analyses in this work was carried out at the Yukawa Institute Computer Facility.

\appendix
\section{Some formulas for 2d CFT  \label{sec:2dCFTformula}}
In the Lorentzian signature, the metric is 
\be
ds^2 = -dt^2 + d\sigma^2 = - dy^+ dy^-.
\ee
where $y^{\pm} = t \pm \sigma$ is a null coordinate. 
The stress energy tensor is 
\be
T_{\mu\nu} = \begin{pmatrix} T_{tt} & T_{t\sigma} \\  T_{\sigma t} & T_{\sigma\sigma}  \end{pmatrix}
\ee
The energy measured in time $t$ is 
\be
E= \int  T_{tt} dx.
\ee
In the light cone coordinate $\omega^{\pm} = t \pm \sigma$, this becomes 
\be
\begin{pmatrix} T_{++} & T_{+-}\\  T_{-+} & T_{--}  \end{pmatrix}
= \f{1}{4}\begin{pmatrix}T_{tt} + 2 T_{t \sigma} + T_{\sigma \sigma} & T_{tt} - T_{\sigma \sigma} \\ T_{tt} - T_{\sigma \sigma}& T_{tt} - 2 T_{t \sigma} + T_{\sigma \sigma} \end{pmatrix}
\ee
Or inversely, 
\be
\begin{pmatrix} T_{tt} & T_{t\sigma}\\  T_{t\sigma} & T_{\sigma\sigma} \end{pmatrix}
= \begin{pmatrix}T_{++} + 2 T_{+-} + T_{--} & T_{++} - T_{--} \\ T_{++} - T_{--} & T_{++} - 2 T_{+-} + T_{--} \end{pmatrix}
\ee
When we perform a conformal transformation $\omega^{\pm} = \omega^{\pm}(y^{\pm})$, the stress tensor is transformed as 
\be
\Big(\f{d\omega^+}{d y^+}\Big)^2T^{(\omega)}_{++} =  T_{++}^{(y)} + \f{c}{24 \pi }\{ \omega^+ , y^+ \},
\ee
where $T^{(\omega)}$ is the stress tensor in the coordinate system $ds^2 = - d\omega ^+ d\omega^-$ whereas $T^{(y)}$ is the stress tensor in the coordinate system $ds^2 = - dy ^+ dy^-$.
Let us think of $\omega^{\pm}$ is a coordinate of Minkowski space and $y^{\pm}$ is that of a cylinder with the length $L$.

\begin{figure}[ht]
\begin{minipage}{0.49\hsize}
\begin{center}
\includegraphics[width=3.75cm]{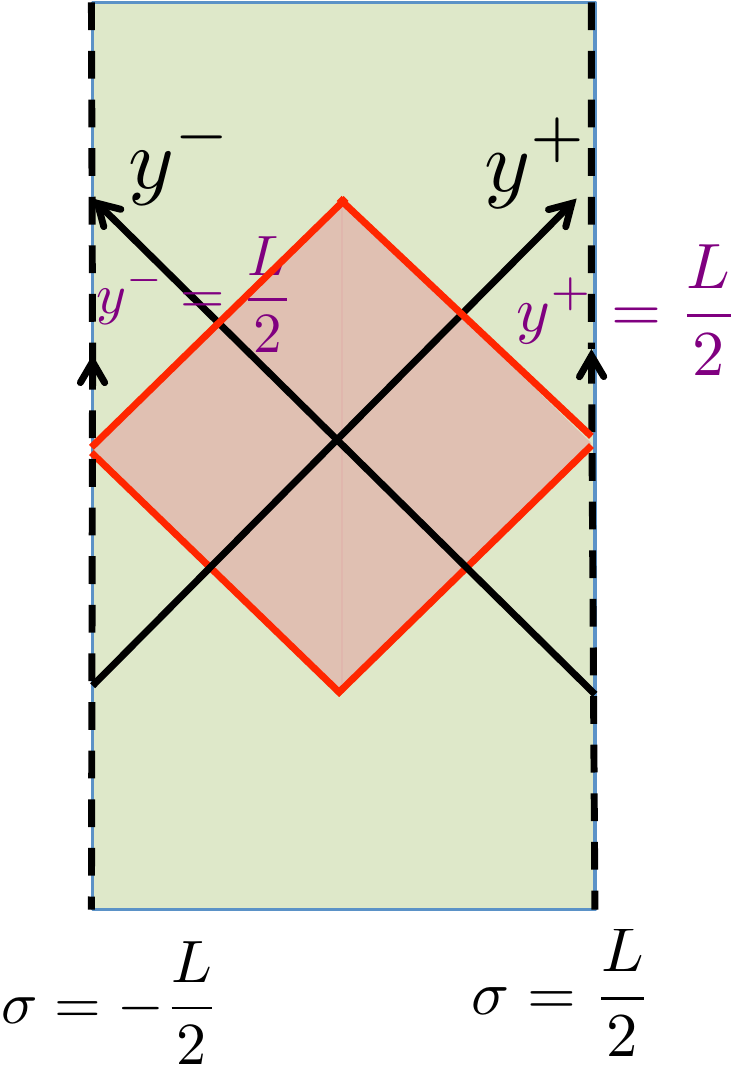}
\end{center}
\end{minipage}
\begin{minipage}{0.49\hsize}
\begin{center}
\includegraphics[width=2.5cm]{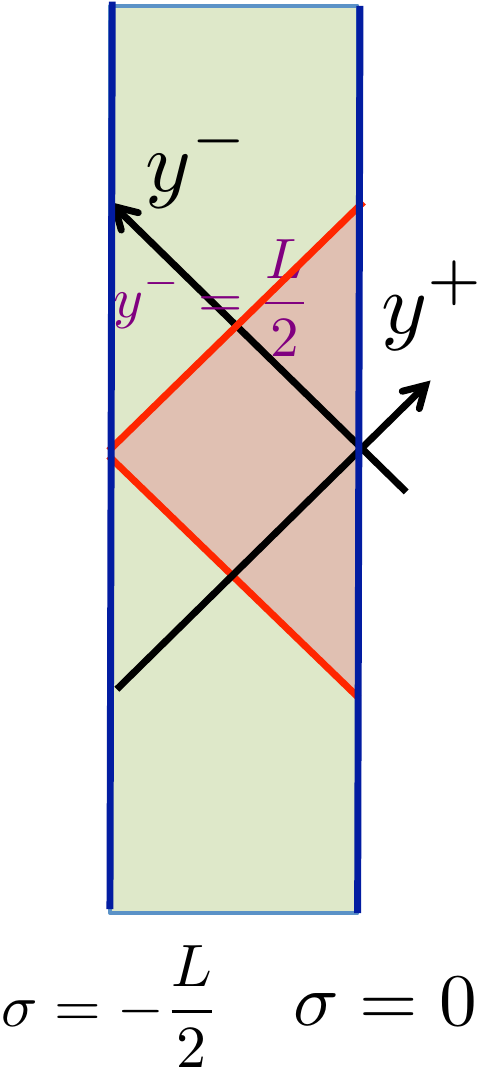}
\end{center}
\end{minipage}
\caption{{\bf Left: }The conformal transformation $\omega^{\pm}(y^{\pm})$ from the cylinder to the Minkowski space. 
The patch $ -\f{L}{2} \le y^{\pm} \le \f{L}{2}$ is mapped to the entire Minkowski space, which is the diamond surrounded by the red solid line.
{\bf Right:} The same conformal diagram with boundaries. The boundaries are located on  $\sigma = -\f{L}{2}$ and $\sigma=0$. 
} 
\label{fig:potential1}
\end{figure}

Then, the conformal transformation $\omega^{\pm}(y^{\pm})$ from the cylinder to the Minkowski space is 
\be
\omega^{\pm}(y^{\pm}) = \f{L}{\pi}\tan \Big(\f{\pi}{L} y^{\pm}\Big).
\ee
The coordinate transformation becomes 
\be
ds^2 = -d\omega^+ d\omega^- = -\f{dy^+dy^-}{\cos^2( \f{\pi}{L}y^+)\cos^2 (\f{\pi}{L}y^-)}
\ee
and the null boundary $y^{\pm} = \f{L}{2},-\f{L}{2}$ is located at infinity.
After the Weyl transformation $ds^2 \to d\tilde{s}^2 = \cos^2( \f{\pi}{L}y^+) \cos^2( \f{\pi}{L}y^-)$, the null boundary $y^{\pm} = \f{L}{2},-\f{L}{2}$ is located at finite distance from the interior and we can continue the geometry beyond them and obtain the cylinder with the metric $d\tilde{s}^2 = - dy^+ dy^- $.
Because on the Minkowski vacuum we do not have the stress tensor expectation value, we obtain the stress energy tensor expectation value on the cylinder with the length $L$:
\be
\braket{T_{++}^{(y)}} =  -\f{c}{24\pi} \{ \omega^+, y^+ \} = -\f{c}{24\pi} \f{2\pi^2}{L^2}.
\ee
Then, we obtain 
\be
T_{tt} = T_{\sigma\sigma} = T_{++}^{(y)} + T_{--}^{(y)} = -\f{c}{12} \f{2\pi}{L^2}.
\ee
Especially, the energy density with $L=2\pi$ is $T_{tt} = -\f{c}{24}$.
The ground state energy of the system is 
\be
E = \int _{0}^{L} T_{tt}d\sigma = -\f{c}{12} \f{2\pi}{L}.
\ee
When the period is $2\pi$, we obtain $E = -\f{c}{12}$.

We can treat boundary CFT in the same manner.
We consider the boundary CFT on a strip with the length $L_{\text{bdy}} = L/2 $.
The geometry is obtained by just cutting the cylinder along $\sigma= 0 , L/2$ and imposing a boundary condition.
We introduce the "same" boundary condition on two boundaries.
The same means that they are CPT conjugate with each others.
The formulas are the same that we obtained in the cylinder case.
The null energy on the strip is 
\be
\braket{T_{++}^{(y)}} =  -\f{c}{24\pi} \{ \omega^+, y^+ \} = -\f{c}{24\pi} \f{2\pi^2}{L^2} = -\f{c}{24\pi} \f{\pi^2}{2L_{\text{bdy}}^2} .
\ee
The energy density and the pressure is 
\be
T_{tt} = T_{\sigma\sigma} = T_{++}^{(y)} + T_{--}^{(y)} = -\f{c}{12} \f{2\pi}{L^2} = -\f{c}{24} \f{\pi}{L_{\text{bdy}}^2}.
\ee
Especially, the energy density for $L_{\text{bdy}} = \pi$ is $T_{tt} =- \f{c}{24\pi}$.
The ground state energy of the system is 
\be
E = \int _{0}^{L_{bdy}} T_{tt}d\sigma = -\f{c}{24} \f{\pi}{L_{\text{bdy}}}.
\ee

\subsection{CFT on AdS$_2$}
We now consider to put the boundary CFT on AdS$_2$.
To compute the stress tensor, we should care about the Weyl anomaly.
The stress tensor is given by
\be
\braket{T_{\mu\nu}^g} = i \f{2}{\s{-g}} \f{\delta}{\delta g^{\mu\nu}} \log Z[g].
\ee
For example, the free scaler case in classical limit, the partition function is 
\be
Z[g] =e^{iS[g]} =\exp \Big( -i\f{1}{2}\int dx^2 \s{-g} g^{\mu\nu}\partial _{\mu}\phi\partial _{\nu}\phi \Big).
\ee
The energy stress tensor is 
\be
T_{\mu\nu}^g = i \f{2}{\s{-g}} \f{\delta}{\delta g^{\mu\nu}} \log Z[g] = \partial_{\mu}\phi\partial_{\nu}\phi - \f{1}{2}g_{\mu\nu}g^{\alpha \beta}\partial_{\alpha}\phi\partial_{\beta}\phi.
\ee
The classical action is invariant under the Weyl transformation $g_{\mu\nu} \to e^{2\omega}g_{\mu\nu}$ since the factor $\s{-g} \to e^{2\omega} \s{-g} $ and $g^{\mu\nu} \to e^{-2\omega}g^{\mu\nu}$ cancel.
However the partition function transforms anomalously as 
\be
Z[g = e^{2\omega} \hat{g}] = \exp \Big\{ i \f{c}{24\pi} \int dx^2 \s{-\hat{g}} [\hat{R} + (\hat{\nabla} \omega)^2 \Big\} Z[\hat{g}]. 
\ee
Therefore, two stress tensors are related by \footnote{In the anomalous term, the first two terms take the form of scalar energy stress tensor and the last two terms is the stress energy tensor for dilaton in JT gravity.}
\ba
\braket{T_{\mu\nu}^g} &= i \f{2}{\s{-g}} \f{\delta}{\delta g^{\mu\nu}} \log Z[g]  \notag \\
&= i \f{2}{e^{2\omega}\s{-g}}\f{\delta}{\delta( e^{-2\omega} g^{\mu\nu})} \log \Big[ \exp \Big\{ i \f{c}{24\pi} \int dx^2 \s{-\hat{g}} [\hat{R} + (\hat{\nabla} \omega)^2 \Big\} Z[\hat{g}]\Big] \notag \\
&= \braket{T^{\hat{g}}_{\mu\nu}} -\f{c}{12 \pi} \Big[\partial_{\mu}\omega \partial_{\nu}\omega  - \f{1}{2} \hat{g}_{\mu\nu} (\hat{\nabla}\omega)^2 - \hat{\nabla}_{\mu}\hat{\nabla}_{\nu} \omega + \hat{g}_{\mu\nu} \hat{\nabla}^2 \omega\Big]
\ea
In AdS$_2$ case, the quantum stress tensor is related to the stress tensor on a strip by
\be
T_{\mu\nu}^{\text{AdS}_2} = \hat{T}_{\mu\nu}^{\text{strip}}  + \f{c}{24\pi} \begin{pmatrix} 1 & 0 \\0 & 1  \end{pmatrix}_{\mu\nu} - \f{c}{24\pi} g_{\mu\nu}. \label{eq:StressTensorAdS2}
\ee
Here the second term means $T_{++} = T_{--} = \f{c}{24\pi}$ and $T_{+-} = T_{-+} = 0$

\subsection{Holographic CFT on AdS$_2$}

We consider holographic CFTs partially on AdS$_2$ for the region $[0,\pi]$ of $[0, \pi(1+a)]$.
To put the theory on the length $\pi(1+a)$, it is convenient to rescale the global AdS$_3$ metric
\be
ds^2 = l_{\text{AdS}_3}^2 \Big[ -\f{4}{(1+a)^2} \cosh^2 \rho dt^2 + d\rho^2 + \f{4}{(1+a)^2}\sinh^2 \rho d\phi^2 \Big].
\ee
Then, for $0 < \phi < \pi$, the UV cutoff surface is given by the condition
\be
e^{-\rho(t,\phi)} = \f{ \epsilon}{1+a} \sin \phi .
\ee
The induced metric on the cutoff surface is 
\ba
ds_{\text{ind}}^2 &=  h_{\mu\nu} dx^{\mu}dx^{\nu} \notag \\
&= - \f{l_{\text{AdS}_3}^2}{\epsilon^2} \Big( \f{1}{\sin \phi} + \f{\epsilon^2}{(1+a)^2} \sin \phi \Big)^2 dt^2 +l_{\text{AdS}_3}^2 \Big[ \f{1}{\tan^2 \phi} +\f{1}{\epsilon^2} \Big( \f{1}{\sin \phi} - \f{\epsilon^2}{(1+a)^2} \sin \phi \Big)^2 \Big] d\phi^2  \notag \\
&= \f{l_{\text{AdS}_3}^2}{\epsilon^2} \Big( \f{-dt^2 + d\phi^2}{\sin^2 \phi} \Big) + \cdots = l_{\text{AdS}_3}^2\f{ds_{\text{AdS}_2}^2}{\epsilon^2} + \cdots .
\ea
The normal vector $n_{A}$ is given by
\be
n_{t} = 0, \qquad n_{\rho} = \f{2l_{\text{AdS}_3}}{\s{4 + \f{(1+a)^2}{\tan^2\phi \cosh^2 \rho}}}, \qquad n_{\phi} = \f{2l_{\text{AdS}_3}}{\tan\phi\s{4 + \f{(1+a)^2}{\tan^2\phi \cosh^2 \rho}}}
\ee
Then, we can calculate $K_{\mu\nu} = P_{\mu}{}^AP_{\nu}{}^B\nabla_{A} n_{B}$ where $P_{\mu}{}^A$ is the projection on to the cutoff surface.
$K_{\mu\nu}$ in $(t,\phi)$ coordinate is 
\ba
K_{tt} &= - l_{\text{AdS}_3} \Big[\f{1}{\epsilon^2 \sin^2 \phi} - \f{1}{2\tan^2{\phi}} \Big] + \mathcal{O}(\epsilon), \notag \\
K_{\phi\phi} &= l_{\text{AdS}_3} \Big[ \f{1}{\epsilon^2 \sin^2 \phi} - 1 +  \f{1}{2\tan^2{\phi}} \Big] + \mathcal{O}(\epsilon) \notag \\
K_{ t\phi} &= K_{ \phi t} = 0.
\ea
and 
\be
K = K_{\mu\nu}h^{\mu\nu} =  \f{1}{ l_{\text{AdS}_3}} \Big(2 - \epsilon^2 \Big) + \mathcal{O}(\epsilon^3).
\ee
The induced metric is expanded as 
\ba
h_{tt}&= l_{\text{AdS}_3}^2\Big[-\f{1}{\epsilon^2 \sin^2 \phi} - \f{2}{(1+a)^2} \Big] + \mathcal{O}(\epsilon) \notag \\
h_{\phi\phi} &= l_{\text{AdS}_3}^2\Big[\f{1}{\epsilon^2 \sin^2 \phi} - \f{2}{(1+a)^2} + \f{1}{\tan^2 \phi}  \Big]+ \mathcal{O}(\epsilon) \notag \\
h_{t\phi} &= h_{\phi t} = 0.
\ea
Therefore, the holographic stress energy tensor \cite{Balasubramanian:1999re} is 
\ba
T_{\mu\nu}dx^{\mu}dx^{\nu} &= -\f{1}{8\pi G_N} \Big( K_{\mu\nu} - K h_{\mu\nu} + \f{1}{l_{\text{AdS}_3}} h_{\mu\nu}  \Big)dx^{\mu}dx^{\nu} \notag \\
&= -\f{l_{\text{AdS}_3}}{8\pi G_N}\Big[ \Big(\f{1}{2\sin^2 \phi} +\f{1}{2} - \f{2}{(1+a)^2} \Big) dt^2 +\Big(-\f{1}{2\sin^2 \phi}+  \f{1}{2} - \f{2}{(1+a)^2} \Big) d\phi^2   \Big] \notag \\
&= -\f{c}{24\pi} \f{(1-a)(a+3)}{(1+a)^2} (dt^2 + d\phi^2) - \f{c}{24\pi} \f{1}{\sin^2\phi} (-dt^2 + d\phi^2) \notag \\
& = -\f{c}{48\pi} \f{(1-a)(a+3)}{(1+a)^2} ((dx^+)^2 +(dx^-)^2 ) - \f{c}{24\pi} g^{\text{AdS}_2}_{\mu\nu}dx^\mu dx^\nu .  
\ea
Here we used the central charges for holographic CFTs $c = \f{3l_{\text{AdS}_3} }{2G_N}$.
This reproduce the stress energy tensor from conformal anomaly argument \eqref{eq:StressTensorAdS2}.
Note that when $a = 1$, the stress tensor vanishes.

\subsection{Some formula for entanglement entropy}
First we consider the system on a infinite line.
Entanglement entropy for single interval $[0,x]$ is 
\be
S_A = \f{c}{3} \log \f{x}{\epsilon} + c_1', \label{eq:EEinterval1}
\ee
where $\epsilon$ is a UV cutoff and $c_1'$ is a non universal constant.

Entanglement entropy for single interval $[0,x]$ on a circle of the length $L$ is 
\be
S_A = \f{c}{3} \log \Big( \f{L}{\pi \epsilon} \sin \f{\pi x}{L} \Big) + c_1'. 
\ee
In free Dirac fermion case, entanglement entropy for two intervals of lengths $l_1,l_2$ with the distance $D$ (i.e. intervals $[-l_1,0]\cup [D,D+l_2]$) between them \cite{Numasawa:2016emc,Caputa:2019avh} is \footnote{When we write the interval as $[x_1,x_2] \cup [x_3,x_4]$, entanglement entropy becomes 
\be
S_A = \f{c}{3} \log \Big(\f{|x_{12}| |x_{34}|}{\epsilon^2}\f{|x_{23}||x_{14}|}{|x_{13}| |x_{24}|} \Big) + 2 c_1 '.
\ee}
\be
S_A = \f{c}{3} \log \Big(\f{l_1 l_2}{\epsilon}\f{D (l_1 + l_2 + D)}{(l_1+D)(l_2 + D)} \Big) + 2 c_1 '. \label{eq:TwoIntervalonLine}
\ee
If the system is in a finite cylinder of the length $L$, entanglement entropy becomes 
\be
S_A = \f{c}{3} \log \Bigg[\Big(\f{L}{\pi \epsilon}\Big)^2\f{\sin\f{\pi l_1}{L}\sin\f{\pi l_2}{L}\sin\f{\pi D}{L}\sin\f{\pi (l_1+l_2+D)}{L}}{\sin\f{\pi (l_1+D)}{L}\sin\f{\pi (l_2+D)}{L}} \Bigg]+ 2 c_1 '. \label{eq:TwoIntervalonCircle}
\ee
Note that in $L\to \infty$ limit entanglement entropy on a circle \eqref{eq:TwoIntervalonCircle} recovers the that on an infinite line \eqref{eq:TwoIntervalonLine}.

In CFT with holographic dual, entanglement entropy for single interval is given by \cite{Headrick:2010zt,Hartman:2013mia}
\be
S_A = \text{min}\{S_1,  S_2 \}.
\ee
where $S_{1}$ and $S_{2}$ are given by
\be
S_{1} = \f{c}{3} \log \f{l_1}{\epsilon} + \f{c}{3} \log \f{l_2}{\epsilon}, \qquad S_{2} =\f{c}{3} \log \f{D}{\epsilon} + \f{c}{3} \log \f{D+l_1+l_2}{\epsilon}.
\ee
For $S_1$, each geodesic connects the end points of each interval whereas  in $S_2$ two geodesics connects the end points of different intervals.
On a circle, entanglement entropy for two intervals is given by 
\be
S_A = \text{min}\{S_1,  S_2 \}.
\ee
with
\ba
S_{1} &= \f{c}{3} \log \bigg[\Big(\f{ L}{\pi \epsilon} \Big)  \sin \f{\pi l_1}{L}\bigg] +\f{c}{3} \log \bigg[\Big(\f{ L}{\pi \epsilon} \Big)  \sin \f{\pi l_2}{L}\bigg], \notag \\
S_{2} &=\f{c}{3} \log \bigg[\Big(\f{ L}{\pi \epsilon} \Big)  \sin \f{\pi D}{L}\bigg]+ \f{c}{3} \log \bigg[\Big(\f{ L}{\pi \epsilon} \Big) \sin\f{\pi (D+l_1+l_2)}{L}\bigg].
\ea

In general CFT, we can express entanglement entropy for two intervals as 
\be
S_A = \f{c}{3} \log \Big(\f{l_1 l_2 \eta}{\epsilon^2}  \Big) + \log F(\eta)  + 2 c_1 ', \qquad \eta = \f{D(l_1+l_2+D)}{(l_1+D)(l_2+D)}.
\ee
with  theory dependent function $F(\eta)$ which always satisfies $F(1) =1$ and $F(0) =1$.
On a circle, entanglement entropy becomes 
\be
S_A = \f{c}{3} \log \bigg[\Big(\f{L}{\pi \epsilon}\Big)^2  \sin \Big(\f{\pi l_1 }{L}\Big)\sin \Big( \f{\pi l_2 }{L}\Big)\eta \bigg]+ \log F(\eta)  + 2 c_1 ', \qquad \eta = \f{\sin\f{\pi D}{L}\sin\f{\pi (l_1+l_2+D)}{L}}{\sin\f{\pi (l_1+D)}{L}\sin\f{\pi (l_2+D)}{L}} .
\ee

Next we consider CFT on a manifold with boundaries on which we impose  a conformal boundary condition.
When system is put on a semi infinite line $[0,\infty )$, entanglement entropy for an interval $[0,x]$ is 
\be
S_A = \f{c}{6}\log \f{2 x }{\epsilon} + \log g+ \f{1}{2}c_1' 
\ee
where $c_1'$ is the same non universal constant in (\ref{eq:EEinterval1}) and   $\log g$ is the boundary entropy that depends on the choice of boundary conditions.
Entanglement entropy between the interval $[0,x]$ on an interval  of the length $L$ with boundaries at $0, L$ is 
\be
S_A = \f{c}{6} \log \Big( \f{2L}{\pi \epsilon} \sin \f{\pi x}{L} \Big) + \f{1}{2}c_1' + \log g.
\ee
In free Dirac fermions, entanglement entropy for single interval of length $l$ with the distance $D$ from the boundary at $x =0$ (i.e. intervals $[D,l + D]$ in a half line $[0,\infty)$) becomes 
\be
S_A = \f{c}{6} \log \Big(\f{l^2}{\epsilon^2}\f{4D(l + D)}{(l + 2D)^2} \Big) +  c_1 '.
\ee
On a finite interval $[0,L]$ with the same boundary condition on each boundary, entanglement entropy becomes 
\be
S_A = \f{c}{6} \log\bigg[ \Big(\f{L}{\pi \epsilon}\Big)^2\f{\sin ^2 \f{\pi l}{L} \sin \f{2\pi D}{L} \sin \f{2\pi (l + D)}{L}}{ \sin^2 \f{\pi (l + 2D)}{L}}  \bigg] + c_1 '.
\ee

In AdS/BCFT models \cite{Takayanagi:2011zk,Fujita:2011fp}, entanglement entropy for single interval is given by
\be
S_A = \text{min}\{S_{con},  S_{dis}  \}.
\ee
where $S_{con}$ and $S_{dis}$ are given by
\be
S_{con} = \f{c}{3} \log \f{\ell}{\epsilon}, \qquad S_{dis} = \f{c}{6} \log \f{2D}{\epsilon} + \f{c}{6} \log \f{2\ell}{\epsilon} + 2 \log g.
\ee
$S_{con}$ is the length of the geodesics that connects two end points, whereas  $S_{dis}$ is the sum of two  disconnected  geodesics that end on the end of the world brane.
On a finite interval, holographic entanglement entropy for single interval becomes 
\be
S_A = \text{min}\{S_{con},  S_{dis}  \}.
\ee
with 
\be
S_{con} = \f{c}{3} \log \bigg[\f{2L}{\pi \epsilon} \sin \f{\pi \ell}{2L}\bigg], \qquad S_{dis} = \f{c}{6} \log \bigg[\f{2L}{\pi \epsilon} \sin \f{\pi D}{L}\bigg] + \f{c}{6} \log \bigg[\f{2L}{\pi \epsilon} \sin \f{\pi (l +D)}{L}\bigg] + 2 \log g.
\ee
For general BCFT, we can express entanglement entropy for two intervals as 
\be
S_A = \f{c}{3} \log \Big(\f{l^2 \eta}{\epsilon^2}  \Big) + \log G(\eta)  +  c_1 ', \qquad \eta = \f{4D(l+D)}{(l+2D)^2}.
\ee
with  theory dependent function $G(\eta)$ which always satisfies $G(1) =1$ and $G(0) = g^2$.
On a circle, entanglement entropy becomes 
\be
S_A = \f{c}{6} \log \bigg[\Big(\f{2L}{\pi \epsilon}\Big)^2  \sin^2\Big(\f{\pi l }{2L}\Big) \eta \bigg]+ \log G(\eta)  + 2 c_1 ', \qquad \eta = \f{\sin\f{\pi D}{L}\sin\f{\pi (l+D)}{L}}{\sin^2\f{\pi (l+2D)}{2L}} .
\ee

\bibliography{4coupledSYK.bib}

\end{document}